\documentclass[12pt, draftclsnofoot, onecolumn]{IEEEtran}
\usepackage{graphicx}
\usepackage{amsmath}
\usepackage{amssymb}
\usepackage{mathrsfs}
\usepackage{stfloats}
\usepackage{dsfont}
\usepackage{setspace}
\usepackage{array}
\usepackage{bbm}
\usepackage[centerlast,small]{caption}
\newtheorem{corollary}{Corollary}
\newtheorem{proposition}{Proposition}
\newtheorem{lemma}{Lemma}
\newtheorem{remark}{Remark}

\newtheorem{approximation}{Approximation}
\begin{document}
%
%
%
%\input COVER.tex
%\newpage
%
%
%
\title{\Large The Intensity Matching Approach: A Tractable Stochastic Geometry Approximation to System-Level Analysis of Cellular Networks}
\author{\normalsize Marco~Di~Renzo,~\IEEEmembership{\normalsize Senior~Member,~IEEE},
\normalsize Wei~Lu,~\IEEEmembership{\normalsize Student~Member,~IEEE}, and
\normalsize Peng~Guan,~\IEEEmembership{\normalsize Student~Member,~IEEE}
\thanks{Manuscript received December 13, 2015; revised April 10, 2016. The authors are with Paris-Saclay University (L2S - CNRS, CentraleSup\'elec, Univ Paris Sud), Paris, France. (e-mail: marco.direnzo@l2s.centralesupelec.fr).}}
%This work was supported in part by the European Commission through the FP7-PEOPLE MITN-CROSSFIRE Project under Grant 317126 and through the H2020-PEOPLE ETN-5Gwireless Project under Grant 641985.
%\thanks{Manuscript received October 17, 2015. The authors are with the Laboratoire des Signaux et Syst\`emes, CNRS, CentraleSup\'elec, Univ Paris Sud, Universit\'e Paris-Saclay, 3 rue Joliot Curie, Plateau du Moulon, 91192, Gif-sur-Yvette, France. (e-mail: marco.direnzo@l2s.centralesupelec.fr).}}
%This work was supported in part by the European Commission through the FP7-PEOPLE MITN-CROSSFIRE Project under Grant 317126 and through the H2020-PEOPLE ETN-5Gwireless Project under Grant 641985.
%
%
\markboth{Submitted for Journal Publication} {M. Di Renzo, W. Lu, and P. Guan: The Intensity Matching Approach: A Tractable Stochastic Geometry Approximation to System-Level Analysis of Cellular Networks}
\maketitle
\vspace{-1.75cm}
\begin{abstract}
\vspace{-0.25cm}
The intensity matching approach for tractable performance evaluation and optimization of cellular networks is introduced. It assumes that the base stations are modeled as points of a Poisson point process and leverages stochastic geometry for system-level analysis. Its rationale relies on observing that system-level performance is determined by the intensity measure of transformations of the underlaying spatial Poisson point process. By approximating the original system model with a simplified one, whose performance is determined by a mathematically convenient intensity measure, tractable yet accurate integral expressions for computing area spectral efficiency and potential throughput are provided. The considered system model accounts for many practical aspects that, for tractability, are typically neglected, \textit{e.g.}, line-of-sight and non-line-of-sight propagation, antenna radiation patterns, traffic load, practical cell associations, general fading channels. The proposed approach, more importantly, is conveniently formulated for unveiling the impact of several system parameters, \textit{e.g.}, the density of base stations and blockages. The effectiveness of this novel and general methodology is validated with the aid of empirical data for the locations of base stations and for the footprints of buildings in dense urban environments.
\end{abstract}
\vspace{-0.25cm}
\begin{IEEEkeywords}
\vspace{-0.25cm}
Ultra-Dense Cellular Networks, Poisson Point Processes, Stochastic Geometry.
\end{IEEEkeywords}
\vspace{-0.25cm}
\section{Introduction} \label{Introduction}
\vspace{-0.25cm}
In the last few years, stochastic geometry has been widely used for system-level modeling, performance evaluation, and optimization of several candidate system architectures, network topologies, and transmission technologies for next-generation cellular networks \cite{AndrewsNov2011}. At present, many tractable mathematical methodologies for analyzing and optimizing (heterogeneous) cellular networks in terms of average rate \cite{MDR_TCOMrate}, coverage \cite{MDR_COMMLPeng} and error probability \cite{MDR_COMML2014}, \cite{Wei_TCOM} exist.

The tractability of currently available methodologies originates from two main assumptions: 1) the network elements are modeled as points of a Poisson Point Process (PPP) and 2) the path-loss, as a function of the distance, is modeled as a power-decaying function with distance-independent parameters \cite{Blaszczyszyn_Infocom2013}. Recent studies based on actual cellular network deployments and building footprints have unveiled that the PPP-based assumption is sufficiently accurate for modeling dense urban environments, \textit{e.g.}, downtown London \cite{ACM_MDR}. They have disclosed, on the other hand, the crucial impact of the path-loss model for system-level analysis and optimization. In particular, coverage and rate estimated by relying on the conventional power-decaying path-loss model are qualitatively and quantitatively different compared with those obtained by using more realistic path-loss models, which, \textit{e.g.}, originate from channel measurements and/or are recommended by standardization working groups for evaluating and comparing different wireless access technologies \cite{3GPP_pathloss}. They have revealed, in addition, the need of taking the radiation pattern of transmit and receive antennas into account, in order to adequately quantify the impact of the other-cell interference and, hence, of network densification, \textit{i.e.}, small cell technology.

Motivated by these considerations, a few researchers have recently generalized the PPP-based approach for modeling cellular networks \cite{AndrewsNov2011}, by assuming more realistic path-loss models \cite{Andrews__LosNlos}-\cite{Lopez-Perez__LosNlos}. These papers have unveiled, \textit{e.g.}, that the impact of network densification depends on the path-loss model being adopted. In \cite{Andrews__LosNlos}, the authors employ a two-slope path-loss model and show that an optimal density of Base Stations (BSs) exists. This finding is in contradiction with \cite{AndrewsNov2011}, which, by assuming a power-decaying path-loss model, proved the density-invariance of interference-limited cellular networks. In \cite{MDR_mmWave}, the author uses a three-state path-loss model that is empirically derived from channel measurements conducted in New York City for transmission in the millimeter wave band. The path-loss model accounts for Line-Of-Sight (LOS), Non-Line-Of-Sight (NLOS), and outage links, whose probability of occurrence is distance-dependent. It is proved that coverage and rate depend on the density of BSs. In \cite{Galiotto__LosNlos} and \cite{Lopez-Perez__LosNlos}, the authors employ a path-loss model that accounts for LOS and NLOS links, whose probability of occurrence is an exponential and a linear function of the distance, respectively. Similar to \cite{Andrews__LosNlos}, it is shown that the density-invariance property does not hold anymore. The authors of \cite{Galiotto__LosNlos} unveil, however, that the impact of network densification depends on the load model being considered: if the densities of Mobile Terminals (MTs) and BSs do not scale at the same rate (\textit{i.e.}, full traffic load), small cell deployments provide better performance compared with the predictions in \cite{Andrews__LosNlos} and \cite{Lopez-Perez__LosNlos}.

The discoveries in \cite{ACM_MDR}, \cite{Andrews__LosNlos}-\cite{Lopez-Perez__LosNlos} bring to light the need of more realistic modeling assumptions in stochastic geometry analysis of cellular networks. They, however, still rely on important simplifications, as well as introduce mathematical methodologies that, although computationally affordable in many cases, lack tractability for general cellular setups. In \cite{Andrews__LosNlos}, no LOS and NLOS links are considered. In \cite{Galiotto__LosNlos}, \cite{Lopez-Perez__LosNlos}, simplified link-state models are assumed. In \cite{Andrews__LosNlos}, \cite{Galiotto__LosNlos}, \cite{Lopez-Perez__LosNlos}, no directional antennas and shadowing for cell association are taken into account, Rayleigh fading for all links is considered, saturated traffic load (except \cite{Galiotto__LosNlos}) is assumed. In spite of that, the frameworks are still formulated in terms of multi-fold integrals, which do not provide direct insight on the impact of key system parameters, \textit{e.g.}, the density of BSs and blockages.

\begin{table}[!t] \footnotesize
\centering
\caption{Summary of main symbols and functions used throughout the paper. \vspace{-0.25cm}}
\newcommand{\tabincell}[2]{\begin{tabular}{@{}#1@{}}#2\end{tabular}}
\begin{tabular}{|l||l|} \hline
\hspace{0.75cm} Symbol/Function & \hspace{4.5cm} Definition \\ \hline \hline
$\mathbb{E}\{\cdot\}$, $\Pr \left\{  \cdot  \right\}$ & Expectation operator, probability measure \\ \hline
${\mathop{\rm Im}\nolimits} \left\{  \cdot  \right\}$, $\mathbbm{j}$ & Imaginary part operator, imaginary unit \\ \hline
$\lambda_{\rm{BS}}$, $\lambda_{\rm{MT}}$ & Density of base stations, mobile terminals \\ \hline
$\Psi_{\rm{BS}}$, $\Psi_{\rm{MT}}$, $\Psi_{\rm{BS}}^{\left( \rm{I} \right)}$ & Poisson point process of base stations, mobile terminals, interfering base stations \\ \hline
$N_{\rm{RB}}$ & Number of resource blocks \\ \hline
$P_{{\rm{BS}}}$, $P_{{\rm{RB}}}$ & Transmit power of base stations, per resource block \\ \hline
$p_{\rm LOS}(\cdot)$, $p_{\rm NLOS}(\cdot)$, $p_{\rm OUT}(\cdot)$ & Probability of line-of-sight, non-line-of-sight, outage \\ \hline
$D_b$, $\mathcal{B}$ & Radius of the $b$th ball of the channel model, number of balls \\ \hline
${q_s^{\left[ {D_{b - 1} ,D_b } \right]} }$ & Link state probability of state $s$ in $\left[ {D_{b - 1} ,D_b } \right)$ \\ \hline
$l_s \left( \cdot \right)$, $\mathcal{X}_s$, $g_s$ & Path-loss, shadowing, fading power gain \\ \hline
$L_s^{(n)}$, $L^{(0)}$ & Inverse average received power of the $n$th link of state $s$, of the intended link \\ \hline
$p_{{\rm{sel}}} \left(  \cdot  \right)$ & Probability that a mobile terminal is scheduled for transmission \\ \hline
$p_{{\rm{off}}} \left(  \cdot  \right)$ & Probability that a base stations is not activated \\ \hline
$G_{\rm{BS}} \left( {\cdot } \right)$, $G_{\rm{MT}} \left( {\cdot } \right)$ & Antenna radiation pattern of base stations, mobile terminals \\ \hline
$G^{(0)}$ & End-to-end antenna gain of the intended link \\ \hline
${\rm K}_q$, ${\gamma_q^{\left( l \right)} }$, $\varphi _q^{\left( {l - 1} \right)}$ & Number of lobes, gain, phase of the antenna radiation pattern \\ \hline
$\sigma_N^2$, $I_{{\rm{agg}}} \left(  \cdot  \right)$ & Noise variance, aggregate other-cell interference \\ \hline
$\Phi_s$ & Poisson point process of the path-loss of state $s$ \\ \hline
$\Lambda _{\Phi _s } \left( {\left[ { \cdot , \cdot } \right)} \right)$, $\Lambda _{\Phi _s }^{\left( 1 \right)} \left( {\left[ { \cdot , \cdot } \right)} \right)$ & Intensity measure of the point process of the path-loss, its first derivative \\ \hline
$f_{X}(\cdot)$, $\mathcal{M}_X(\cdot)$ & Probability density function, moment generating function of random variable $X$ \\ \hline
$\mathbbm{1}_{\left[ {x,y} \right]} \left( \cdot \right)$ or $\mathbbm{1}\left( \cdot \right)$ & Indicator function \\ \hline
${}_2F_1 \left( { \cdot , \cdot , \cdot , \cdot } \right)$ &  Gauss hypergeometric function \\ \hline
$\delta(\cdot)$, $\Gamma(\cdot)$ & Dirac delta function, gamma function \\ \hline
$\left\|  \cdot  \right\|_F^2$ & Frobenius norm \\ \hline
$\mathcal{H}(\cdot)$, $\overline {\mathcal{H}}$ & Heaviside function, complementary Heaviside function \\ \hline
$\mathcal{A}_s$ or $\Upsilon _s \left(  \cdot  \right)$ & Probability of being in state $s$ \\ \hline
SINR, ASE, PT & Signal-to-interference+noise-ratio, area spectral efficiency, potential throughput \\ \hline
$\mathcal{R}$, $\mathcal{C}(\cdot)$ & Shannon rate, coverage probability \\ \hline
\end{tabular}
\label{Table_Notation} \vspace{-0.65cm}
\end{table}
In the present paper, a novel methodology to leverage stochastic geometry for modeling, evaluating, and optimizing cellular networks in a tractable yet accurate manner is introduced. The proposed approach accounts for several important aspects that are overlooked in previous works \cite{Andrews__LosNlos}-\cite{Lopez-Perez__LosNlos}, and, more importantly, it provides direct insight on the impact of key system parameters. For example, it allows us to prove that a local optimum for the density of BSs exists and that it depends on the density of blockages. Notably, we introduce a new and mathematically tractable link state model and prove, with the aid of empirical data, that it is flexible enough for approximating several link state models widely adopted in the literature. We propose, in addition, a general approach for estimating the parameters of the new link state model in order to closely match empirical propagation and blockage models. Compared with other mathematical approaches currently available in the literature, it leads to a simpler yet accurate mathematical formulation of key performance indicators for cellular network design, as well as direct insight on the impact of several system parameters. The details of the proposed approach and the complete set of design guidelines that emerge from it are discussed in Sections \ref{IntensityMatching} and \ref{Trends}.

The remainder of the present paper is organized as follows. In Section \ref{SystemModel}, the system model is summarized. In Section \ref{IntensityMatching}, the proposed Intensity Matching (IM)-based methodology is introduced and its rationale is discussed. In Section \ref{Performance}, the mathematical frameworks for computing Area Spectral Efficiency (ASE) and Potential Throughput (PT) are reported. In Section \ref{Trends}, performance trends and design guidelines for system optimization are elaborated. In Section \ref{Results}, the IM-based approach is substantiated with the aid of empirical data for the locations of BSs and for the footprints of buildings \cite{ACM_MDR}. Finally, Section \ref{Conclusion} concludes this paper.

\textit{Notation}: For the convenience of the readers, a summary of the main symbols and functions used throughout the present paper is provided in Table \ref{Table_Notation}.
\vspace{-0.5cm}
\section{System Model} \label{SystemModel} \vspace{-0.25cm}
\subsection{PPP-Based Abstraction Modeling} \label{PPP_CellularModeling} \vspace{-0.25cm}
A downlink (single-tier) cellular network is considered. The BSs are modeled as points of a homogeneous PPP, denoted by $\Psi_{\rm{BS}}$, of density $\lambda_{\rm{BS}}$. The MTs are modeled as another homogeneous PPP, denoted by $\Psi_{\rm{MT}}$, of density $\lambda_{\rm{MT}}$. $\Psi_{\rm{BS}}$ and $\Psi_{\rm{MT}}$ are independent. Each BS has $N_{\rm{RB}}$ orthogonal Resource Blocks (RBs) for serving the MTs, \textit{i.e.}, $N_{\rm{RB}}$ MTs can be served, at most, by any BSs without intra-cell interference. Each BS transmits with constant power in each RB. Let $P_{\rm{BS}}$ be the power budget of each BS and $P_{\rm{RB}}$ be the transmit power per RB. $P_{\rm{BS}}$ is equally distributed among the RBs, \textit{i.e.}, $P_{{\rm{RB}}}  = {{P_{{\rm{BS}}} } \mathord{\left/ {\vphantom {{P_{{\rm{BS}}} } {N_{{\rm{RB}}} }}} \right. \kern-\nulldelimiterspace} {N_{{\rm{RB}}} }}$, regardless of the number of RBs that are actually used by each BS. Generalizations where the BSs transmit with unequal power are left to future research. The mathematical frameworks are developed for the typical MT, denoted by ${\rm{MT}}^{\left( 0 \right)}$, that is located at the origin (Slivnyak theorem \cite[Th. 1.4.5]{BaccelliBook2009}). The BS serving ${\rm{MT}}^{\left( 0 \right)}$ is denoted by ${\rm{BS}}^{\left( 0 \right)}$ and the set of interfering BSs on a RB is denoted by $\Psi_{\rm{BS}} ^{\left( \rm{I} \right)}$.
\begin{table}[!t] \footnotesize
\centering
\caption{Widely used link state models. The parameters $a_{(\cdot)}$, $b_{(\cdot)}$ and $c_{(\cdot)}$ are environmental-dependent. \vspace{-0.25cm}}
\newcommand{\tabincell}[2]{\begin{tabular}{@{}#1@{}}#2\end{tabular}}
\begin{tabular}{|c||c|c|c|} \hline
 & $p_{\rm LOS}(r)$& $p_{\rm NLOS}(r)$& $p_{\rm OUT}(r)$\\ \hline \hline
 \tabincell{c}{3GPP \cite{3GPP_pathloss}} &$\min \left\{ {\frac{{a_{\rm{3G}}}}{r},c_{\rm{3G}}} \right\}\left( {1 - {e^{ - \frac{r}{{b_{\rm{3G}}}}}}} \right) + {e^{ - \frac{r}{{b_{\rm{3G}}}}}}$ &$1-p_{\rm LOS}(r)$& 0\\ \hline
\tabincell{c}{Random Shape \cite{Heath__Blockage}}&${{a_{\rm RS}}\exp \left( { - {b_{\rm RS}}r} \right)}$&$1-p_{\rm LOS}(r)$& 0\\ \hline
\tabincell{c}{Linear \cite{Lopez-Perez__LosNlos}}& $1-p_{\rm NLOS}(r)$&$\min \left\{ {{a_{\rm L}}r + {b_{\rm L}},{c_{\rm L}}} \right\}$& 0\\ \hline
\tabincell{c}{Empirical mmWave \cite{MDR_mmWave}}&$\left( {1 - {p_{{\rm{OUT}}}}\left( r \right)} \right){e^{ - {a_{{\rm{mm}}}}r}}$&$1-p_{\rm LOS}(r)-p_{\rm OUT}(r)$&$\max \left\{ {0,1 - {e^{ - {b_{{\rm{mm}}}}r + {c_{{\rm{mm}}}}}}} \right\}$\\ \hline
\tabincell{c}{Two-ball mmWave \cite{MDR_mmWave}} & \multicolumn{3}{c|}{{\rm{see \eqref{Eq_1} with}} $\mathcal{S}=3$, $s = \left\{ {{\rm{LOS}},{\rm{NLOS}},{\rm{OUT}}} \right\}$, $\mathcal{B}=2$}\\ \hline
\end{tabular}
\label{LinkStateModels} \vspace{-0.25cm}
\end{table}
\vspace{-0.5cm}
\subsection{Link State Modeling} \label{LinkStateModeling} 
\begin{figure}[!t]
\centering
\includegraphics[width=0.9\columnwidth]{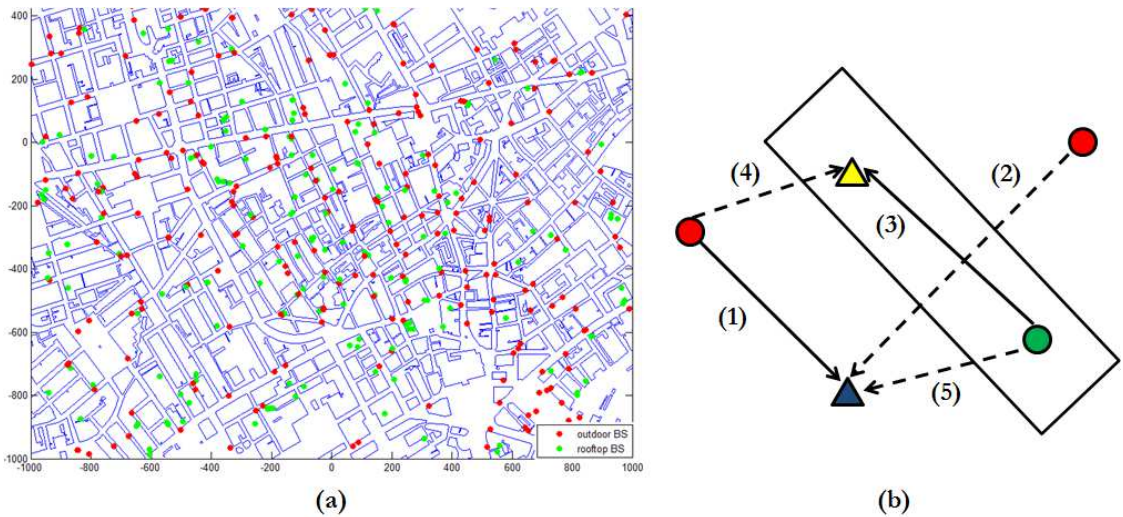}
\caption{Considered cellular network (a) and multi-state blockage model (b). (a) It is obtained from the dataset in \cite{ACM_MDR} and represents a dense urban environment in downtown London. The red and green dots represent BSs that are located outside and inside the buildings, respectively. The blue shapes represent the buildings in the considered region. (b) It provides a sketched illustration of a multi-state blockage model (dots: BSs, triangles: MTs): (1) outdoor LOS link, (2) outdoor NLOS link, (3) indoor LOS link, (4) outdoor-to-indoor link, (5) indoor-to-outdoor link. Every type of link has a different probability of occurrence and the corresponding channels have different parameters. \vspace{-0.65cm}} \label{Fig_SystemModel}
\end{figure}

\vspace{-0.25cm}
Consider an arbitrary link of length $r$, \textit{i.e.}, the distance from a BS to a MT is equal to $r$. Due to large-scale environmental-dependent blockages \cite[Slide 98]{MDR_Tutorial}, each link can be in $\mathcal{S}$ different states. Let $\mathsf{S}$ denote the set of $\mathcal{S}$ states. The probability of being in state $s \in \mathsf{S}$ is denoted by $p_{s} \left( \cdot \right)$, which is a function of $r$ and of the environment. By definition, $\sum\nolimits_{s \in \mathsf{S}} {p_s \left( r \right)}  = 1$ for every $r$. Examples of two-state ($\mathcal{S}=2$) and three-state ($\mathcal{S}=3$) link models are constituted by micro-wave and millimeter-wave outdoor links, which, because of the presence of buildings, can be either in LOS or NLOS \cite{3GPP_pathloss}, \cite{Galiotto__LosNlos}, \cite{Lopez-Perez__LosNlos}, and in LOS, NLOS or outage (OUT) \cite{MDR_mmWave}, respectively. Table \ref{LinkStateModels} provides link state models that are often used for system-level performance evaluation. In Fig. \ref{Fig_SystemModel}, we provide an illustration of the system model under analysis that accounts for the location of cellular BSs and for the presence of buildings. In particular, Fig. 1(a) is obtained by using the empirical dataset in \cite{ACM_MDR}, to which the readers are referred for further details. In Fig. 1(b), we provide a sketched representation of a typical urban environment where a multi-state link model emerges. Further details and illustrations are available in \cite[Slides 39, 98]{MDR_Tutorial}.

In the present paper, we adopt the so-called multi-ball link state model as the constituent building block of the proposed IM-based approach. The reason of this choice is twofold: 1) its mathematical tractability and 2) its flexibility for approximating other link state models. Further details are provided in Section \ref{IntensityMatching}. The accuracy of the multi-ball link state model has been experimentally validated in \cite{ACM_MDR}. In mathematical terms, $p_{s} \left( \cdot \right)$ can be formulated as follows: \vspace{-0.1cm}
\begin{equation}
\label{Eq_1}
\begin{array}{l}
p_s \left( r \right) = \sum\nolimits_{b = 1}^{{\mathcal{B}} + 1} {q_s^{\left[ {D_{b - 1} ,D_b } \right]} \mathbbm{1}_{\left[ {D_{b - 1} ,D_b } \right]} \left( r \right)} \quad {\rm{with}}\quad \sum\nolimits_{s \in {\mathsf{S}}} {q_s^{\left[ {D_{b - 1} ,D_b } \right]} }  = 1; b = 1,2, \ldots ,{\mathcal{B}} + 1
\end{array} \vspace{-0.1cm}
\end{equation}
\noindent where $\mathcal{B}$ denotes the number of balls, $D_b$ is the radius of the $b$th ball with $D_0 = 0$ and $D_{\mathcal{B}+1} = \infty$, ${q_s^{\left[ {D_{b - 1} ,D_b } \right]} }$ is the probability that the link is in state $s$ if $r \in \left[ {D_{b - 1} ,D_b } \right)$, $\mathbbm{1}_{\left[ {x,y} \right]} \left( r \right)$ is the indicator function defined as $\mathbbm{1}_{\left[ {x,y} \right]} \left( r \right) = 1$ if $r \in \left[ x, y \right)$ and $0$ otherwise, $\sum\nolimits_{s \in {\mathsf{S}}} {q_s^{\left[ {D_{b - 1} ,D_b } \right]} }  = 1$ holds by definition of probability. An illustration for $\mathcal{B}=3$ is reported in \cite[Slide 109]{MDR_Tutorial}.

The probability that a link is in state $s$ is independent of the other links. From the thinning theorem of PPPs \cite{BaccelliBook2009}, the BSs whose links are in state $s$ constitute a non-homogeneous PPP of density $\lambda _{{\rm{BS}},s} \left( r \right) = \lambda _{{\rm{BS}}} p_s \left( r \right)$. This PPP is denoted by $\Psi _{{\rm{BS}},s}$ and $\bigcup\nolimits_{s \in {\mathsf{S}}} {\Psi _{{\rm{BS}},s} }  = \Psi _{{\rm{BS}}}$ holds.
\vspace{-0.5cm}
\subsection{Channel Modeling} \label{ChannelModeling} \vspace{-0.25cm}
Path-loss, shadowing and fast-fading are considered, whose probability distribution depends on the link state. All links in the same state are independent and identically distributed (i.i.d.). Intended and interfering links are denoted by the superscripts $^{(0)}$ and $^{(i)}$, respectively.
\paragraph{Path-Loss}
Consider a link of length $r$ in state $s \in \mathsf{S}$. The distance-dependent path-loss model is $l_s \left( r \right) = \kappa _s r^{\alpha _s }$, where $\kappa_s$ is the path-loss constant and $\alpha_s$ is the path-loss slope.
\paragraph{Shadowing}
Consider a link in state $s \in \mathsf{S}$. Shadowing follows a log-normal distribution with mean equal to $\mu_s$ (in dB) and standard deviation equal to  $\sigma_s$ (in dB). It is denoted by $\mathcal{X}_s$, and its Probability Density Function (PDF) is equal to $f_{{\mathcal{X}}_s } \left( x \right) = \frac{{10}}{{\ln \left( {10} \right)}}\frac{1}{{\sqrt {2\pi } \sigma _s x}}\exp \left( { - \frac{{\left( {10\log _{10} x - \mu _s } \right)^2 }}{{2\sigma _s^2 }}} \right)$.
\paragraph{Fading}
Consider a link in state $s \in \mathsf{S}$. The power gain due to small-scale fading follows a gamma distribution with fading parameter $m_s$ and mean $\Omega_s$. It is denoted by $g_s$, and its PDF is $f_{g_s } \left( x \right) = \frac{{m_s^{m_s } x^{m_s  - 1} }}{{\Omega_s ^{m_s } \Gamma \left( {m_s } \right)}}\exp \left( { - \frac{{m_s x}}{\Omega_s }} \right)$, where $\Gamma (\cdot)$ is the gamma function. The gamma model is chosen due to its tractability and the wide range of fading severities that can be handled, \textit{e.g.}, LOS, NLOS ($m_s \approx 1$) and no fading ($m_s \to \infty$) links. Also, it allows one to analyze and compare multiple-antenna transmission schemes over Rayleigh fading channels \cite[Slide 91]{MDR_Tutorial}.
\begin{remark} \label{Remark__Pathloss}
The proposed approach can be generalized to account for the bounded path-loss model $l_s \left( r \right) = \kappa _s \left( {\max \left\{ {\tilde r_{s} ,r} \right\}} \right)^{\alpha_s}$, where $\tilde r_s \ge 0$ avoids the singularity at the origin. For typical cellular network deployments, however, the condition $D_0 < \tilde r_s < D_1$ holds. As a result, the final formulas are more analytically involving, but the inherent performance trends are not affected. Numerical examples based on actual cellular network deployments are reported in \cite{ACM_MDR}. In this paper, for this reason, we have decided to report the mathematical formulas only for $\tilde r_s = 0$. \hfill $\Box$
\end{remark}
\vspace{-0.5cm}
\subsection{Cell Association Modeling} \label{CellAssociationModeling} \vspace{-0.25cm}
A cell association criterion based on the average highest received power is assumed. Let the superscript $^{(n)}$ identify a generic BS-to-MT link. The serving BS, ${\rm{BS}}^{\left( 0 \right)}$, is obtained as follows:
\begin{equation}
\label{Eq_2}
\begin{array}{l}
{\rm{BS}}^{\left( 0 \right)}  = \arg \max _{s \in {\mathsf{S}},{\rm{BS}}^{\left( n \right)}  \in \Psi _{{\rm{BS}},s} } \left\{ {{{{\mathcal{X}}_s^{\left( n \right)}  } \mathord{\left/
 {\vphantom {{{\mathcal{X}}_s^{\left( n \right)} g_s^{\left( n \right)} } {l_s \left( {r^{\left( n \right)} } \right)}}} \right.
 \kern-\nulldelimiterspace} {l_s \left( {r^{\left( n \right)} } \right)}}} \right\} = \arg \min _{s \in {\mathsf{S}},{\rm{BS}}^{\left( n \right)}  \in \Psi _{{\rm{BS}},s} } \left\{ {L_s^{\left( n \right)} } \right\}
\end{array}
\end{equation}
\noindent where $L_s^{\left( n \right)}  = {{l_s \left( {r^{\left( n \right)} } \right)} \mathord{\left/ {\vphantom {{l_s \left( {r^{\left( n \right)} } \right)} {{\mathcal{X}}_s^{\left( n \right)} }}} \right. \kern-\nulldelimiterspace} {{\mathcal{X}}_s^{\left( n \right)} }}$ denotes the inverse of the average received power of the $n$th link in $\Psi _{{\rm{BS}},s}$. As for the intended link, we have $L^{\left( 0 \right)}  = \min _{s \in {\mathsf{S}},{\rm{BS}}^{\left( n \right)}  \in \Psi _{{\rm{BS}},s} } \left\{ {L_s^{\left( n \right)} } \right\}$.
\vspace{-0.5cm}
\subsection{Load Modeling} \label{LoadModeling} \vspace{-0.25cm}
To account for arbitrary triplets $\left\{ {\lambda _{{\rm{BS}}} ,\lambda _{{\rm{MT}}} ,N_{{\rm{RB}}} } \right\}$, we use an approach similar to that in \cite{WiOPT_2013}. We, however, generalize it for modeling the setup $N_{{\rm{RB}}}  > 1$. It is worth mentioning that the approach in \cite{WiOPT_2013} is applicable to cellular networks whose coverage regions (cells) constitute a Voronoi tessellation. The distribution of the area of the Voronoi cells is, however, still obtained by using simulations. The approach in \cite{WiOPT_2013}, hence, is applicable to a cell association criterion based on the shortest distance. It cannot be used, on the other hand, if the cell association in \eqref{Eq_2} is employed. Some illustrations are available in \cite[Slides 118-121]{MDR_Tutorial}. To the best of the authors knowledge, there are no empirical results for the distribution of the area of the coverage regions that originate from the cell association in \eqref{Eq_2}. For this reason and for mathematical tractability, we rely on a first-order moment matching approach for approximating the latter distribution \cite{Load_Harpreet}. The rationale and the mathematical foundation behind this approximation can be found in \cite{Load_Sarabjot}. The proposed modeling approximation, more precisely, relies on the following \textit{Lemma \ref{Lemma__MeanArea}}.
\begin{lemma} \label{Lemma__MeanArea}
Consider the link state model, the path-loss model and the cell association criterion introduced in Sections \ref{LinkStateModeling}, \ref{ChannelModeling} and \ref{CellAssociationModeling}, respectively. Let $\Pr^0 \left\{  \cdot  \right\}$ and ${\mathbb{E}}^0\left\{  \cdot  \right\}$ denote the probability and the expectation operators under the Palm probability \cite{BaccelliBook2009}. The mean value (average) of the area of the associated coverage regions (cells) can be formulated as follows:
\begin{equation}
\label{Eq_3-pre}
\begin{array}{l}
\hspace{-0.35cm} {\mathbb{E}}^0\left\{ {{\rm{area}}} \right\}\mathop  = \limits^{\left( a \right)} 2\pi \sum\nolimits_{s \in {\mathsf{S}}} {\int\nolimits_0^\infty  {\Pr^0 \left\{ {r \in {\mathsf C}_s \left( {{\rm{BS}}^{\left( 0 \right)} } \right)} \right\}p_s \left( r \right)rdr} } \\
\hspace{-0.35cm} \mathop  = \limits^{\left( b \right)} 2\pi \sum\nolimits_{s \in {\mathsf{S}}} {\int\nolimits_0^\infty  {\left( {\prod\nolimits_{\tilde s \in {\mathsf{S}}} {\Pr^0 \left\{ {\frac{{\kappa _{\tilde s} \tilde r^{\alpha _{\tilde s} } }}{{{\cal X}_{\tilde s} }} > \frac{{\kappa _s r^{\alpha _s } }}{{{\cal X}_s^{\left( 0 \right)} }}} \right\}} } \right)p_s \left( r \right)rdr} }  \\
 \hspace{-0.35cm} \mathop  = \limits^{\left( c \right)} 2\pi \sum\nolimits_{s \in {\mathsf{S}}} {\int\nolimits_0^\infty  {\left( {{\mathbb{E}}_{{\cal X}_s^{\left( 0 \right)} } \left\{ {\exp \left( { - \sum\nolimits_{s \in {\mathsf{S}}} {2\pi \lambda _{{\rm{BS}}} {\mathbb{E}}_{{\cal X}_{\tilde s} } \left\{ {\int\nolimits_0^{c\left( {\tilde r,{\cal X}_{\tilde s} ,{\cal X}_s^{\left( 0 \right)} } \right)} {p_{\tilde s} \left( {\tilde r} \right)\tilde rd\tilde r} } \right\}} } \right)} \right\}} \right)p_s \left( r \right)rdr} } \\
 \hspace{-0.35cm} \mathop  = \limits^{\left( d \right)} \sum\nolimits_{s \in {\mathsf{S}}} {{{{\mathcal{A}}_s } \mathord{\left/
 {\vphantom {{{\mathcal{A}}_s } {\lambda _{{\rm{BS}}} }}} \right.
 \kern-\nulldelimiterspace} {\lambda _{{\rm{BS}}} }}} \mathop  = \limits^{\left( e \right)} {1 \mathord{\left/
 {\vphantom {1 {\lambda _{{\rm{BS}}} }}} \right.
 \kern-\nulldelimiterspace} {\lambda _{{\rm{BS}}} }}
\end{array}
\end{equation}
\noindent where ${\mathbb{E}}\left\{  \cdot  \right\}$ is the expectation, ${{\mathsf C}_s \left( {{\rm{BS}}^{\left( 0 \right)} } \right)}$ is the ``cell'' of ${{\rm{BS}}^{\left( 0 \right)} }$ in state $s$, \textit{i.e.}, the set of points in the plane that are served by ${{\rm{BS}}^{\left( 0 \right)} }$ and whose links are in state $s$, ${{\cal X}_s^{\left( 0 \right)} }$ is the shadowing of a point at distance $r$ from ${{\rm{BS}}^{\left( 0 \right)} }$ whose link is in state $s$, ${{\cal X}_{\tilde s}}$ is the shadowing of the same point at distance $\tilde r$ from another (generic) BS different from ${{\rm{BS}}^{\left( 0 \right)} }$ whose link is in state $\tilde s$, $c\left( {\tilde r,{\cal X}_{\tilde s} ,{\cal X}_s^{\left( 0 \right)} } \right) = \left( {\left( {{{{\cal X}_{\tilde s} } \mathord{\left/ {\vphantom {{{\cal X}_{\tilde s} } {{\cal X}_s^{\left( 0 \right)} }}} \right. \kern-\nulldelimiterspace} {{\cal X}_s^{\left( 0 \right)} }}} \right)\left( {{{\kappa _s } \mathord{\left/ {\vphantom {{\kappa _s } {\kappa _{\tilde s} }}} \right. \kern-\nulldelimiterspace} {\kappa _{\tilde s} }}} \right)\tilde r^{\alpha _s } } \right)^{{1 \mathord{\left/ {\vphantom {1 {\alpha _{\tilde s} }}} \right. \kern-\nulldelimiterspace} {\alpha _{\tilde s} }}}$, and ${\mathcal{A}}_s$ is the association probability of state $s$, \textit{i.e.}, the probability that ${{\rm{MT}}^{\left( 0 \right)} }$ is served by a BS whose link with it is in state $s$, as follows:
\begin{equation}
\label{Eq_3-preBIS}
\begin{array}{l}
\hspace{-0.25cm}{\mathcal{A}}_s  = {\mathbb{E}}_{L_s^{\left( 0 \right)} } \left\{ {\prod\limits_{\tilde s \ne s,\,\tilde s \in {\mathsf{S}}} {\Pr \left\{ {\left. {L_{\tilde s}^{\left( 0 \right)}  > L_s^{\left( 0 \right)} } \right|L_s^{\left( 0 \right)} } \right\}} } \right\} = \int\nolimits_0^\infty  {\prod\limits_{\tilde s \ne s,\,\tilde s \in {\mathsf{S}}} {\Pr \left\{ {\left. {L_{\tilde s}^{\left( 0 \right)}  > x} \right|x} \right\}f_{L_s^{\left( 0 \right)} } \left( x \right)dx} }
\end{array}
\end{equation}
\noindent where $L_s^{\left( 0 \right)}  = \min_{{\rm{BS}}^{\left( n \right)}  \in \Psi _{{\rm{BS}},s} } \left\{ {L_s^{\left( n \right)} } \right\}$ and $f_{L_s^{\left( 0 \right)} } \left(  \cdot  \right)$ is the PDF of $L_s^{\left( 0 \right)}$ for $s \in \mathsf{S}$.

\emph{Proof}: It follows from the definition of mean (average) area in \cite{Load_Sarabjot}, by using \eqref{Eq_1} and \eqref{Eq_2}, and by taking into account that the considered PPPs are independent and non-homogeneous. In particular: (a) follows by definition of mean area, (b) from the definition of cell association in \eqref{Eq_2}, (c) by using the same steps as in \cite[Lemma 2]{Load_Sarabjot}, (d) by computing \eqref{Eq_3-preBIS} with the aid of \eqref{Eq_18}-\eqref{Eq_20} and comparing it with (c), and (e) by definition of association probability, \textit{i.e.}, $\sum\nolimits_{s \in {\mathsf{S}}} {{\mathcal{A}}_s }  = 1$. \hfill $\Box$
\end{lemma}

Based on \textit{Lemma \ref{Lemma__MeanArea}}, the following approximation for the PDF of the area of the cells is used.
\begin{approximation} \label{Approximation_CellSizeDistribution}
Consider a cellular network with PPP-distributed BSs of density $\lambda_{\rm{BS}}$ and the system model in Sections \ref{LinkStateModeling}-D. The PDF of the area of the cells is approximated as:
\begin{equation}
\label{Eq_3}
\begin{array}{l}
f_{{\rm{area}}} \left( x \right) \approx \left( {\frac{{3.5}}{{{\mathbb{E}}\left\{ {{\rm{area}}} \right\}}}} \right)^{3.5} \frac{{x^{2.5} }}{{\Gamma \left( {3.5} \right)}}\exp \left( { - \frac{{3.5}}{{{\mathbb{E}}\left\{ {{\rm{area}}} \right\}}}x} \right)
\end{array}
\end{equation}
\noindent which is a gamma random variable with parameters $\left( {m,\Omega } \right) = \left( {3.5,{\mathbb{E}}\left\{ {{\rm{area}}} \right\}} \right)$ \cite[Eq. (1)]{WiOPT_2013}. \hfill $\Box$
\end{approximation}

The modeling assumption in \eqref{Eq_3} foresees to approximate the actual PDF of the area of the cells originating from \eqref{Eq_2} with the PDF of the area of a Voronoi tessellation having the same average area. Since ${\mathbb{E}}\left\{ {{\rm{area}}} \right\} = {1 \mathord{\left/ {\vphantom {1 {\lambda _{{\rm{BS}}} }}} \right. \kern-\nulldelimiterspace} {\lambda _{{\rm{BS}}} }}$ in \eqref{Eq_3-pre}, the PDF in \eqref{Eq_3} coincides with the PDF corresponding to a Voronoi tessellation. In Section \ref{Results}, this approach is shown to be accurate for various setups.
\begin{remark} \label{Remark__IntuitionApproximation}
At the time of writing, we have no simple and intuitive explanation for the fact that the PDF of the area of the cells in \eqref{Eq_3} coincides with that of a Voronoi tessellation, even though the cell association considered in the present paper is not based on the shortest distance criterion. We can, however, safely affirm the following. Let us consider, as an example, a two-state link model with, \textit{e.g.}, LOS and NLOS links, without shadowing. The corresponding cell association criterion based on the smallest path-loss partitions the plane into cells that, according to \eqref{Eq_3-pre}, have the same average area as those of a Voronoi tessellation, \textit{i.e.}, ${1 \mathord{\left/ {\vphantom {1 {\lambda _{{\rm{BS}}} }}} \right. \kern-\nulldelimiterspace} {\lambda _{{\rm{BS}}} }}$. They are, however, constituted of points in the plane that are not necessarily contiguous, \textit{e.g.}, see \cite[Slide 120]{MDR_Tutorial}. Even though \eqref{Eq_3} implies that the distribution of the area of the cells is the same as that of a Voronoi tessellation, the shape of the cells is different. We note, in addition, that this is different from \cite{Load_Harpreet} and \cite{Load_Sarabjot}, where multi-tier cellular networks are considered. A single-tier cellular network with LOS and NLOS links is, from the cell association standpoint, not exactly the same as a two-tier cellular network. All the BSs, in fact, are homogeneous in terms of deployment density and transmit power. Each cell, more precisely, is constituted by all the possible points in the plane that are either in LOS or NLOS, but still experience the smallest path-loss in the downlink. It is reasonable to expect that, for typical path-loss exponents, the cells are constituted by spatial locations that are in LOS with respect to the BS. In the limiting regime where all the points are in LOS, the cell association would boil down to the shortest distance criterion. In this limiting case, \eqref{Eq_3} would be exact. This provides a somehow intuitive justification for \eqref{Eq_3}. \hfill $\Box$
\end{remark}

As better described in Section \ref{ProblemFormulation}, ASE and PT on a generic RB depend on: 1) the probability that the typical MT is scheduled for transmission in a RB and 2) the probability that a generic BS does not transmit in a RB. The first probability accounts for the fact that a number of MTs greater than $N_{{\rm{RB}}}$ may be associated to a BS. Thus, at most $N_{{\rm{RB}}}$ MTs can be served and the others are blocked. The second probability accounts for the fact that fewer MTs than $N_{{\rm{RB}}}$ may be associated to a BS, which implies that a BS may not be active in some of its RBs. These probabilities are denoted by $p_{{\rm{sel}}} \left(  \cdot  \right)$ and $p_{{\rm{off}}} \left(  \cdot  \right)$, respectively, and both depend on the triplet $\left\{ {\lambda _{{\rm{BS}}} ,\lambda _{{\rm{MT}}} ,N_{{\rm{RB}}} } \right\}$. They are formulated in the following two lemmas under the assumption that the MTs associated to a BS are randomly chosen for transmission in an arbitrary RB.
\begin{lemma} \label{Lemma__SelectionMT}
Consider the triplet $\left\{ {\lambda _{{\rm{BS}}} ,\lambda _{{\rm{MT}}} ,N_{{\rm{RB}}} } \right\}$. $p_{{\rm{sel}}} \left(  \cdot  \right)$ can be formulated as follows: \vspace{-0.1cm}
\begin{equation}
\label{Eq_4}
\begin{array}{l}
p_{{\rm{sel}}} \left( {\lambda _{{\rm{BS}}} ,\lambda _{{\rm{MT}}} ,N_{{\rm{RB}}} } \right) = 1 - f_{{\rm{sel}}}^{\left( a \right)} \left( {f_{{\rm{sel}}}^{\left( b \right)} - f_{{\rm{sel}}}^{\left( c \right)} } \right)
\end{array} \vspace{-0.1cm}
\end{equation}
\noindent where ${}_2F_1 \left( { \cdot , \cdot , \cdot , \cdot } \right)$ is the Gauss hypergeometric function and:
\begin{equation}
\label{Eq_5}
\begin{array}{l}
 f_{{\rm{sel}}}^{\left( a \right)} = f_{{\rm{sel}}}^{\left( a \right)} \left( {\lambda _{{\rm{BS}}} ,\lambda _{{\rm{MT}}} ,N_{{\rm{RB}}} } \right) = \frac{{3.5^{4.5} \Gamma \left( {4.5 + N_{{\rm{RB}}} } \right)}}{{\Gamma \left( {4.5} \right)}}\left( {{{\lambda _{{\rm{MT}}} } \mathord{\left/
 {\vphantom {{\lambda _{{\rm{MT}}} } {\lambda _{{\rm{BS}}} }}} \right.
 \kern-\nulldelimiterspace} {\lambda _{{\rm{BS}}} }}} \right)^{N_{{\rm{RB}}} } \left( {\frac{1}{{3.5 + {{\lambda _{{\rm{MT}}} } \mathord{\left/
 {\vphantom {{\lambda _{{\rm{MT}}} } {\lambda _{{\rm{BS}}} }}} \right.
 \kern-\nulldelimiterspace} {\lambda _{{\rm{BS}}} }}}}} \right)^{4.5 + N_{{\rm{RB}}} }  \\
 f_{{\rm{sel}}}^{\left( b \right)} = f_{{\rm{sel}}}^{\left( b \right)} \left( {\lambda _{{\rm{BS}}} ,\lambda _{{\rm{MT}}} ,N_{{\rm{RB}}} } \right) = \frac{1}{{\Gamma \left( {1 + N_{{\rm{RB}}} } \right)}}{}_2F_1 \left( {1,4.5 + N_{{\rm{RB}}} ,1 + N_{{\rm{RB}}} ,\frac{{{{\lambda _{{\rm{MT}}} } \mathord{\left/
 {\vphantom {{\lambda _{{\rm{MT}}} } {\lambda _{{\rm{BS}}} }}} \right.
 \kern-\nulldelimiterspace} {\lambda _{{\rm{BS}}} }}}}{{3.5 + {{\lambda _{{\rm{MT}}} } \mathord{\left/
 {\vphantom {{\lambda _{{\rm{MT}}} } {\lambda _{{\rm{BS}}} }}} \right.
 \kern-\nulldelimiterspace} {\lambda _{{\rm{BS}}} }}}}} \right) \\
 f_{{\rm{sel}}}^{\left( c \right)} =  f_{{\rm{sel}}}^{\left( c \right)} \left( {\lambda _{{\rm{BS}}} ,\lambda _{{\rm{MT}}} ,N_{{\rm{RB}}} } \right) = \frac{{N_{{\rm{RB}}} }}{{\Gamma \left( {2 + N_{{\rm{RB}}} } \right)}}{}_2F_1 \left( {1,4.5 + N_{{\rm{RB}}} ,2 + N_{{\rm{RB}}} ,\frac{{{{\lambda _{{\rm{MT}}} } \mathord{\left/
 {\vphantom {{\lambda _{{\rm{MT}}} } {\lambda _{{\rm{BS}}} }}} \right.
 \kern-\nulldelimiterspace} {\lambda _{{\rm{BS}}} }}}}{{3.5 + {{\lambda _{{\rm{MT}}} } \mathord{\left/
 {\vphantom {{\lambda _{{\rm{MT}}} } {\lambda _{{\rm{BS}}} }}} \right.
 \kern-\nulldelimiterspace} {\lambda _{{\rm{BS}}} }}}}} \right)
 \end{array}
\end{equation}

\emph{Proof}: Let $N^{'}$ be the number of other MTs associated to a generic BS conditioned on a generic MT being associated to the BS. By definition,  $p_{{\rm{sel}}} \left( {\lambda _{{\rm{BS}}} ,\lambda _{{\rm{MT}}} ,N_{{\rm{RB}}} } \right) = \sum\nolimits_{n = 0}^{N_{{\rm{RB}}}  - 1} {1\Pr \left\{ {N^{'}  = n} \right\}}  + \sum\nolimits_{n = N_{{\rm{RB}}} }^{ + \infty } {\left( {{{N_{{\rm{RB}}} } \mathord{\left/ {\vphantom {{N_{{\rm{RB}}} } {\left( {n + 1} \right)}}} \right. \kern-\nulldelimiterspace} {\left( {n + 1} \right)}}} \right)\Pr \left\{ {N^{'}  = n} \right\}}$. With the aid of \textit{Approximation \ref{Approximation_CellSizeDistribution}}, ${\Pr \left\{ {N^{'}  = n} \right\}}$ follows from \cite{WiOPT_2013}. The proof concludes by formulating the summations in terms of ${}_2F_1 \left(  \cdot  \right)$ functions. \hfill $\Box$
\end{lemma}
\begin{lemma} \label{Lemma__InactiveBS}
Consider the triplet $\left\{ {\lambda _{{\rm{BS}}} ,\lambda _{{\rm{MT}}} ,N_{{\rm{RB}}} } \right\}$. $p_{{\rm{off}}} \left(  \cdot  \right)$ can be formulated as follows: \vspace{-0.1cm}
\begin{equation}
\label{Eq_6}
\begin{array}{l}
p_{{\rm{off}}} \left( {\lambda _{{\rm{BS}}} ,\lambda _{{\rm{MT}}} ,N_{{\rm{RB}}} } \right) = 1 - {{\lambda _{{\rm{MT}}} } \mathord{\left/
 {\vphantom {{\lambda _{{\rm{MT}}} } {\left( {\lambda _{{\rm{BS}}} N_{{\rm{RB}}} } \right)}}} \right.
 \kern-\nulldelimiterspace} {\left( {\lambda _{{\rm{BS}}} N_{{\rm{RB}}} } \right)}} - p_{{\rm{off}}}^{\left( a \right)}  + p_{{\rm{off}}}^{\left( b \right)}  + p_{{\rm{off}}}^{\left( c \right)}
\end{array} \vspace{-0.1cm}
\end{equation}
\noindent where $p_{{\rm{off}}}^{\left( x \right)}  = p_{{\rm{off}}}^{\left( x \right)} \left( {\lambda _{{\rm{BS}}} ,\lambda _{{\rm{MT}}} ,N_{{\rm{RB}}} } \right)$ for $x=\{a,b,c,\}$ are as follows:
\begin{equation}
\label{Eq_7}
\begin{array}{l}
 p_{{\rm{off}}}^{\left( a \right)} = \frac{{3.5^{3.5} \Gamma \left( {4.5 + N_{{\rm{RB}}} } \right)}}{{\Gamma \left( {3.5} \right)\Gamma \left( {2 + N_{{\rm{RB}}} } \right)}}\frac{{\left( {{{\lambda _{{\rm{MT}}} } \mathord{\left/
 {\vphantom {{\lambda _{{\rm{MT}}} } {\lambda _{{\rm{BS}}} }}} \right.
 \kern-\nulldelimiterspace} {\lambda _{{\rm{BS}}} }}} \right)^{1 + N_{{\rm{RB}}} } }}{{\left( {3.5 + {{\lambda _{{\rm{MT}}} } \mathord{\left/
 {\vphantom {{\lambda _{{\rm{MT}}} } {\lambda _{{\rm{BS}}} }}} \right.
 \kern-\nulldelimiterspace} {\lambda _{{\rm{BS}}} }}} \right)^{4.5 + N_{{\rm{RB}}} } }}{}_2F_1 \left( {1,4.5 + N_{{\rm{RB}}} ,2 + N_{{\rm{RB}}} ,\frac{{{{\lambda _{{\rm{MT}}} } \mathord{\left/
 {\vphantom {{\lambda _{{\rm{MT}}} } {\lambda _{{\rm{BS}}} }}} \right.
 \kern-\nulldelimiterspace} {\lambda _{{\rm{BS}}} }}}}{{{{3.5 + \lambda _{{\rm{MT}}} } \mathord{\left/
 {\vphantom {{3.5 + \lambda _{{\rm{MT}}} } {\lambda _{{\rm{BS}}} }}} \right.
 \kern-\nulldelimiterspace} {\lambda _{{\rm{BS}}} }}}}} \right) \\
 p_{{\rm{off}}}^{\left( b \right)}  = \frac{{3.5^{3.5} \Gamma \left( {4.5 + N_{{\rm{RB}}} } \right)}}{{\Gamma \left( {3.5} \right)N_{{\rm{RB}}} \Gamma \left( {1 + N_{{\rm{RB}}} } \right)}}\frac{{\left( {{{\lambda _{{\rm{MT}}} } \mathord{\left/
 {\vphantom {{\lambda _{{\rm{MT}}} } {\lambda _{{\rm{BS}}} }}} \right.
 \kern-\nulldelimiterspace} {\lambda _{{\rm{BS}}} }}} \right)^{1 + N_{{\rm{RB}}} } }}{{\left( {{{3.5 + \lambda _{{\rm{MT}}} } \mathord{\left/
 {\vphantom {{3.5 + \lambda _{{\rm{MT}}} } {\lambda _{{\rm{BS}}} }}} \right.
 \kern-\nulldelimiterspace} {\lambda _{{\rm{BS}}} }}} \right)^{4.5 + N_{{\rm{RB}}} } }}{}_2F_1 \left( {1,4.5 + N_{{\rm{RB}}} ,2 + N_{{\rm{RB}}} ,\frac{{{{\lambda _{{\rm{MT}}} } \mathord{\left/
 {\vphantom {{\lambda _{{\rm{MT}}} } {\lambda _{{\rm{BS}}} }}} \right.
 \kern-\nulldelimiterspace} {\lambda _{{\rm{BS}}} }}}}{{{{3.5 + \lambda _{{\rm{MT}}} } \mathord{\left/
 {\vphantom {{3.5 + \lambda _{{\rm{MT}}} } {\lambda _{{\rm{BS}}} }}} \right.
 \kern-\nulldelimiterspace} {\lambda _{{\rm{BS}}} }}}}} \right) \\
 p_{{\rm{off}}}^{\left( c \right)} = \frac{{3.5^{3.5} \Gamma \left( {5.5 + N_{{\rm{RB}}} } \right)}}{{\Gamma \left( {3.5} \right)N_{{\rm{RB}}} \Gamma \left( {3 + N_{{\rm{RB}}} } \right)}}\frac{{\left( {{{\lambda _{{\rm{MT}}} } \mathord{\left/
 {\vphantom {{\lambda _{{\rm{MT}}} } {\lambda _{{\rm{BS}}} }}} \right.
 \kern-\nulldelimiterspace} {\lambda _{{\rm{BS}}} }}} \right)^{2 + N_{{\rm{RB}}} } }}{{\left( {{{3.5 + \lambda _{{\rm{MT}}} } \mathord{\left/
 {\vphantom {{3.5 + \lambda _{{\rm{MT}}} } {\lambda _{{\rm{BS}}} }}} \right.
 \kern-\nulldelimiterspace} {\lambda _{{\rm{BS}}} }}} \right)^{5.5 + N_{{\rm{RB}}} } }}{}_2F_1 \left( {2,5.5 + N_{{\rm{RB}}} ,3 + N_{{\rm{RB}}} ,\frac{{{{\lambda _{{\rm{MT}}} } \mathord{\left/
 {\vphantom {{\lambda _{{\rm{MT}}} } {\lambda _{{\rm{BS}}} }}} \right.
 \kern-\nulldelimiterspace} {\lambda _{{\rm{BS}}} }}}}{{{{3.5 + \lambda _{{\rm{MT}}} } \mathord{\left/
 {\vphantom {{3.5 + \lambda _{{\rm{MT}}} } {\lambda _{{\rm{BS}}} }}} \right.
 \kern-\nulldelimiterspace} {\lambda _{{\rm{BS}}} }}}}} \right)
 \end{array}
\end{equation}

\emph{Proof}: Let $N$ be the number of MTs associated to a BS. By definition, $p_{{\rm{off}}} \left( {\lambda _{{\rm{BS}}} ,\lambda _{{\rm{MT}}} ,N_{{\rm{RB}}} } \right) = \sum\nolimits_{n = 0}^{N_{{\rm{RB}}} } {\left( {1 - {n \mathord{\left/ {\vphantom {n {N_{{\rm{RB}}} }}} \right. \kern-\nulldelimiterspace} {N_{{\rm{RB}}} }}} \right)\Pr \left\{ {N = n} \right\}}$. With the aid of \textit{Approximation \ref{Approximation_CellSizeDistribution}}, ${\Pr \left\{ {N  = n} \right\}}$ follows from \cite{WiOPT_2013}. The proof is concluded by formulating the summations in terms of ${}_2F_1 \left(  \cdot  \right)$ functions. \hfill $\Box$
\end{lemma}

Under the assumption that the network is either dense (\textit{i.e.}, ${{\lambda _{{\rm{BS}}} } \mathord{\left/ {\vphantom {{\lambda _{{\rm{BS}}} } {\lambda _{{\rm{MT}}} }}} \right. \kern-\nulldelimiterspace} {\lambda _{{\rm{MT}}} }} \gg 1$) or sparse (\textit{i.e.}, ${{\lambda _{{\rm{BS}}} } \mathord{\left/ {\vphantom {{\lambda _{{\rm{BS}}} } {\lambda _{{\rm{MT}}} }}} \right. \kern-\nulldelimiterspace} {\lambda _{{\rm{MT}}} }} \ll 1$), the lemmas can be simplified as summarized in the following two corollaries.
\begin{corollary} \label{Corollary__DenseNets}
Consider ${{\lambda _{{\rm{BS}}} } \mathord{\left/ {\vphantom {{\lambda _{{\rm{BS}}} } {\lambda _{{\rm{MT}}} }}} \right. \kern-\nulldelimiterspace} {\lambda _{{\rm{MT}}} }} \gg 1$. $p_{{\rm{sel}}} \left(  \cdot  \right)$ and $p_{{\rm{off}}} \left(  \cdot  \right)$ in \eqref{Eq_4} and \eqref{Eq_6} simplify as follows:
\begin{equation}
\label{Eq_8}
\begin{array}{l}
 p_{{\rm{sel}}} \left( {\lambda _{{\rm{BS}}} ,\lambda _{{\rm{MT}}} ,N_{{\rm{RB}}} } \right) \to 1 - \frac{{\Gamma \left( {4.5 + N_{{\rm{RB}}} } \right)}}{{\Gamma \left( {4.5} \right)\Gamma \left( {1 + N_{{\rm{RB}}} } \right)}}\frac{1}{{1 + N_{{\rm{RB}}} }}\left( {\frac{{{{\lambda _{{\rm{MT}}} } \mathord{\left/
 {\vphantom {{\lambda _{{\rm{MT}}} } {\lambda _{{\rm{BS}}} }}} \right.
 \kern-\nulldelimiterspace} {\lambda _{{\rm{BS}}} }}}}{{3.5}}} \right)^{N_{{\rm{RB}}} }  \\
 p_{{\rm{off}}} \left( {\lambda _{{\rm{BS}}} ,\lambda _{{\rm{MT}}} ,N_{{\rm{RB}}} } \right) \to 1 - {{\lambda _{{\rm{MT}}} } \mathord{\left/
 {\vphantom {{\lambda _{{\rm{MT}}} } {\left( {\lambda _{{\rm{BS}}} N_{{\rm{RB}}} } \right)}}} \right.
 \kern-\nulldelimiterspace} {\left( {\lambda _{{\rm{BS}}} N_{{\rm{RB}}} } \right)}}
\end{array}
\end{equation}

\emph{Proof}: It follows from \eqref{Eq_4}-\eqref{Eq_7}, using asymptotic approximations for the ${}_2F_1 \left(  \cdot  \right)$ function. \hfill $\Box$
\end{corollary}
\begin{corollary} \label{Corollary__SparseNets}
Consider ${{\lambda _{{\rm{BS}}} } \mathord{\left/ {\vphantom {{\lambda _{{\rm{BS}}} } {\lambda _{{\rm{MT}}} }}} \right. \kern-\nulldelimiterspace} {\lambda _{{\rm{MT}}} }} \ll 1$. $p_{{\rm{sel}}} \left(  \cdot  \right)$ and $p_{{\rm{off}}} \left(  \cdot  \right)$ in \eqref{Eq_4} and \eqref{Eq_6} simplify as follows:
\begin{equation}
\label{Eq_9}
\begin{array}{l}
 p_{{\rm{sel}}} \left( {\lambda _{{\rm{BS}}} ,\lambda _{{\rm{MT}}} ,N_{{\rm{RB}}} } \right) \to N_{{\rm{RB}}} \left( {{{\lambda _{{\rm{MT}}} } \mathord{\left/
 {\vphantom {{\lambda _{{\rm{MT}}} } {\lambda _{{\rm{BS}}} }}} \right.
 \kern-\nulldelimiterspace} {\lambda _{{\rm{BS}}} }}} \right)^{ - 1}  \\
 p_{{\rm{off}}} \left( {\lambda _{{\rm{BS}}} ,\lambda _{{\rm{MT}}} ,N_{{\rm{RB}}} } \right) \to \frac{4}{{63}}\frac{{3.5^{3.5} }}{{\Gamma \left( {3.5} \right)}}\frac{{\Gamma \left( {4.5 + N_{{\rm{RB}}} } \right)}}{{\Gamma \left( {1 + N_{{\rm{RB}}} } \right)}}\left( {\frac{{\lambda _{{\rm{MT}}} }}{{\lambda _{{\rm{BS}}} }}} \right)^{ - 3.5}
\end{array}
\end{equation}

\emph{Proof}: It follows from \eqref{Eq_4}-\eqref{Eq_7}, using asymptotic approximations for the ${}_2F_1 \left(  \cdot  \right)$ function. \hfill $\Box$
\end{corollary}

Based on this load model, from the thinning theorem of PPPs \cite{BaccelliBook2009}, the set of BSs that are active on a RB and whose links are in state $s$ constitute a non-homogeneous PPP of density $\lambda _{{\rm{BS}},s}^{\left( {\rm{I}} \right)} \left( r \right) = \left( {1 - p_{{\rm{off}}} } \right)\lambda _{{\rm{BS}},s} \left( r \right) = \left( {1 - p_{{\rm{off}}} } \right)\lambda _{{\rm{BS}}} p_s \left( r \right) = \lambda _{{\rm{BS}}}^{\left( {\rm{I}} \right)} p_s \left( r \right)$, where $\lambda _{{\rm{BS}}}^{\left( {\rm{I}} \right)}  = \left( {1 - p_{{\rm{off}}} } \right)\lambda _{{\rm{BS}}}$ is the density of active BSs in a RB. This PPP is denoted by $\Psi _{{\rm{BS}},s}^{\left( {\rm{I}} \right)}$ and $\bigcup\nolimits_{s \in {\rm{S}}} {\Psi _{{\rm{BS}},s}^{\left( {\rm{I}} \right)} }  = \Psi _{{\rm{BS}}}^{\left( {\rm{I}} \right)}$ holds.
\begin{remark} \label{Remark__Load_vs_Galiotto}
Compared with \cite{Galiotto__LosNlos}, where, for tractability, it is assumed that the MTs connect to their closest BS (see footnote 3 therein), the proposed mathematical approach is more rigorous thanks to \eqref{Eq_3-pre} and to the aid of \cite{Load_Sarabjot}. It is, in addition, formulated in a more general manner, since an arbitrary number of RBs and the selection probability in \eqref{Eq_4} are taken into account. \hfill $\Box$
\end{remark}
\begin{table}[!t] \footnotesize
\centering
\caption{Examples of antenna radiation patterns. The notation is provided in footnote 1. \vspace{-0.25cm}}
\newcommand{\tabincell}[2]{\begin{tabular}{@{}#1@{}}#2\end{tabular}}
\begin{tabular}{|c||c|} \hline
 & $G_q \left( {\theta _q } \right)$ \\ \hline \hline
Omni-directional & 1 \\ \hline
3GPP \cite{Antenna_3GPP} & $\gamma_q^{({\rm{3GPP}})} 10^{ - \left( {{6 \mathord{\left/ {\vphantom {6 5}} \right. \kern-\nulldelimiterspace} 5}} \right)\left( {{{\theta _q } \mathord{\left/ {\vphantom {{\theta _q } {\phi _q^{({\rm{3dB}})} }}} \right. \kern-\nulldelimiterspace} {\phi _q^{({\rm{3dB}})} }}} \right)^2 } \mathbbm{1}_{\left[ {0,\phi _q^{\left( {3{\rm{GPP}}} \right)} } \right]} \left( {\left| {\theta _q } \right|} \right) + \gamma_q^{\left( {{\rm{3GPP}}} \right)} 10^{ - {{A_q } \mathord{\left/ {\vphantom {{A_q } {10}}} \right. \kern-\nulldelimiterspace} {10}}} \mathbbm{1}_{\left[ {\phi _q^{\left( {3{\rm{GPP}}} \right)} ,\pi } \right]} \left( {\left| {\theta _q } \right|} \right)$ \\ \hline
UWLA \cite{Antenna_UWLA} & $\gamma_q^{\left( {{\rm{UWLA}}} \right)} \left| {N_q^{ - 1} \sin \left( {N_q \pi \nu ^{ - 1} \cos \left( {\theta _q } \right)d_q } \right)\sin ^{ - 1} \left( {\pi \nu ^{ - 1} \cos \left( {\theta _q } \right)d_q } \right)} \right|^2$ \\ \hline
Three-Sector \cite{Antenna_Sectored} & $\begin{array}{l} \gamma_q^{\left( {{\rm{1}}{\rm{,sec}}} \right)} {\mathbbm{1}}_{\left[ {0,\phi _q^{\left( {{\rm{1}}{\rm{,sec}}} \right)} } \right]} \left( {\left| {\theta _q } \right|} \right) + \gamma_q^{\left( {{\rm{2}}{\rm{,sec}}} \right)} {\mathbbm{1}}_{\left[ {\phi _q^{\left( {{\rm{3}}{\rm{,sec}}} \right)} ,\pi } \right]} \left( {\left| {\theta _q } \right|} \right) \\
  + \gamma_q^{\left( {{\rm{1}}{\rm{,sec}}} \right)} \left( {1 - {{\left( {\left| {\theta _q } \right| - \phi _q^{\left( {{\rm{1}}{\rm{,sec}}} \right)} } \right)} \mathord{\left/
 {\vphantom {{\left( {\left| {\theta _q } \right| - \phi _q^{\left( {{\mathbbm{1}}{\rm{,sec}}} \right)} } \right)} {\epsilon _q }}} \right.
 \kern-\nulldelimiterspace} {\epsilon _q }}} \right){\mathbbm{1}}_{\left[ {\phi _q^{\left( {{\rm{2}}{\rm{,sec}}} \right)} ,\phi _q^{\left( {{\rm{2}}{\rm{,sec}}} \right)} } \right]} \left( {\left| {\theta _q } \right|} \right) \\
  + \left( {2g_q^{\left( {{\rm{2}}{\rm{,sec}}} \right)} {{\left( {\left| {\theta _q } \right| - \phi _q^{\left( {{\rm{2}}{\rm{,sec}}} \right)} } \right)} \mathord{\left/
 {\vphantom {{\left( {\left| {\theta _q } \right| - \phi _q^{\left( {{\rm{2}}{\rm{,sec}}} \right)} } \right)} {\epsilon _q }}} \right.
 \kern-\nulldelimiterspace} {\epsilon _q }}} \right){\mathbbm{1}}_{\left[ {\phi _q^{\left( {{\rm{2}}{\rm{,sec}}} \right)} ,\phi _q^{\left( {{\rm{3}}{\rm{,sec}}} \right)} } \right]} \left( {\left| {\theta _q } \right|} \right) \end{array}$ \\ \hline
Two-lobe \cite{MDR_mmWave} & see \eqref{Eq_10} with ${{\rm K}_q } = 2$ \\ \hline
\end{tabular}
\label{AntennaRadiationPatterns} \vspace{-0.65cm}
\end{table}
\vspace{-0.5cm}
\subsection{Antenna Radiation Pattern} \label{AntennaPattern} \vspace{-0.25cm}
In \cite{ACM_MDR}, it is empirically shown that the antenna radiation pattern greatly affects the performance of cellular networks. Different radiation patterns are typically used for system-level performance evaluation. Notable examples are provided in Table \ref{AntennaRadiationPatterns}\footnote{Notation of Table \ref{AntennaRadiationPatterns} -- 3GPP (6 sectors): $\phi _q^{\left( {3{\rm{dB}}} \right)}  = 35$ degrees, $A_q  = 23$, $\phi _q^{\left( {{\rm{3GPP}}} \right)}  = 48.46$ degrees, $\gamma_q^{\left( {{\rm{3GPP}}} \right)}  = 9.33$. UWLA (Uniformly Weighted Linear Array): $N_q  = 8$ is the number of antenna-elements, $d_q  = \nu /2$ is the uniform spacing between them and $\nu$ is the wavelength, $\gamma_q^{\left( {{\rm{UWLA}}} \right)}  = 12.1631$. Three-Sector: $\gamma_q^{({\rm{1}}{\rm{,sec}})}  = \left( {2\pi  - \left( {2\pi  - 3\epsilon _q /2 - \varpi _q } \right)\gamma_q^{({\rm{1}}{\rm{,sec}})} } \right)/\varpi _q$, $\theta _q^{\left( {{\rm{1}}{\rm{,sec}}} \right)}  = \left( {\varpi _q  - \epsilon _q } \right)/2$, $\theta _q^{\left( {{\rm{2}}{\rm{,sec}}} \right)}  = \left( {\varpi _q  + \epsilon _q } \right)/2$, $\theta _q^{\left( {{\rm{3}}{\rm{,sec}}} \right)}  = \varpi _q /2 + \epsilon _q$, $g_q^{(2,\sec )}  = 0.05$, $\varpi _q  = 70$ degrees, $\epsilon_q  = 10$ degrees.}.

With the exception of the two-lobe model \cite{MDR_mmWave}, the antenna radiation patterns in Table \ref{AntennaRadiationPatterns} are, usually, mathematically intractable. The proposed IM-based approach, thus, relies on a generalized version of the two-lobe model, which is referred to as the multi-lobe model. In \cite{ACM_MDR}, it has been proved to be sufficiently accurate with the aid of numerical simulations.

Let $\theta _q  \in \left[ { - \pi ,\pi } \right)$ for $q \in \left\{ {{\rm{BS}},{\rm{MT}}} \right\}$ be the angle towards the boresight direction. The multi-lobe antenna radiation pattern of BSs and MTs may be different and can be formulated as:
\begin{equation}
\label{Eq_10}
\begin{array}{l}
G_q \left( {\theta _q } \right) = \sum\nolimits_{l = 1}^{{\rm K}_q } {\gamma_q^{\left( l \right)} \mathbbm{1}_{\left[ {\varphi _q^{\left( {l - 1} \right)} ,\varphi _q^{\left( l \right)} } \right]} } \left( {\left| {\theta _q } \right|} \right)
\end{array}
\end{equation}
\noindent where ${\rm K}_q$ is the number of lobes, ${\gamma_q^{\left( l \right)} }$ is the gain of the $l$th lobe, $\varphi _q^{\left( 0 \right)}  = 0 < \varphi _q^{\left( 1 \right)}  < \varphi _q^{\left( 2 \right)}  <  \cdots  < \varphi _q^{\left( {{\rm K}_q  - 1} \right)}  < \varphi _q^{\left( {{\rm K}_q } \right)}  = \pi$ are the angles associated with the lobes, and $\int\nolimits_{ - \pi }^{ \pi } {G_q \left( \theta  \right)d\theta }  = 2\pi$.

The antenna radiation pattern in \eqref{Eq_10} is not only mathematically tractable, but it provides an accurate step-wise approximation of other antenna radiation patterns as well \cite{ACM_MDR}, \textit{e.g.}, those in Table \ref{AntennaRadiationPatterns}. Consider a generic antenna radiation pattern $G_q^{(X)} \left(  \cdot  \right)$. Its multi-lobe approximation in \eqref{Eq_10} can be found by solving the minimization problem as follows ($\left\|  \cdot  \right\|_F^2$ is the Frobenius norm):
\begin{equation}
\label{Eq_11}
\begin{array}{l}
\mathop {\arg \min }\nolimits_{\left\{ {\gamma_q^{\left( l \right)} } \right\},\left\{ {\varphi _q^{\left( l \right)} } \right\}} \left\{ {\left\| {\log _{10} \left( {G_q^{\left( X \right)} \left( {\theta } \right)} \right) - \log _{10} \left( {G_q \left( {\theta } \right)} \right)} \right\|_F^2 } \right\}
\end{array}
\end{equation}

The larger the number of lobes is, the more accurate but more complex the multi-lobe approximation is. Table \ref{AntennaRadiationPatterns_Fitting} provides the five-lobe (${\rm K}_q  = 5$) approximation of some antenna radiation patterns in Table \ref{AntennaRadiationPatterns}. ${\rm K}_q  = 5$ yields a good trade-off between complexity and accuracy.

\begin{table}[!t] \footnotesize
\centering
\caption{Five-lobe approximation of the antenna radiation patterns in Table \ref{AntennaRadiationPatterns} based on \eqref{Eq_10} and \eqref{Eq_11}. \vspace{-0.25cm}}
\newcommand{\tabincell}[2]{\begin{tabular}{@{}#1@{}}#2\end{tabular}}
\begin{tabular}{|c||c|c|c|c|c|c|c|c|c|} \hline
 & $\varphi _q^{\left( 1 \right)}$ & $\varphi _q^{\left( 2 \right)}$ & $\varphi _q^{\left( 3 \right)}$ & $\varphi _q^{\left( 4 \right)}$ & $\gamma_q^{\left( 1 \right)}$ & $\gamma_q^{\left( 2 \right)}$ & $\gamma_q^{\left( 3 \right)}$ & $\gamma_q^{\left( 4 \right)}$ & $\gamma_q^{\left( 5 \right)}$ \\ \hline \hline
3GPP & 0.2114 &	0.4229 &	0.6343 &	0.8457 &	8.3951 &	4.4863 &	1.2797 &	0.1943 &	0.0468 \\ \hline
UWLA & 0.1115 &	0.2524 &	2.8892 &	3.0301 &	9.9251 &	1.9782 &	0.1405 &	1.9782 &	9.9251 \\ \hline
Tree-Sector & 0.5236 &	0.6109 &	0.6981 &	0.7854 &	4.9464 &	3.7022 &	1.2366 &	0.0248 &	0.05 \\ \hline
\end{tabular}
\label{AntennaRadiationPatterns_Fitting} \vspace{-0.65cm}
\end{table}
For brevity, no pointing errors on the intended link are considered. This assumption can be removed as shown in \cite{MDR_mmWave}. Thus, the directivity gain of the intended link is $G^{\left( 0 \right)}  = G_{{\rm{BS}}} \left( 0 \right)G_{{\rm{MT}}} \left( 0 \right)$. Since, on the other hand, the interfering BSs focus their beams towards their intended MTs and they are both randomly deployed, the radiation patterns of all non-intended links are randomly oriented with respect to each other and uniformly distributed in $\left[ {-\pi,\pi } \right)$. So, the directivity gain of a generic interfering link, $G^{\left( i \right)}  = G_{{\rm{BS}}} \left( {\theta _{{\rm{BS}}}^{\left( i \right)} } \right)G_{{\rm{MT}}} \left( {\theta _{{\rm{MT}}}^{\left( i \right)} } \right)$, has the following PDF: \vspace{-0.25cm}
\begin{equation}
\label{Eq_12}
\begin{array}{l}
f_{G^{\left( i \right)} } \left( \gamma \right) = \sum\nolimits_{l_1  = 1}^{{\rm K}_{{\rm{BS}}} } {\sum\nolimits_{l_2  = 1}^{{\rm K}_{{\rm{MT}}} } {\frac{{\omega _{{\rm{BS}}}^{\left( {l_1 } \right)} }}{{2\pi }}\frac{{\omega _{{\rm{MT}}}^{\left( {l_2 } \right)} }}{{2\pi }}\delta \left( {\gamma - \gamma_{{\rm{BS}}}^{\left( {l_1 } \right)} \gamma_{{\rm{MT}}}^{\left( {l_2 } \right)} } \right)} }
\end{array} \vspace{-0.25cm}
\end{equation}
\noindent where $\omega _q^{\left( l \right)}  = 2\left( {\varphi _q^{\left( l \right)}  - \varphi _q^{\left( {l - 1} \right)} } \right)$ for $q \in \left\{ {{\rm{BS}},{\rm{MT}}} \right\}$ and $\delta \left(  \cdot  \right)$ is the Dirac delta function.
\begin{remark} \label{Remark__AntennaGain_DirectLink}
With no pointing errors, the multi-lobe approximation is necessary only to compute the distribution of the other-cell interference. It is not needed, on the other hand, for obtaining $G^{\left( 0 \right)}$. Thus, we assume $G^{\left( 0 \right)}  = G_{{\rm{BS}}}^{(X)} \left( 0 \right)G_{{\rm{MT}}}^{(X)} \left( 0 \right)$ for every antenna radiation pattern $X$. \hfill $\Box$
\end{remark}
\vspace{-0.5cm}
\subsection{Problem Formulation} \label{ProblemFormulation} \vspace{-0.25cm}
The performance metrics of interest are ASE and PT, which are expressed in $\rm{bps/Hz/m^2}$. They can be formulated as follows (for simplicity, we use the short-hand $p_{{\rm{sel}}}  = p_{{\rm{sel}}} \left( {\lambda _{{\rm{BS}}} ,\lambda _{{\rm{MT}}} ,N_{{\rm{RB}}} } \right)$): \vspace{-0.55cm}
\begin{equation}
\label{Eq_13}
\begin{array}{l}
{\rm{ASE}} = \left( {{{\lambda _{{\rm{MT}}} p_{{\rm{sel}}} } \mathord{\left/
 {\vphantom {{\lambda _{{\rm{MT}}} p_{{\rm{sel}}} } {\ln \left( 2 \right)}}} \right.
 \kern-\nulldelimiterspace} {\ln \left( 2 \right)}}} \right)\ln \left( {1 + {\rm{SINR}}} \right){\rm{;}}\quad {\rm{PT}} = \lambda _{{\rm{MT}}} p_{{\rm{sel}}} \log _2 \left( {1 + {\rm{T}}} \right)\Pr \left\{ {{\rm{SINR}} \ge {\rm{T}}} \right\}
\end{array} \vspace{-0.15cm}
\end{equation}
\noindent where SINR denotes the Signal-to-Interference+Noise-Ratio at ${\rm{MT}}^{\left( 0 \right)}$ and ${\rm{T}}$ is the minimum SINR threshold for successful decoding. The ASE is obtained from the Shannon rate ${\mathcal{R}} = \ln \left( {1 + {\rm{SINR}}} \right)$ and the PT depends on the coverage probability ${\mathcal{C}}\left( {\rm{T}} \right) = \Pr \left\{ {{\rm{SINR}} \ge {\rm{T}}} \right\}$.
\begin{remark} \label{Remark__ShannonFormulas}
The definition of ASE and PT in \eqref{Eq_13} is based on information-theoretic arguments. This implies that encoding, decoding and the related assumptions on the channel state information that need to be fulfilled at the transmitter and receiver are those stated in \cite[Sec. IV]{AndrewsNov2011}. \hfill $\Box$
\end{remark}

In particular, ${\rm{SINR}} = {\rm{SINR}}\left( {\lambda _{{\rm{BS}}} ,\lambda _{{\rm{MT}}} ,N_{{\rm{RB}}} ,p_{{\rm{off}}} } \right)$ and it can be formulated as follows:
\begin{equation}
\label{Eq_14}
\begin{array}{l}
{\rm{SINR}} = {\rm{SINR}}\left( {L^{\left( 0 \right)} } = {L_s^{\left( 0 \right)} } \right) = {\frac{{{{P_{{\rm{RB}}} G^{\left( 0 \right)} g_s^{\left( 0 \right)} } \mathord{\left/
 {\vphantom {{P_{{\rm{RB}}} G^{\left( 0 \right)} g_s^{\left( 0 \right)} } {L_s^{\left( 0 \right)} }}} \right.
 \kern-\nulldelimiterspace} {L_s^{\left( 0 \right)} }}}}{{\sigma _N^2  + I_{{\rm{agg}}} \left( {L_s^{\left( 0 \right)} } \right)}}} \quad \quad {\rm{if}} \quad {L^{\left( 0 \right)} } = {L_s^{\left( 0 \right)} }
\end{array}
\end{equation}
\noindent where ${\sigma _N^2 }$ is the noise power in a RB, $L_s^{\left( 0 \right)}  = \min_{{\rm{BS}}^{\left( n \right)}  \in \Psi _{{\rm{BS}},s} } \left\{ {L_s^{\left( n \right)} } \right\}$, $I_{{\rm{agg}}} \left(  \cdot  \right)$ is the interference: \vspace{-0.25cm}
\begin{equation}
\label{Eq_15}
\begin{array}{l}
I_{{\rm{agg}}} \left( {L^{\left( 0 \right)} } \right) = \sum\nolimits_{\tilde s \in {\mathsf{S}}} {\sum\nolimits_{{\rm{BS}}^{\left( i \right)}  \in \Psi^{(\rm{I})} _{{\rm{BS}},\tilde s} } {\left( {{{P_{{\rm{RB}}} G^{\left( i \right)} g_{\tilde s}^{\left( i \right)} } \mathord{\left/
 {\vphantom {{P_{{\rm{RB}}} G^{\left( i \right)} g_{\tilde s}^{\left( i \right)} } {L_{\tilde s}^{\left( i \right)} }}} \right.
 \kern-\nulldelimiterspace} {L_{\tilde s}^{\left( i \right)} }}} \right)} \mathbbm{1}\left( {L_{\tilde s}^{\left( i \right)}  > L^{\left( 0 \right)} } \right)}
\end{array} \vspace{-0.25cm}
\end{equation}
\noindent where, for simplicity, the short-hand ${\mathbbm{1}}\left( y > x \right) = {\mathbbm{1}}_{\left[ x ,\infty \right]} \left( y \right) = {\mathbbm{1}}_{\left[0, y \right]} \left( x \right)$ is used.
\begin{remark} \label{Remark__GeneralizationMIMO}
The SINR in \eqref{Eq_14} is inherently formulated for application to single-input-single-output systems. It can find application, however, to system setups where multiple-input-multiple-output transmission schemes are used and for which the end-to-end power gains of intended and interfering links can be formulated in terms of a gamma random variable. Further details about this generalization are available in \cite{MDR_COMMLPeng}. Relevant examples of multiple-input-multiple-output transmission schemes where this equivalency holds are illustrated in \cite[Slides 90, 91]{MDR_Tutorial}. \hfill $\Box$
\end{remark}

By using the Moment Generating Function (MGF) approach in \cite{MDR_TCOMrate}, $\mathcal{R}$ can be formulated as:
\begin{equation}
\label{Eq_16}
\begin{array}{l}
{\mathcal{R}} = \sum\nolimits_{s \in {\mathsf{S}}} {{\mathbb{E}}_{L_s^{\left( 0 \right)} } \left\{ {{\mathbb{E}}\left\{ {\left. {\ln \left( {1 + {\rm{SINR}}\left( {L_s^{\left( 0 \right)} } \right)} \right)} \right|L_s^{\left( 0 \right)} } \right\}\Pr \left\{ {L^{\left( 0 \right)}  = L_s^{\left( 0 \right)} } \right\}} \right\}} \\
= \sum\nolimits_{s \in {\mathsf{S}}} {{\mathbb{E}}_{L_s^{\left( 0 \right)} } \left\{ {\left( {\int\nolimits_0^{\infty } {\exp \left( { - \sigma _N^2 z} \right){\mathcal{\bar M}}_{g_s^{\left( 0 \right)} } \left( {\left. {\frac{{P_{{\rm{RB}}} G^{\left( 0 \right)} z}}{{L_s^{\left( 0 \right)} }}} \right|L_s^{\left( 0 \right)} } \right){\mathcal{M}}_{I_{{\rm{agg}}} \left( {L_s^{\left( 0 \right)} } \right)} \left( {\left. z \right|L_s^{\left( 0 \right)} } \right)\frac{{dz}}{z}} } \right)\Upsilon _s \left( {L_s^{\left( 0 \right)} } \right)} \right\}}  \\
 = \sum\nolimits_{s \in {\mathsf{S}}} {\int\nolimits_0^{\infty } {\int\nolimits_0^{\infty } {\exp \left( { - \sigma _N^2 z} \right){\mathcal{\bar M}}_{g_s^{\left( 0 \right)} } \left( {\left. {\frac{{P_{{\rm{RB}}} G^{\left( 0 \right)} z}}{x}} \right|x} \right){\mathcal{M}}_{I_{{\rm{agg}}} \left( x \right)} \left( {\left. z \right|x} \right)} \Upsilon _s \left( x \right)f_{L_s^{\left( 0 \right)} } \left( x \right)\frac{{dzdx}}{z}} }
\end{array}
\end{equation}
\noindent where ${\mathcal{\bar M}}_{g_s^{\left( 0 \right)} } \left( {\left. z \right|x} \right) = 1 - {\mathcal{M}}_{g_s^{\left( 0 \right)} } \left( {\left. z \right|x} \right)$, $f_{L_s^{\left( 0 \right)} } \left(  \cdot  \right)$ is the PDF of $L_s^{\left( 0 \right)}$, $\Upsilon _s \left( {L_s^{\left( 0 \right)} } \right) = \Pr \left\{ {L^{\left( 0 \right)}  = L_s^{\left( 0 \right)} } \right\} = \prod\nolimits_{r \ne s \in {\mathsf{S}}} {\Pr \left\{ {\left. {L_r^{\left( 0 \right)}  > L_s^{\left( 0 \right)} } \right|L_s^{\left( 0 \right)} } \right\}}$ follows from the independence of $\Psi _{{\rm{BS}},s}$ for ${s \in {\mathsf{S}}}$, ${\mathcal{M}}_{g_s^{\left( 0 \right)} } \left( {\left. {{z \mathord{\left/ {\vphantom {z {L_s^{\left( 0 \right)} }}} \right. \kern-\nulldelimiterspace} {L_s^{\left( 0 \right)} }}} \right|x} \right)$ $= {\mathbb{E}}_{g_s^{\left( 0 \right)} } \left\{ {\left. {\exp \left( { - z\left( {{{g_s^{\left( 0 \right)} } \mathord{\left/ {\vphantom {{g_s^{\left( 0 \right)} } {L_s^{\left( 0 \right)} }}} \right. \kern-\nulldelimiterspace} {L_s^{\left( 0 \right)} }}} \right)} \right)} \right|L_s^{\left( 0 \right)} } \right\}$ is the MGF of ${g_s^{\left( 0 \right)} }$ conditioned on ${L_s^{\left( 0 \right)} }$, ${\mathcal{M}}_{I_{{\rm{agg}}} \left( {L_s^{\left( 0 \right)} } \right)} \left( {\left. z \right|L_s^{\left( 0 \right)} } \right)$ $= {\mathbb{E}}\left\{ {\left. {\exp \left( { - zI_{{\rm{agg}}} \left( {L_s^{\left( 0 \right)} } \right)} \right)} \right|L_s^{\left( 0 \right)} } \right\}$ is the MGF of ${I_{{\rm{agg}}} \left( {L^{\left( 0 \right)} = L_s^{\left( 0 \right)} } \right)}$ in \eqref{Eq_15} given ${L^{\left( 0 \right)} =L_s^{\left( 0 \right)} }$.

By using the Gil-Pelaez approach in \cite{MDR_COMMLPeng}, ${\mathcal{C}}\left( \cdot \right)$ can be formulated as follows:
\begin{equation}
\label{Eq_17}
\begin{array}{l}
{\mathcal{C}}\left( {\rm{T}} \right) = \sum\nolimits_{s \in {\mathsf{S}}} {{\mathbb{E}}_{L_s^{\left( 0 \right)} } \left\{ {\Pr \left\{ {\left. {{\rm{SINR}}\left( {L_s^{\left( 0 \right)} } \right) \ge {\rm{T}}} \right|L_s^{\left( 0 \right)} } \right\}\Pr \left\{ {L^{\left( 0 \right)}  = L_s^{\left( 0 \right)} } \right\}} \right\}} \\
= \sum\nolimits_{s \in {\mathsf{S}}} {{\mathbb{E}}_{L_s^{\left( 0 \right)} } \left\{ {\Pr \left\{ {\left. {I_{{\rm{agg}}} \left( {L_s^{\left( 0 \right)} } \right) \le \frac{1}{{\rm{T}}}\frac{{P_{{\rm{RB}}} G^{\left( 0 \right)} g_s^{\left( 0 \right)} }}{{L_s^{\left( 0 \right)} }} - \sigma _N^2 } \right|L_s^{\left( 0 \right)} } \right\}\Upsilon _s \left( {L_s^{\left( 0 \right)} } \right)} \right\}}  \\
= \sum\limits_{s \in {\mathsf{S}}} {\int\limits_0^\infty  {\left( {\frac{1}{2} - \frac{1}{\pi }\int\limits_0^\infty  {{\mathop{\rm Im}\nolimits} \left\{ {\exp \left( {{\mathbbm{j}}z\sigma _N^2 } \right){\mathcal{M}}_{g_s^{\left( 0 \right)} } \left( {\left. {{\mathbbm{j}}z\frac{{P_{{\rm{RB}}} G^{\left( 0 \right)} }}{{{\rm{T}}x}}} \right|x} \right){\mathcal{M}}_{I_{{\rm{agg}}} \left( x \right)} \left( {\left. {{\mathbbm{j}}z} \right|x} \right)} \right\}\frac{{dz}}{z}} } \right)\Upsilon _s \left( x \right)f_{L_s^{\left( 0 \right)} } \left( x \right)dx} }
\end{array}
\end{equation}
\noindent where ${\mathop{\rm Im}\nolimits} \left\{  \cdot  \right\}$ is the imaginary part operator and $\mathbbm{j}$ is the imaginary unit.

The expressions in \eqref{Eq_16} and \eqref{Eq_17} are general and, in particular, are applicable to any link state, channel, and radiation pattern models. Depending on the chosen models, however, their computation may not be either mathematically or numerically possible. Usually, the distribution of ${g_s^{\left( 0 \right)} }$ does not pose any relevant issues. If, for example, ${g_s^{\left( 0 \right)} }$ follows a gamma distribution (Section \ref{ChannelModeling}): ${\mathcal{M}}_{g_s^{\left( 0 \right)} } \left( z \right) = \left( {1 + z\left( {{{\Omega _s } \mathord{\left/ {\vphantom {{\Omega _s } {m_s }}} \right. \kern-\nulldelimiterspace} {m_s }}} \right)} \right)^{ - m_s }$ in \eqref{Eq_16} and \eqref{Eq_17}. The IM-based approach provides a methodology for efficiently computing \eqref{Eq_16} and \eqref{Eq_17} under general system models. Also, it leads to mathematical frameworks that provide insight for system design. For brevity, in the sequel, we focus our attention only on the ASE. Similar comments apply to the PT.
\vspace{-0.50cm}
\section{The Intensity Matching Approach} \label{IntensityMatching} \vspace{-0.25cm}
Besides ${\mathcal{M}}_{g_s^{\left( 0 \right)} } \left( {\left.  \cdot  \right| \cdot } \right)$, three functions are needed for computing $\mathcal{R}$ in \eqref{Eq_16}: $f_{L_s^{\left( 0 \right)} } \left(  \cdot  \right)$, $\Upsilon _s \left(  \cdot  \right)$, and ${\mathcal{M}}_{I_{{\rm{agg}}} \left(  \cdot  \right)} \left( {\left.  \cdot  \right| \cdot } \right)$. For arbitrary link state, channel, and radiation pattern models, they can be formulated in mathematical terms by invoking the displacement theorem of PPPs \cite{Blaszczyszyn_Infocom2013}. In simple but general terms, it can be formulated as follows. Let ${\Psi _{{\rm{BS}},s} }$ for $s \in \mathsf{S}$ be non-homogeneous and independent PPPs of density $\lambda _{{\rm{BS}},s} \left( r \right) = \lambda _{{\rm{BS}}} p_s \left( r \right)$. The sets of path-loss $\Phi _s  = \left\{ {L_s^{\left( n \right)} } \right\}_{{\rm{BS}}^{\left( n \right)}  \in \Psi _{{\rm{BS}},s} }$ for $s \in \mathsf{S}$ are non-homogeneous and independent PPPs on ${\mathbb{R}}^ +$ with intensity measure:
\begin{equation}
\label{Eq_18}
\begin{array}{l}
\hspace{-0.70cm} \Lambda _{\Phi _s } \left( {\left[ {0,\xi } \right)} \right) = {\mathbb{E}}\left\{ {\Phi _s \left( {\left[ {0,\xi } \right)} \right)} \right\} = {\mathbb{E}}\left\{ {\sum\nolimits_{{\rm{BS}}^{\left( n \right)}  \in \Psi _{{\rm{BS}},s} } {\mathbbm{1}_{\left[ {0,\xi } \right]} \left( {L_s^{\left( n \right)} } \right)} } \right\} \\
\hspace{-0.70cm} \mathop  = \limits^{\left( a \right)} 2\pi \lambda _{{\rm{BS}}} {\mathbb{E}}_{{\mathcal{X}}_s } \left\{ {\int\nolimits_0^\infty  {\Pr \left\{ {{{l_s \left( r \right)} \mathord{\left/
 {\vphantom {{l_s \left( r \right)} {{\mathcal{X}}_s }}} \right.
 \kern-\nulldelimiterspace} {{\mathcal{X}}_s }} \in \left[ {0,\xi } \right)} \right\}p_s \left( r \right)rdr} } \right\} = 2\pi \lambda _{{\rm{BS}}} {\mathbb{E}}_{{\mathcal{X}}_s } \left\{ {\int\nolimits_0^{\left( {{{{\mathcal{X}}_s \xi } \mathord{\left/
 {\vphantom {{{\mathcal{X}}_s \xi } {\kappa _s }}} \right.
 \kern-\nulldelimiterspace} {\kappa _s }}} \right)^{{1 \mathord{\left/
 {\vphantom {1 {\alpha _s }}} \right.
 \kern-\nulldelimiterspace} {\alpha _s }}} } {p_s \left( r \right)rdr} } \right\}
\end{array}
\end{equation}
\noindent where (a) directly follows from the displacement theorem of PPPs \cite[Th. 1.3.9]{Blaszczyszyn_Infocom2013}.

From \eqref{Eq_18}, $f_{L_s^{\left( 0 \right)} } \left(  \cdot  \right)$ and $\Upsilon _s \left(  \cdot  \right)$ follow from the void probability of PPPs \cite[Th. 1.1.5]{BaccelliBook2009}:
\begin{equation}
\label{Eq_19}
\begin{array}{l}
 f_{L_s^{\left( 0 \right)} } \left( x \right) =  - {{d\Pr \left\{ {L_s^{\left( 0 \right)}  > x} \right\}} \mathord{\left/
 {\vphantom {{d\Pr \left\{ {L_s^{\left( 0 \right)}  > x} \right\}} {dx}}} \right.
 \kern-\nulldelimiterspace} {dx}} =  - {{d\Pr \left\{ {\Phi _s \left( {\left[ {0,x} \right)} \right) = 0} \right\}} \mathord{\left/
 {\vphantom {{d\Pr \left\{ {\Phi _s \left( {\left[ {0,x} \right)} \right) = 0} \right\}} {dx}}} \right.
 \kern-\nulldelimiterspace} {dx}} \\
 \hspace{1.5cm}\mathop  = \limits^{\left( a \right)}  - {{d\exp \left( { - \Lambda _{\Phi _s } \left( {\left[ {0,x} \right)} \right)} \right)} \mathord{\left/
 {\vphantom {{d\exp \left( { - \Lambda _{\Phi _s } \left( {\left[ {0,x} \right)} \right)} \right)} {dx}}} \right.
 \kern-\nulldelimiterspace} {dx}} = \Lambda _{\Phi _s }^{\left( 1 \right)} \left( {\left[ {0,x} \right)} \right)\exp \left( { - \Lambda _{\Phi _s } \left( {\left[ {0,x} \right)} \right)} \right)
\end{array}
\end{equation}
\begin{equation}
\label{Eq_20}
\begin{array}{l}
 \Upsilon _s \left( x \right) = \prod\nolimits_{r \ne s \in {\mathsf{S}}} {\Pr \left\{ {\left. {L_r^{\left( 0 \right)}  > x} \right|x} \right\}}  = \prod\nolimits_{r \ne s \in {\mathsf{S}}} {\Pr \left\{ {\left. {\Phi _r \left( {\left[ {0,x} \right)} \right) = 0} \right|x} \right\}}  \\
 \hspace{1.2cm} \mathop  = \limits^{\left( a \right)} \prod\nolimits_{r \ne s \in {\mathsf{S}}} {\exp \left( { - \Lambda _{\Phi _r } \left( {\left[ {0,x} \right)} \right)} \right)}  = \exp \left( { - \sum\nolimits_{r \ne s \in {\mathsf{S}}} {\Lambda _{\Phi _r } \left( {\left[ {0,x} \right)} \right)} } \right)
\end{array}
\end{equation}
\noindent where (a) originates from \cite[Th. 1.1.5]{BaccelliBook2009} and $\Lambda _{\Phi _s }^{\left( 1 \right)} \left( {\left[ { \cdot , \cdot } \right)} \right)$ is the first derivative of $\Lambda _{\Phi _s } \left( {\left[ { \cdot , \cdot } \right)} \right)$.

From \eqref{Eq_18}, ${\mathcal{M}}_{I_{{\rm{agg}}} \left(  \cdot  \right)} \left( {\left.  \cdot  \right| \cdot } \right)$ follows from the Laplace functional of PPPs \cite[Prop. 1.2.2]{BaccelliBook2009}:
\begin{equation}
\label{Eq_21}
\begin{array}{l}
 {\mathcal{M}}_{I_{{\rm{agg}}} \left( x \right)} \left( {\left. z \right|x} \right) = {\mathbb{E}}\left\{ {\exp \left( { - zI_{{\rm{agg}}} \left( x \right)} \right)} \right\} \\
  \hspace{2.65cm} = {\mathbb{E}}\left\{ {\exp \left( { - z\sum\nolimits_{\tilde s \in {\mathsf{S}}} {\sum\nolimits_{{\rm{BS}}^{\left( i \right)}  \in \Psi^{(\rm{I})} _{{\rm{BS}},\tilde s} } {\left( {{{P_{{\rm{RB}}} G^{\left( i \right)} g_{\tilde s}^{\left( i \right)} } \mathord{\left/
 {\vphantom {{P_{{\rm{RB}}} G^{\left( i \right)} g_{\tilde s}^{\left( i \right)} } {L_{\tilde s}^{\left( i \right)} }}} \right.
 \kern-\nulldelimiterspace} {L_{\tilde s}^{\left( i \right)} }}} \right)} \mathbbm{1}\left( {L_{\tilde s}^{\left( i \right)}  > x} \right)} } \right)} \right\} \\
 \hspace{2.65cm} \mathop  = \limits^{\left( a \right)} \prod\nolimits_{\tilde s \in {\mathsf{S}}} {{\mathbb{E}}\left\{ {\exp \left( { - z\sum\nolimits_{{\rm{BS}}^{\left( i \right)}  \in \Psi^{(\rm{I})} _{{\rm{BS}},\tilde s} } {\left( {{{P_{{\rm{RB}}} G^{\left( i \right)} g_{\tilde s}^{\left( i \right)} } \mathord{\left/
 {\vphantom {{P_{{\rm{RB}}} G^{\left( i \right)} g_{\tilde s}^{\left( i \right)} } {L_{\tilde s}^{\left( i \right)} }}} \right.
 \kern-\nulldelimiterspace} {L_{\tilde s}^{\left( i \right)} }}} \right)} \mathbbm{1}\left( {L_{\tilde s}^{\left( i \right)}  > x} \right)} \right)} \right\}}  \\
 \hspace{2.65cm} \mathop  = \limits^{\left( b \right)} \prod\nolimits_{\tilde s \in {\mathsf{S}}} {\exp \left( { - \left( {1 - p_{{\rm{off}}} } \right){\mathbb{E}}_{G^{\left( i \right)} ,g_{\tilde s}^{\left( i \right)} } \left\{ {{\rm Z} _{\tilde s} \left( {P_{{\rm{RB}}} z,x;G^{\left( i \right)} ,g_{\tilde s}^{\left( i \right)} } \right)} \right\}} \right)} \\
\hspace{2.65cm} \mathop  = \limits^{\left( c \right)} \prod\nolimits_{\tilde s \in {\mathsf{S}}} {\exp \left( { - \left( {1 - p_{{\rm{off}}} } \right)\int\nolimits_x^\infty  {\left( {1 - \Xi _{\tilde s} \left( {P_{{\rm{RB}}} z,y} \right)} \right)\Lambda _{\Phi _{\tilde s} }^{\left( 1 \right)} \left( {\left[ {0,y} \right)} \right)dy} } \right)}
\end{array}
\end{equation}
\noindent where the following short-hands are introduced:
\begin{equation}
\label{Eq_22}
\begin{array}{l}
 {\rm Z} _{\tilde s} \left( {P_{{\rm{RB}}} z,x;G^{\left( i \right)} ,g_{\tilde s}^{\left( i \right)} } \right) = \int\nolimits_x^\infty  {\left( {1 - \exp \left( { - z{{P_{{\rm{RB}}} G^{\left( i \right)} g_{\tilde s}^{\left( i \right)} } \mathord{\left/
 {\vphantom {{P_{{\rm{RB}}} G^{\left( i \right)} g_{\tilde s}^{\left( i \right)} } y}} \right.
 \kern-\nulldelimiterspace} y}} \right)} \right)\Lambda _{\Phi _{\tilde s} }^{\left( 1 \right)} \left( {\left[ {0,y} \right)} \right)dy}  \\
 \Xi _{\tilde s} \left( {P_{{\rm{RB}}} z,y} \right) = {\mathbb{E}}_{G^{\left( i \right)} ,g_{\tilde s}^{\left( i \right)} } \left\{ {\exp \left( { - z{{P_{{\rm{RB}}} G^{\left( i \right)} g_{\tilde s}^{\left( i \right)} } \mathord{\left/
 {\vphantom {{P_{{\rm{RB}}} G^{\left( i \right)} g_{\tilde s}^{\left( i \right)} } y}} \right.
 \kern-\nulldelimiterspace} y}} \right)} \right\}
\end{array}
\end{equation}
\noindent and (a) follows from the independence of ${\Psi^{(\rm{I})} _{{\rm{BS}},\tilde s} }$ for $\tilde s \in \mathsf{S}$, (b) originates from \cite[Prop. 1.2.2]{BaccelliBook2009}, (c) is the same as (b) but the expectation ${\mathbb{E}}_{G^{\left( i \right)} ,g_{\tilde s}^{\left( i \right)} } \left\{  \cdot  \right\}$ is moved inside the integral. The $\left( {1 - p_{{\rm{off}}} } \right)$ factor in \eqref{Eq_21} accounts for the active interfering BSs on a RB based on the load model in Section \ref{LoadModeling}, \textit{i.e.}, $\Lambda _{\Phi _s } \left( {\left[ {0,\xi } \right)} \right) \mapsto \left( {1 - p_{{\rm{off}}} } \right)\Lambda _{\Phi _s } \left( {\left[ {0,\xi } \right)} \right)$ to account for the active interfering BSs.
\vspace{-0.5cm}
\subsection{Motivation} \label{IM_Motivation} \vspace{-0.25cm}
It is possible, in principle, to plug \eqref{Eq_18}-\eqref{Eq_22} in \eqref{Eq_16} and \eqref{Eq_17}, and to obtain a general and exact mathematical approach for computing relevant performance indicators for cellular network design. The resulting expressions are, however, formulated in terms of multi-fold integrals, which are typically numerically intractable, and, more importantly, neither shed light on performance trends nor provide design insight. To overcome this issue, two options are possible: i) to simplify the system model, in order to get mathematically tractable expressions for \eqref{Eq_18}-\eqref{Eq_22} and ii) to introduce approximations in \eqref{Eq_18}-\eqref{Eq_22}, in order to make their computation analytically tractable. The first option has been widely adopted. In particular, the main simplifying assumption that makes \eqref{Eq_18}-\eqref{Eq_22} tractable relies on considering a single-state ($S=1$) link model. With this simplifying assumption, $\Lambda _{\Phi _s } \left( {\left[ { \cdot , \cdot } \right)} \right)$ can be formulated in closed-form and the integral in \eqref{Eq_22} is usually (\textit{e.g.}, for omni-directional antennas) computable in closed-form as well. The details can be found in \cite{Blaszczyszyn_Infocom2013}, \cite[Slide 107]{MDR_Tutorial}. Recently, however, it has been shown that making simplistic assumptions on link state and path-loss models lead to inaccurate predictions of the impact of key design parameters \cite{Andrews__LosNlos}-\cite{Lopez-Perez__LosNlos}. Motivated by these considerations, we introduce an approximation that provides closed-form expressions for \eqref{Eq_18}-\eqref{Eq_22}, making \eqref{Eq_16} and \eqref{Eq_17} computationally affordable, as well as that offers design insight for system-level optimization.
\vspace{-0.5cm}
\subsection{Rationale} \label{IM_Ratoinale} \vspace{-0.25cm}
The rationale behind the IM-based approach originates from direct inspection of \eqref{Eq_16}-\eqref{Eq_22}. \\
\textbf{1)} The two-fold integrals in \eqref{Eq_16} and \eqref{Eq_17} are, usually, unlikely amenable to simplifications without reducing the generality of the system model or without considering specific parameters. \\
\textbf{2)} The integral $\widetilde \Lambda _{\Phi _s } \left( {\left. {\left[ {0,\xi } \right)} \right|{\mathcal{X}}_s } \right) = \widetilde \Lambda _{\Phi _s } \left( { {\left[ {0,{\mathcal{X}}_s \xi } \right)} } \right) = \int\nolimits_0^{\left( {{{{\mathcal{X}}_s \xi } \mathord{\left/ {\vphantom {{{\mathcal{X}}_s \xi } {\kappa _s }}} \right. \kern-\nulldelimiterspace} {\kappa _s }}} \right)^{{1 \mathord{\left/ {\vphantom {1 {\alpha _s }}} \right. \kern-\nulldelimiterspace} {\alpha _s }}} } {p_s \left( r \right)rdr}$ in \eqref{Eq_18} is usually computable in closed-form for typical link state models. Table \ref{IntensityMeasures} provides it for the case studies available in Table \ref{LinkStateModels}\footnote{Notation of Table \ref{IntensityMeasures} -- $\eta _s \left( \xi  \right) = \left( {{\mathcal{X}}_s \xi /\kappa _s } \right)^{1/\alpha _s }$, $s \in \left\{ {{\rm{LOS}},{\rm{NLOS}}} \right\}$. RS = Random Shape, L = Linear, mmW = Empirical mmWave. $\widetilde\Lambda _{{\rm{NLOS}}}^{\left( X \right)} \left( {\left[ {0,{\mathcal{X}}_{{\rm{NLOS}}} \xi } \right)} \right) = \left( {1/2} \right)\eta _{{\rm{NLOS}}}^2 \left( \xi  \right) - \widetilde\Lambda _{{\rm{LOS}}}^{\left( X \right)} \left( {\eta _{{\rm{NLOS}}} \left( \xi  \right),x_{{\rm{NLOS}}}^{\left( X \right)} } \right)$, $X = \left\{ {{\rm{3GPP}},{\rm{L}}} \right\}$; $\widetilde\Lambda _{{\rm{NLOS}}}^{\left( {{\rm{RS}}} \right)} \left( {\left[ {0,{\mathcal{X}}_{{\rm{NLOS}}} \xi } \right)} \right) = \left( {1/2} \right)\eta _{{\rm{NLOS}}}^2 \left( \xi  \right) - \widetilde\Lambda _{{\rm{LOS}}}^{\left( {{\rm{RS}}} \right)} \left( {\eta _{{\rm{NLOS}}} \left( \xi  \right)} \right)$. $x_s^{\left( {{\rm{3GPP}}} \right)}  = \kappa _s \left( {a_{{\rm{3G}}} /c_{{\rm{3G}}} } \right)^{\alpha _s }$, $x_s^{\left( {\rm{L}} \right)}  = \kappa _s \left( {c_{\rm{L}} /a_{\rm{L}}  - b_{\rm{L}} /a_{\rm{L}} } \right)^{\alpha _s }$. ${\cal H}\left( x \right) = 1$ if $x\ge0$, ${\cal H}\left( x \right) = 0$ if $x<0$; $\overline {\cal H} \left( x \right) = 1 - {\cal H}\left( x \right)$. ${\cal K}_1  = a_{{\rm{mm}}}^{ - 2}$, ${\cal K}_2  = e^{c_{{\rm{mm}}} } \left( {a_{{\rm{mm}}}  + b_{{\rm{mm}}} } \right)^{ - 2}$, $R = a_{{\rm{mm}}} b_{{\rm{mm}}}^{ - 1} c_{{\rm{mm}}}$, $W = \left( {a_{{\rm{mm}}}  + b_{{\rm{mm}}} } \right)b_{{\rm{mm}}}^{ - 1} c_{{\rm{mm}}}$,$Q_s  = a_{{\rm{mm}}} \kappa _s^{ - 1/\alpha _s }$, $T_s  = b_{{\rm{mm}}} \kappa _s^{ - 1/\alpha _s }$, $V_s  = \left( {a_{{\rm{mm}}}  + b_{{\rm{mm}}} } \right)\kappa _s^{ - 1/\alpha _s }$, $Z_s  = \kappa _s \left( {b_{{\rm{mm}}}^{ - 1} c_{{\rm{mm}}} } \right)^{\alpha _s }$. $\widetilde\Lambda _{{\rm{OUT}}}^{\left( {{\rm{mmW}}} \right)} \left( {\left[ {0,{\mathcal{X}}_{{\rm{OUT}}} \xi } \right)} \right) = 0$.}. The issue is the computation of the expectation with respect to the shadowing, \textit{i.e.}, $\mathcal{X}_s$. This is due to the intractable PDF of the log-normal distribution (see Section \ref{ChannelModeling}). It is worth mentioning that, if a multi-state path-loss model is considered, the impact of shadowing cannot be taken into account, differently from \cite{Blaszczyszyn_Infocom2013}, by simply scaling the density of BSs. \\
\textbf{3)} The expectation with respect to small-scale fading, \textit{i.e.}, ${g_{\tilde s}^{\left( i \right)} }$, in \eqref{Eq_21} or \eqref{Eq_22} is usually computable for several fading models, \textit{e.g.}, the gamma distribution. The expectation with respect to the antenna radiation pattern, \textit{i.e.}, ${G^{\left( i \right)} }$, is usually difficult to be computed for general antenna models. Even if it was computable, the resulting integral over $y$ would not be, in general, solvable. The use of the multi-ball (approximated) model in \eqref{Eq_11} allows one, on the other hand, to compute the expectation without affecting the computation of the resulting integral over $y$. \\
\textbf{4)} The integral over $y$ in \eqref{Eq_21} or \eqref{Eq_22} is, in general, not computable in closed-form because of the complicated expression of the intensity measure in \eqref{Eq_18}, which is not even available in closed-form due to the need of taking into account the shadowing for cell association (see \eqref{Eq_2}). \\
\textbf{5)} Given antenna radiation pattern and small-scale fading models, the direct inspection of \eqref{Eq_16}-\eqref{Eq_22} brings to our attention that ASE and PT are uniquely determined by the intensity measure in \eqref{Eq_18}. Given two different system models, in other words, the resulting ASE and PT would be the same if they happen to have, for every link state, the same intensity measure in \eqref{Eq_18}.

\begin{table}[!t] \footnotesize
\centering
\caption{$\widetilde \Lambda _{\Phi _s } \left( { {\left[ {\cdot, \cdot} \right)} } \right)$ of the link state models in Table \ref{LinkStateModels}. The notation is provided in footnote 2. \vspace{-0.25cm}}
\newcommand{\tabincell}[2]{\begin{tabular}{@{}#1@{}}#2\end{tabular}}
\begin{tabular}{|c|} \hline
$\begin{array}{l}
 \widetilde\Lambda _{{\rm{LOS}}}^{\left( {{\rm{3GPP}}} \right)} \left( {\left[ {0,{\mathcal{X}}_{{\rm{LOS}}} \xi } \right)} \right) = \widetilde\Lambda _{{\rm{LOS}}}^{\left( {{\rm{3GPP}}} \right)} \left( {\eta _{{\rm{LOS}}} \left( \xi  \right),x_{{\rm{LOS}}}^{\left( {{\rm{3GPP}}} \right)} } \right) \\
 = \overline {\cal H} \left( {x - x_{{\rm{LOS}}}^{\left( {{\rm{3GPP}}} \right)} } \right) \left( {b_{{\rm{3G}}}^2 \left( {c_{{\rm{3G}}}  - 1} \right)\left( {e^{ - \eta _{{\rm{LOS}}} \left( \xi  \right)/b_{{\rm{3G}}} }  - 1} \right) + b_{{\rm{3G}}} \left( {c_{{\rm{3G}}}  - 1} \right)e^{ - \eta _{{\rm{LOS}}} \left( \xi  \right)/b_{{\rm{3G}}} } \eta _{{\rm{LOS}}} \left( \xi  \right) + \left( {1/2} \right)c_{{\rm{3G}}} \eta _{{\rm{LOS}}}^2 \left( \xi  \right)} \right) \\
 + {\cal H}\left( {x - x_{{\rm{LOS}}}^{\left( {{\rm{3GPP}}} \right)} } \right)\left( { - a_{{\rm{3G}}}^2 /c_{{\rm{3G}}}  + \left( {b_{{\rm{3G}}}^2  + a_{{\rm{3G}}} b_{{\rm{3G}}} /c_{{\rm{3G}}}  - a_{{\rm{3G}}} b_{{\rm{3G}}} } \right)e^{ - a_{{\rm{3G}}} /\left( {b_{{\rm{3G}}} c_{{\rm{3G}}} } \right)} } \right) \\
 + {\cal H}\left( {x - x_{{\rm{LOS}}}^{\left( {{\rm{3GPP}}} \right)} } \right)\left( { - b_{{\rm{3G}}} e^{ - \eta _{{\rm{LOS}}} \left( \xi  \right)/b_{{\rm{3G}}} } \left( {b_{{\rm{3G}}}  - a_{{\rm{3G}}}  + \eta _{{\rm{LOS}}} \left( \xi  \right)} \right) + a_{{\rm{3G}}} \eta _{{\rm{LOS}}} \left( \xi  \right)} \right)
 \end{array}$ \\ \hline
$\widetilde\Lambda _{{\rm{LOS}}}^{\left( {{\rm{RS}}} \right)} \left( {\left[ {0,{\mathcal{X}}_{{\rm{LOS}}} \xi } \right)} \right) = \widetilde\Lambda _{{\rm{LOS}}}^{\left( {{\rm{RS}}} \right)} \left( \eta _{{\rm{LOS}}} \left( \xi  \right) \right) = a_{{\rm{RS}}} \left( {b_{{\rm{RS}}}^{ - 2}  - b_{{\rm{RS}}}^{ - 2} \left( {1 + b_{{\rm{RS}}} \eta _{{\rm{LOS}}} \left( \xi  \right)} \right)e^{ - b_{{\rm{RS}}} \eta _{{\rm{LOS}}} \left( \xi  \right)} } \right)$ \\ \hline
$\begin{array}{l}
 \widetilde\Lambda _{{\rm{LOS}}}^{\left( {\rm{L}} \right)} \left( {\left[ {0,{\mathcal{X}}_{{\rm{LOS}}} \xi } \right)} \right) = \widetilde\Lambda _{{\rm{LOS}}}^{\left( {\rm{L}} \right)} \left( {\eta _{{\rm{LOS}}} \left( \xi  \right),x_{{\rm{LOS}}}^{\left( {\rm{L}} \right)} } \right) = \overline {\cal H} \left( {x - x_{{\rm{LOS}}}^{\left( {\rm{L}} \right)} } \right)\left( {1/2 - b_{\rm{L}} /2 - a_{\rm{L}} \eta _{{\rm{LOS}}} \left( \xi  \right)/3} \right)\eta _{{\rm{LOS}}}^2 \left( \xi  \right) \\
 \hspace{2.55cm} + 1/\left( {6a_{\rm{L}}^2 } \right)\left( {\left( {c_{\rm{L}}  - b_{\rm{L}} } \right)^3  + 3a_{\rm{L}}^2 \eta _{{\rm{LOS}}}^2 \left( \xi  \right)\left( {1 - c_{\rm{L}} } \right)} \right){\cal H}\left( {x - x_{{\rm{LOS}}}^{\left( {\rm{L}} \right)} } \right)
  \end{array}$ \\ \hline
$\begin{array}{l}
\widetilde\Lambda _{{\rm{LOS}}}^{\left( {{\rm{mmW}}} \right)} \left( {\left[ {0,{\mathcal{X}}_{{\rm{LOS}}} \xi } \right)} \right) = \widetilde \Upsilon _0 \left( {{\mathcal{X}}_{{\rm{LOS}}} \xi ;{\rm{LOS}}} \right); \quad \widetilde\Lambda _{{\rm{NLOS}}}^{\left( {{\rm{mmW}}} \right)} \left( {\left[ {0,{\mathcal{X}}_{{\rm{NLOS}}} \xi } \right)} \right) = \widetilde \Upsilon _1 \left( {{\mathcal{X}}_{{\rm{NLOS}}} \xi ;{\rm{NLOS}}} \right) - \widetilde \Upsilon _0 \left( {{\mathcal{X}}_{{\rm{NLOS}}} \xi ;{\rm{NLOS}}} \right) \\
 \widetilde \Upsilon _0 \left( {x;s} \right) = {\cal K}_2 \left( {e^{ - W}  + We^{ - W}  - e^{ - V_s x^{1/\alpha _s } }  - V_s x^{1/\alpha _s } e^{ - V_s x^{1/\alpha _s } } } \right){\cal H}\left( {x - Z_s } \right) \\
 \hspace{1.27cm} + {\cal K}_1 \left( {1 - e^{ - Q_x x^{1/\alpha _s } }  - Q_s x^{1/\alpha _s } e^{ - Q_s x^{1/\alpha _s } } } \right)\overline {\cal H} \left( {x - Z_s } \right) + {\cal K}_1 \left( {1 - e^{ - R}  - Re^{ - R} } \right){\cal H}\left( {x - Z_s } \right) \\
  \widetilde \Upsilon _1 \left( {x;s} \right) = \left( {1/2} \right)\kappa _s^{ - 2/\alpha _s } x^{2/\alpha _s } \overline {\cal H} \left( {x - Z_s } \right) + \left( {1/2} \right)\left( {b_{{\rm{mm}}}^{ - 1} c_{{\rm{mm}}} } \right)^2 {\cal H}\left( {x - Z_s } \right) \\
 \hspace{1.27cm} + b_{{\rm{mm}}}^{ - 2} e^{c_{{\rm{mm}}} } \left( {e^{ - c_{{\rm{mm}}} }  + e^{ - c_{{\rm{mm}}} } c_{{\rm{mm}}}  - e^{ - T_s x^{1/\alpha _s } }  - T_s x^{1/\alpha _s } e^{ - T_s x^{1/\alpha _s } } } \right){\cal H}\left( {x - Z_s } \right)
 \end{array}$ \\ \hline
\end{tabular}
\label{IntensityMeasures} \vspace{-0.65cm}
\end{table}
Moving from these considerations, the IM-based approach is based on the following. i) No attempt for simplifying \eqref{Eq_16} and \eqref{Eq_17} is made. In Section \ref{Trends}, however, asymptotic expressions that offer design guidelines in relevant operating regimes (\textit{e.g.}, dense and sparse networks) are provided. ii) For every link state, the intensity measure in \eqref{Eq_18} is approximated with another intensity measure that is more suitable for mathematical analysis and that, under the general system model of Section \ref{SystemModel}, leads to closed-form expressions for \eqref{Eq_19}-\eqref{Eq_22}. With the aid of this methodology, ASE and PT are formulated in a tractable (and insightful) two-fold integral.
\vspace{-0.5cm}
\subsection{Proposed Methodology} \label{IM_Methodology} \vspace{-0.25cm}
Based on this rationale, the IM-based approach requires the choice of: i) an approximated intensity measure that, based on \eqref{Eq_18}-\eqref{Eq_22}, is suitable for mathematical analysis and ii) a criterion for computing, based on the exact intensity measure in \eqref{Eq_18}, the set of its constituent parameters. In addition, the choice of the approximated intensity and of the matching criterion need to be formulated in a way that the impact of all relevant design parameters can be still identified. In the remainder of this paper, to avoid ambiguity, all the parameters of the system model corresponding to the IM-based approximation are identified by adding $\widehat {(\cdot)}$ to the original parameter.
\setcounter{paragraph}{0}
\paragraph{Approximated Intensity Measure} It is chosen so that it corresponds to a system model where: i) the link state model is the multi-ball model in \eqref{Eq_1}, \textit{i.e.}, $p_s \left( r \right) \mapsto \widehat p_s \left( r \right)$, ii) $\mathcal{X}_s \mapsto \mathcal{\widehat X}_s = 1$ for $s \in \mathsf{S}$, and iii) $\lambda _{{\rm{BS}},s} \left( r \right) \mapsto \widehat \lambda _{{\rm{BS}},s} \left( r \right) = \lambda _{{\rm{BS}},s} \left( r \right){\mathbb{E}}\left\{ {{\mathcal{X}}_s^{{2 \mathord{\left/ {\vphantom {2 {\alpha _s }}} \right. \kern-\nulldelimiterspace} {\alpha _s }}} } \right\}\mathop  = \limits^{\left( a \right)} \lambda _{{\rm{BS}},s} \left( r \right)\exp \left( {{{2 \widetilde \mu _s } \mathord{\left/ {\vphantom {{2 \widetilde \mu _s } {\alpha _s }}} \right. \kern-\nulldelimiterspace} {\alpha _s }} + {{2 \widetilde \sigma _s^2 } \mathord{\left/ {\vphantom {{2 \widetilde \sigma _s^2 } {\alpha _s^2 }}} \right. \kern-\nulldelimiterspace} {\alpha _s^2 }}} \right)$ for $s \in \mathsf{S}$, where (a) follows from the fractional moments of a log-normal distribution with $\widetilde \mu _s  = \mu _s {{\ln \left( {10} \right)} \mathord{\left/ {\vphantom {{\ln \left( {10} \right)} {10}}} \right. \kern-\nulldelimiterspace} {10}}$ and $\widetilde \sigma _s  = \sigma _s {{\ln \left( {10} \right)} \mathord{\left/ {\vphantom {{\ln \left( {10} \right)} {10}}} \right. \kern-\nulldelimiterspace} {10}}$. Accordingly, the approximated intensity measure, $\widehat \Lambda _{\Phi _s } \left( {\left[ { \cdot , \cdot } \right)} \right)$, of \eqref{Eq_18} and its first derivative, $\widehat \Lambda _{\Phi _s }^{\left( 1 \right)} \left( {\left[ { \cdot , \cdot } \right)} \right)$, can be formulated as follows:
\begin{equation}
\label{Eq_23}
\begin{array}{l}
 \Lambda _{\Phi _s } \left( {\left[ {0,\xi } \right)} \right) \approx \widehat \Lambda _{\Phi _s } \left( {\left[ {0,\xi } \right)} \right) = 2\pi \lambda _{{\rm{BS}}} \exp \left( {{{2\widetilde \mu _s } \mathord{\left/
 {\vphantom {{2\widetilde \mu _s } {\alpha _s }}} \right.
 \kern-\nulldelimiterspace} {\alpha _s }} + {{2\widetilde \sigma _s^2 } \mathord{\left/
 {\vphantom {{2\widetilde \sigma _s^2 } {\alpha _s^2 }}} \right.
 \kern-\nulldelimiterspace} {\alpha _s^2 }}} \right)\int\nolimits_0^{\left( {{\xi  \mathord{\left/
 {\vphantom {\xi  {\kappa _s }}} \right.
 \kern-\nulldelimiterspace} {\kappa _s }}} \right)^{{1 \mathord{\left/
 {\vphantom {1 {\alpha _s }}} \right.
 \kern-\nulldelimiterspace} {\alpha _s }}} } {\widehat p_s \left( r \right)rdr}  \\
 \hspace{1.0cm} \mathop  = \limits^{\left( a \right)} 2\pi \lambda _{{\rm{BS}}} \exp \left( {{{2\widetilde \mu _s } \mathord{\left/
 {\vphantom {{2\widetilde \mu _s } {\alpha _s }}} \right.
 \kern-\nulldelimiterspace} {\alpha _s }} + {{2\widetilde \sigma _s^2 } \mathord{\left/
 {\vphantom {{2\widetilde \sigma _s^2 } {\alpha _s^2 }}} \right.
 \kern-\nulldelimiterspace} {\alpha _s^2 }}} \right)\int\nolimits_0^{\left( {{\xi  \mathord{\left/
 {\vphantom {\xi  {\kappa _s }}} \right.
 \kern-\nulldelimiterspace} {\kappa _s }}} \right)^{{1 \mathord{\left/
 {\vphantom {1 {\alpha _s }}} \right.
 \kern-\nulldelimiterspace} {\alpha _s }}} } {\left( {\sum\nolimits_{b = 1}^{{\mathcal{\widehat B}} + 1} {\widehat q_s^{\left[ {\widehat D_{b - 1} ,\widehat D_b } \right]} {\mathbbm{1}}_{\left[ {\widehat D_{b - 1} ,\widehat D_b } \right]} \left( r \right)} } \right)rdr}  \\
 \hspace{1.0cm} \mathop  = \limits^{\left( b \right)} \pi \lambda _{{\rm{BS}}} \Theta _s \sum\nolimits_{b = 1}^{\widehat{\cal B}} {\hat q_s^{\left[ {\widehat D_{b - 1} ,\widehat D_b } \right]} \overline {\cal H} \left( {\xi  - \kappa _s \widehat D_b^{\alpha _s } } \right){\cal H}\left( {\xi  - \kappa _s \widehat D_{b - 1}^{\alpha _s } } \right)\left( {\left( {{\xi  \mathord{\left/
 {\vphantom {\xi  {\kappa _s }}} \right.
 \kern-\nulldelimiterspace} {\kappa _s }}} \right)^{2/\alpha _s }  - \widehat D_{b - 1}^2 } \right)}  \\
 \hspace{1.0cm} + \pi \lambda _{{\rm{BS}}} \Theta _s \sum\nolimits_{b = 1}^{\widehat{\cal B}} {\hat q_s^{\left[ {\widehat D_{b - 1} ,\widehat D_b } \right]} {\cal H}\left( {\xi  - \kappa _s \widehat D_b^{\alpha _s } } \right)\left( {\widehat D_b^2  - \widehat D_{b - 1}^2 } \right)} \\
 \hspace{1.0cm} + \pi \lambda _{{\rm{BS}}} \Theta _s \hat q_s^{\left[ {\widehat D_{\widehat{\cal B}} ,\infty } \right]} {\cal H}\left( {\xi  - \kappa _s \widehat D_{\widehat{\cal B}}^{\alpha _s } } \right)\left( {\left( {{\xi  \mathord{\left/
 {\vphantom {\xi  {\kappa _s }}} \right.
 \kern-\nulldelimiterspace} {\kappa _s }}} \right)^{2/\alpha _s }  - \widehat D_{\widehat{\cal B}}^2 } \right)
\end{array}
\end{equation}
\begin{equation}
\label{Eq_24}
\begin{array}{l}
 \widehat\Lambda _{\Phi _s }^{\left( 1 \right)} \left( {\left[ {0,\xi } \right)} \right) = \pi \lambda _{{\rm{BS}}} \Theta _s \left( {{2 \mathord{\left/
 {\vphantom {2 {\alpha _s }}} \right.
 \kern-\nulldelimiterspace} {\alpha _s }}} \right)\kappa _s^{ - 2/\alpha _s } \xi ^{2/\alpha _s  - 1} \sum\nolimits_{b = 1}^{\widehat{\cal B}} {\hat q_s^{\left[ {\widehat D_{b - 1} ,\widehat D_b } \right]} \overline {\cal H} \left( {\xi  - \kappa _s \widehat D_b^{\alpha _s } } \right){\cal H}\left( {\xi  - \kappa _s \widehat D_{b - 1}^{\alpha _s } } \right)}  \\
 \hspace{2.05cm} + \pi \lambda _{{\rm{BS}}} \Theta _s \left( {{2 \mathord{\left/
 {\vphantom {2 {\alpha _s }}} \right.
 \kern-\nulldelimiterspace} {\alpha _s }}} \right)\kappa _s^{ - 2/\alpha _s } \xi ^{2/\alpha _s  - 1} \hat q_s^{\left[ {\widehat D_{\widehat{\cal B}} ,\infty } \right]} {\cal H}\left( {\xi  - \kappa _s \widehat D_{\widehat{\cal B}}^{\alpha _s } } \right)
\end{array}
\end{equation}
\noindent where (a) follows from \eqref{Eq_1}, (b) by computing the resulting integral, $\Theta _s  = \exp \left( {{{2\tilde \mu _s } \mathord{\left/ {\vphantom {{2\tilde \mu _s } {\alpha _s }}} \right. \kern-\nulldelimiterspace} {\alpha _s }} + {{2\tilde \sigma _s^2 } \mathord{\left/ {\vphantom {{2\tilde \sigma _s^2 } {\alpha _s^2 }}} \right. \kern-\nulldelimiterspace} {\alpha _s^2 }}} \right)$.
\begin{remark} \label{Remark__IM_a}
The assumption $\mathcal{X}_s \mapsto \mathcal{\widehat X}_s = 1$ for $s \in \mathsf{S}$ is primarily made for mathematical tractability, due to the intractable expression of the PDF of the log-normal distribution. It, however, does not imply that log-normal shadowing is neglected in the approximated system model. The impact of shadowing, in fact, \textit{explicitly} appears in $\widehat \lambda _{{\rm{BS}},s} \left( \cdot \right)$ and \textit{implicitly} appears in the set of parameters $\left\{ {{\mathcal{\widehat B}},\widehat D_b ,\widehat q_s^{\left[ { \cdot , \cdot } \right]} } \right\}$ for $b=1, 2, \ldots, {\mathcal{\widehat B}}$, \textit{i.e.}, $\left\{ {{\mathcal{\widehat B}},\widehat D_b ,\hat q_s^{\left[ { \cdot , \cdot } \right]} } \right\} = \Big\{ {\mathcal{\widehat B}}\left( {\mu _s ,\sigma _s , \kappa_s, \alpha _s} \right),$ $\widehat D_b \left( {\mu _s ,\sigma _s ,\kappa_s, \alpha _s } \right),\widehat q_s^{\left[ { \cdot , \cdot } \right]} \left( {\mu _s ,\sigma _s ,\kappa_s, \alpha _s} \right) \Big\}$. Stated differently, the main impact of shadowing explicitly appears in $\widehat \lambda _{{\rm{BS}},s} \left( \cdot \right)$, while its secondary (remaining) impact is absorbed into the modified set of parameters $\left\{ {{\mathcal{\widehat B}},\widehat D_b ,\widehat q_s^{\left[ { \cdot , \cdot } \right]} } \right\}$ for $b=1, 2, \ldots, {\mathcal{\widehat B}}$. This is apparent in \eqref{Eq_25} shown below, where the criterion for obtaining the parameters of the approximation is formally stated. In the rest of the present paper, for simplicity, we avoid this heavy notation and assume it implicitly. \hfill $\Box$
\end{remark}
\begin{remark} \label{Remark__IM_b}
Even though, based on \textit{Remark \ref{Remark__IM_a}}, the impact of shadowing seems to disappear in the mathematical framework, we prove, in Section \ref{Trends}, that it can be clearly identified, in the final expressions of ASE and PT, as a function of the parameters $\left\{ {\widehat D_b ,\widehat q_s^{\left[ { \cdot , \cdot } \right]} ,\exp \left( {{{2 \widetilde \mu _s } \mathord{\left/ {\vphantom {{2 \widetilde \mu _s } {\alpha _s }}} \right. \kern-\nulldelimiterspace} {\alpha _s }} + {{2 \widetilde \sigma _s^2 } \mathord{\left/ {\vphantom {{2 \widetilde \sigma _s^2 } {\alpha _s^2 }}} \right. \kern-\nulldelimiterspace} {\alpha _s^2 }}} \right)} \right\}$. ${{\mathcal{\widehat B}}}$, on the other hand, is decided a priori to keep the computational complexity under control. \hfill $\Box$
\end{remark}
\begin{remark} \label{Remark__IM_c}
The choice of $\widehat \lambda _{{\rm{BS}},s} \left( \cdot \right)$, and, in particular, the scaling factor $\exp \left( {{{2\widetilde \mu _s } \mathord{\left/ {\vphantom {{2\widetilde \mu _s } {\alpha _s }}} \right. \kern-\nulldelimiterspace} {\alpha _s }} + {{2\widetilde \sigma _s^2 } \mathord{\left/ {\vphantom {{2\widetilde \sigma _s^2 } {\alpha _s^2 }}} \right. \kern-\nulldelimiterspace} {\alpha _s^2 }}} \right)$, allows our approach to encompass, as a special case, the (exact) mathematical framework in \cite{Blaszczyszyn_Infocom2013}, which is applicable to the analysis of single-state link and unbounded path-loss models. \hfill $\Box$
\end{remark}
\paragraph{Criterion for ``Matching'' the Intensities} Let ${{\mathcal{\widehat B}}}$ be given (see \textit{Remark \ref{Remark__IM_b}}). $\widehat \Lambda _{\Phi _s } \left( {\left[ {\cdot,\cdot } \right)} \right)$ requires the estimation of $\left\{ {\widehat D_b ,\widehat q_s^{\left[ { \cdot , \cdot } \right]} } \right\}$ for $b=1, 2, \ldots, {\mathcal{\widehat B}}$. The adopted criterion is as follows:
\begin{equation}
\label{Eq_25}
\begin{array}{l}
\hspace{-0.30cm} \mathop {\arg \min }\nolimits_{\left\{ {\widehat D_b } \right\},\left\{ {\widehat q_s^{\left[ { \cdot , \cdot } \right]} } \right\}} \left\{ {\left\| {\ln \left( {\left( {2\pi \lambda _{{\rm{BS}}} } \right)^{ - 1} \Lambda _{\Phi _s } \left( {\left[ {0,x_{{\rm{IM}}} } \right)} \right)} \right) - \ln \left( {\left( {2\pi \lambda _{{\rm{BS}}} } \right)^{ - 1} \widehat \Lambda _{\Phi _s } \left( {\left[ {0,x_{{\rm{IM}}} } \right)} \right)} \right)} \right\|_F^2 } \right\}
\end{array}
\end{equation}
\noindent where ${x_{{\rm{IM}}} }$ is chosen sufficiently large in order to approximate the entire body of the intensity measure and the logarithm function is used to better control the accuracy of the approximation. In particular, $\Lambda _{\Phi _s } \left( {\left[ {0,\xi } \right)} \right) = 2\pi \lambda _{{\rm{BS}}} \int\nolimits_0^{ \infty } {\widetilde \Lambda _{\Phi _s } \left( {\left[ {0,x\xi } \right)} \right)f_{{\mathcal{X}}_s } \left( x \right)dx}$, where $\widetilde \Lambda _{\Phi _s } \left( {\left[ { \cdot , \cdot } \right)} \right)$ is one of the functions available in Table \ref{IntensityMeasures} and $\widehat \Lambda _{\Phi _s } \left( {\left[ { \cdot , \cdot } \right)} \right)$ is the intensity measure in \eqref{Eq_23}.
\begin{remark} \label{Remark__IM_d}
From \eqref{Eq_25}, all the parameters in $\widehat \Lambda _{\Phi _s } \left( {\left[ { \cdot , \cdot } \right)} \right)$ of \eqref{Eq_23} are independent of $\lambda _{{\rm{BS}}}$. \hfill $\Box$
\end{remark}
\begin{remark} \label{Remark__IM_e}
If the link state model is based on empirical data, \textit{e.g.}, on actual footprints of buildings \cite{ACM_MDR}, \eqref{Eq_25} tells us that we can avoid to estimate $p_s \left(  \cdot  \right)$ and can directly use empirical estimates of the associated intensity measure, which, besides the locations and shapes of buildings, depends on path-loss and shadowing models as well. How to compute $\Lambda _{\Phi _s } \left( {\left[ {\cdot,\cdot } \right)} \right)$ based on empirical data is discussed in \cite[Eq. (18)]{ACM_MDR}. $\Lambda _{\Phi _s } \left( {\left[ {\cdot,\cdot } \right)} \right)$ may be even provided by telecommunication operators, which could compute it based on specific path-loss and shadowing models tailored to particular urban cities. With this input, \eqref{Eq_16} and \eqref{Eq_17} can be exploited for system-level optimization as a function of many important system parameters (besides path-loss and blockage models). \hfill $\Box$
\end{remark}

It is worth mentioning, finally, the relevance that \textit{tractable but accurate approximations} are recently gaining in the context of stochastic geometry analysis of realistic but intractable network models \cite{Haenggi__Approximation}. The IM-based approach provides a contribution to these research activities. The proposed approach, in particular, is aimed to yield a tractable approximation for taking into account important link-level characteristics, \textit{e.g.}, multi-state links, that, if neglected, may lead to erroneous conclusions about the performance of cellular networks. The BSs are, however, still assumed to be distributed according to a PPP. Current research activities on modeling the locations of cellular BSs with the aid of point processes different from the PPP, \textit{e.g.}, \cite{Haenggi__Approximation}, are complementary to the proposed IM-based approach. The generalization of the IM-based approach to account for non-PPP models for the locations of cellular BSs is a research issue currently being investigated by the authors. It is, however, beyond the scope of the present paper.
\vspace{-0.25cm}
\section{Area Spectral Efficiency and Potential Throughput} \label{Performance} \vspace{-0.25cm}
Based on the IM-based approach, the following two propositions provide tractable (\textit{i.e.}, easy to be computed numerically) mathematical expressions for ASE and PT. Short-hands: ${\mathcal{L}}_s \left( x \right) = \widehat\Lambda _{\Phi _s }^{\left( 1 \right)} \left( {\left[ {0,x} \right)} \right)\exp \left( { - \sum\nolimits_{r \in {\mathsf{S}}} {\widehat\Lambda _{\Phi _r } \left( {\left[ {0,x} \right)} \right)} } \right)$, ${\cal M}_{g_s^{\left( 0 \right)} } \left( x \right) = \left( {1 + {{x\Omega _s } \mathord{\left/ {\vphantom {{x\Omega _s } {m_s }}} \right. \kern-\nulldelimiterspace} {m_s }}} \right)^{ - m_s }$, $\bar {\cal M}_{g_s^{\left( 0 \right)} } \left( x \right) = 1 - {\cal M}_{g_s^{\left( 0 \right)} } \left( x \right)$, ${\cal F}_s \left( x \right) = 1 - \sum\nolimits_{l_1  = 1}^{{\rm{K}}_{{\rm{BS}}} } {\sum\nolimits_{l_2  = 1}^{{\rm{K}}_{{\rm{MT}}} } {\frac{{\omega _{{\rm{BS}}}^{\left( {l_1 } \right)} }}{{2\pi }}\frac{{\omega _{{\rm{MT}}}^{\left( {l_2 } \right)} }}{{2\pi }}} } {}_2F_1 \left( {m_s , - \frac{2}{{\alpha _s }},1 - \frac{2}{{\alpha _s }}, - x\frac{{\Omega _s }}{{m_s }}\gamma _{{\rm{BS}}}^{\left( {l_1 } \right)} \gamma _{{\rm{MT}}}^{\left( {l_2 } \right)} } \right)$.
\begin{proposition} \label{Proposition__ASE}
Consider the approximated intensity measure in \eqref{Eq_23}. The ASE is the following:
\begin{equation}
\label{Eq_26}
\begin{array}{l}
{\rm{ASE}} = \frac{{\lambda _{{\rm{MT}}} p_{{\rm{sel}}} }}{{\ln \left( 2 \right)}}\sum\nolimits_{s \in \mathsf{S}} {\int\nolimits_0^\infty  {\int\nolimits_0^\infty  {\exp \left( { -\frac{zx{\sigma _N^2 }}{{G^{\left( 0 \right)} P_{{\rm{RB}}} }}} \right){\cal M}_{I_{\rm{agg}} \left( x \right)} \left( {\frac{z}{{G^{\left( 0 \right)} }}\left| x \right.} \right)\bar {\cal M}_{g_s^{\left( 0 \right)} } \left( z \right)\mathcal{L} _s \left( x \right) \frac{{dzdx}}{z}} } }
\end{array}
\end{equation}
\noindent where ${\cal M}_{I_{\rm{agg}} \left( x \right)} \left( {z\left| x \right.} \right) = \exp \left( {\sum\nolimits_{r \in {\mathsf{S}}} {\widehat {\cal T}_r \left( {z,x} \right)} } \right)$ and $\widehat {\cal T}_r \left( { \cdot , \cdot } \right)$ is defined as follows:
\begin{equation}
\label{Eq_27}
\begin{array}{l}
 \widehat {\cal T}_r \left( {z,x} \right) = \pi \lambda _{{\rm{BS}}}^{\left( {\rm{I}} \right)} \Theta _r \left( {\frac{x}{{\kappa _r }}} \right)^{{2 \mathord{\left/
 {\vphantom {2 {\alpha _r }}} \right.
 \kern-\nulldelimiterspace} {\alpha _r }}} {\cal F}_r \left( z \right)\sum\limits_{b = 1}^{\widehat{\cal B}} {\widehat q_r^{\left[ {\widehat D_{b - 1} ,\widehat D_b } \right]} \left( {\overline {\cal H} \left( {x - \kappa _r \widehat D_b^{\alpha _r } } \right){\cal H}\left( {x - \kappa _r \widehat D_{b - 1}^{\alpha _r } } \right)} \right)}  \\
  + \pi \lambda _{{\rm{BS}}}^{\left( {\rm{I}} \right)} \Theta _r \sum\limits_{b = 1}^{\widehat{\cal B}} {\widehat q_r^{\left[ {\widehat D_{b - 1} ,\widehat D_b } \right]} \left( {\widehat D_{b - 1}^2 {\cal F}_r \left( {\frac{{xz}}{{\kappa _r \widehat D_{b - 1}^{\alpha _r } }}} \right)\overline {\cal H} \left( {x - \kappa _r \widehat D_{b - 1}^{\alpha _r } } \right) - \widehat D_b^2 {\cal F}_r \left( {\frac{{xz}}{{\kappa _r \widehat D_b^{\alpha _r } }}} \right)\overline {\cal H} \left( {x - \kappa _r \widehat D_b^{\alpha _r } } \right)} \right)}  \\
  + \pi \lambda _{{\rm{BS}}}^{\left( {\rm{I}} \right)} \Theta _r \widehat q_r^{\left[ {\widehat D_{\widehat{\cal B}} ,\infty } \right]} \left( {\widehat D_{\widehat{\cal B}}^2 {\cal F}_r \left( {\frac{{xz}}{{\kappa _r \widehat D_{\widehat{\cal B}}^{\alpha _r } }}} \right)\overline {\cal H} \left( {x - \kappa _r \widehat D_{\widehat{\cal B}}^{\alpha _r } } \right) + \left( {\frac{x}{{\kappa _r }}} \right)^{{2 \mathord{\left/
 {\vphantom {2 {\alpha _r }}} \right.
 \kern-\nulldelimiterspace} {\alpha _r }}} {\cal F}_r \left( z \right){\cal H}\left( {x - \kappa _r \widehat D_{\widehat{\cal B}}^{\alpha _r } } \right)} \right)
\end{array}
\end{equation}

\emph{Proof}: It follows from \eqref{Eq_16}, inserting \eqref{Eq_23}, \eqref{Eq_24} in \eqref{Eq_19}-\eqref{Eq_22} and computing the integrals. \hfill $\Box$
\end{proposition}
\begin{proposition} \label{Proposition__PT}
Consider the approximated intensity measure in \eqref{Eq_23}. Let the same definitions as in \textit{Proposition \ref{Proposition__ASE}} hold. The PT is ${\rm{PT = }}\lambda _{{\rm{MT}}} p_{{\rm{sel}}} \log _2 \left( {1 + {\rm{T}}} \right)\left( {{1 \mathord{\left/ {\vphantom {1 2}} \right. \kern-\nulldelimiterspace} 2} - {{{\mathcal{\bar C}}\left( {\rm{T}} \right)} \mathord{\left/ {\vphantom {{{\rm{\bar C}}\left( {\rm{T}} \right)} \pi }} \right. \kern-\nulldelimiterspace} \pi }} \right)$, where:
\begin{equation}
\label{Eq_28}
\begin{array}{l}
\hspace{-0.12cm}{\mathcal{\bar C}}\left( {\rm{T}} \right) = \sum\nolimits_{s \in {\mathsf{S}}} {\int \nolimits_0^\infty  {\int \nolimits_0^\infty  {{\mathop{\rm Im}\nolimits} \left\{ {\exp \left( {\frac{{{\mathbbm{j}}zx\sigma _N^2 {\rm{T}}}}{{G^{\left( 0 \right)} P_{{\rm{RB}}} }}} \right){\cal M}_{g_s^{\left( 0 \right)} } \left( {{\mathbbm{j}}z} \right){\cal M}_{I_{\rm{agg}} \left( x \right)} \left( { - \frac{{{\mathbbm{j}}z{\rm{T}}}}{{G^{\left( 0 \right)} }}\left| x \right.} \right)} \right\}\mathcal{L} _s \left( x \right)\frac{{dzdx}}{z}} } }
\end{array}
\end{equation}

\emph{Proof}: It follows from \eqref{Eq_17}, inserting \eqref{Eq_23}, \eqref{Eq_24} in \eqref{Eq_19}-\eqref{Eq_22} and computing the integrals. \hfill $\Box$
\end{proposition}

It is worth mentioning that the mathematical tractability of ASE and PT in \eqref{Eq_26} and \eqref{Eq_28}, respectively, originates from the adopted multi-ball link state model (see Section \ref{LinkStateModeling}) and from the IM-based approximation introduced in Section \ref{IM_Methodology}. These are the main novelties of the proposed approach, which make the analytical expressions of ASE and PT in \textit{Proposition \ref{Proposition__ASE}} and \textit{Proposition \ref{Proposition__PT}} along with their associated mathematical derivations unique.

To enable easier understanding of the impact of the link state model on system design and optimization, the following corollary provides a simplified framework under the assumption of a two-state ($\mathcal{S}=2$) and single-ball ($\mathcal{B}=1$) blockage model. In Section \ref{Trends}, it is used to discuss performance trends and to provide guidelines for system-level optimization. To be concrete and clear, we assume that the two states, $s_1$ and $s_2$, correspond to LOS and NLOS links, respectively. For ease of understanding, thus, the notation $s_1  \mapsto {\mathop{\rm LOS}\nolimits}$ and $s_2  \mapsto {\mathop{\rm NLOS}\nolimits}$ is adopted.
\begin{corollary} \label{Corollary__ASE_PT}
If $\mathcal{S}=2$ and $\mathcal{B}=1$, the expressions of ASE in \eqref{Eq_26} and PT in \eqref{Eq_28} still hold, but ${\cal M}_{I_{{\rm{agg}}} \left( x \right)} \left( {z\left| x \right.} \right) = \exp \left( {\widehat {\cal T}_{{\rm{LOS}}} \left( {z,x} \right) + \widehat {\cal T}_{{\rm{NLOS}}} \left( {z,x} \right)} \right)$, and \eqref{Eq_23}, \eqref{Eq_24}, \eqref{Eq_27} simplify as follows:
\begin{equation}
\label{Eq_29}
\begin{array}{l}
 \widehat\Lambda _{\Phi _s } \left( {\left[ {0,\xi } \right)} \right) = \pi \lambda _{{\rm{BS}}} \Theta _s \widehat q_s^{\left[ {0,\widehat D_1 } \right]} \left( {\left( {{\xi  \mathord{\left/
 {\vphantom {\xi  {\kappa _s }}} \right.
 \kern-\nulldelimiterspace} {\kappa _s }}} \right)^{{2 \mathord{\left/
 {\vphantom {2 {\alpha _s }}} \right.
 \kern-\nulldelimiterspace} {\alpha _s }}} \overline {\cal H} \left( {\xi  - \kappa _s \widehat D_1^{\alpha _s } } \right) + \widehat D_1^2 {\cal H}\left( {\xi  - \kappa _s \widehat D_1^{\alpha _s } } \right)} \right) \\
 \hspace{2.05cm} + \pi \lambda _{{\rm{BS}}} \Theta _s \widehat q_s^{\left[ {\widehat D_1 ,\infty } \right]} \left( {\left( {{\xi  \mathord{\left/
 {\vphantom {\xi  {\kappa _s }}} \right.
 \kern-\nulldelimiterspace} {\kappa _s }}} \right)^{{2 \mathord{\left/
 {\vphantom {2 {\alpha _s }}} \right.
 \kern-\nulldelimiterspace} {\alpha _s }}}  - \widehat D_1^2 } \right){\cal H}\left( {\xi  - \kappa _s \widehat D_1^{\alpha _s } } \right)
\end{array}
\end{equation}
\begin{equation}
\label{Eq_30}
\begin{array}{l}
 \widehat\Lambda _{\Phi _s }^{\left( 1 \right)} \left( {\left[ {0,\xi } \right)} \right) = \pi \lambda _{{\rm{BS}}} \Theta _s \widehat q_s^{\left[ {0,\widehat D_1 } \right]} \left( {{2 \mathord{\left/
 {\vphantom {2 {\alpha _s }}} \right.
 \kern-\nulldelimiterspace} {\alpha _s }}} \right)\kappa _s^{ - {2 \mathord{\left/
 {\vphantom {2 {\alpha _s }}} \right.
 \kern-\nulldelimiterspace} {\alpha _s }}} \xi ^{{2 \mathord{\left/
 {\vphantom {2 {\alpha _s }}} \right.
 \kern-\nulldelimiterspace} {\alpha _s }} - 1} \overline {\cal H} \left( {\xi  - \kappa _s \widehat D_1^{\alpha _s } } \right) \\
 \hspace{2.05cm} + \pi \lambda _{{\rm{BS}}} \Theta _s \widehat q_s^{\left[ {\widehat D_1 ,\infty } \right]} \left( {{2 \mathord{\left/
 {\vphantom {2 {\alpha _s }}} \right.
 \kern-\nulldelimiterspace} {\alpha _s }}} \right)\kappa _s^{ - {2 \mathord{\left/
 {\vphantom {2 {\alpha _s }}} \right.
 \kern-\nulldelimiterspace} {\alpha _s }}} \xi ^{{2 \mathord{\left/
 {\vphantom {2 {\alpha _s }}} \right.
 \kern-\nulldelimiterspace} {\alpha _s }} - 1} {\cal H}\left( {\xi  - \kappa _s \widehat D_1^{\alpha _s } } \right)
\end{array}
\end{equation}
\begin{equation}
\label{Eq_31}
\begin{array}{l}
\hspace{-0.25cm} \widehat {\cal T}_r \left( {z,x} \right) = \pi \lambda _{{\rm{BS}}}^{\left( {\rm{I}} \right)} \Theta _r \widehat q_r^{\left[ {0,\widehat D_1 } \right]} \left( {\left( {\frac{x}{{\kappa _r }}} \right)^{{2 \mathord{\left/
 {\vphantom {2 {\alpha _r }}} \right.
 \kern-\nulldelimiterspace} {\alpha _r }}} {\cal F}_r \left( z \right)\overline {\cal H} \left( {x - \kappa _r \widehat D_1^{\alpha _r } } \right) - \widehat D_1^2 {\cal F}_r \left( {\frac{{xz}}{{\kappa _r \widehat D_1^{\alpha _r } }}} \right)\overline {\cal H} \left( {x - \kappa _r \widehat D_1^{\alpha _r } } \right)} \right) \\
 \hspace{1.28cm} + \pi \lambda _{{\rm{BS}}}^{\left( {\rm{I}} \right)} \Theta _r \widehat q_r^{\left[ {\widehat D_1 ,\infty } \right]} \left( {\left( {\frac{x}{{\kappa _r }}} \right)^{{2 \mathord{\left/
 {\vphantom {2 {\alpha _r }}} \right.
 \kern-\nulldelimiterspace} {\alpha _r }}} {\cal F}_r \left( z \right){\cal H}\left( {x - \kappa _r \widehat D_1^{\alpha _r } } \right) + \widehat D_1^2 {\cal F}_r \left( {\frac{{xz}}{{\kappa _r \widehat D_1^{\alpha _r } }}} \right)\overline {\cal H} \left( {x - \kappa _r \widehat D_1^{\alpha _r } } \right)} \right)
\end{array}
\end{equation}

\emph{Proof}: It follows by setting $\mathcal{S}=2$ and $\mathcal{B}=1$, some algebra and simplifications. \hfill $\Box$
\end{corollary}
\vspace{-0.5cm}
\section{Performance Trends and Design Insights} \label{Trends} \vspace{-0.25cm}
In this section, based on the mathematical frameworks in Section \ref{Performance}, we study the impact of several system parameters on the performance of cellular networks. Due to space limitations, we focus our attention only on the ASE. By using a similar methodology of analysis, the same study can be conducted for the PT. To gain the most of the insight for cellular networks design, the mathematical framework in \textit{Corollary \ref{Corollary__ASE_PT}} constitutes the departing point of our analysis.

\begin{table}[!t] \footnotesize
\centering
\caption{Auxiliary functions in \eqref{Eq_32}. The notation is provided in footnote 3. \vspace{-0.25cm}}
\newcommand{\tabincell}[2]{\begin{tabular}{@{}#1@{}}#2\end{tabular}}
\begin{tabular}{|c|} \hline
$\begin{array}{l}
 \widehat {\cal T}_{{\rm{in}}} \left( {z,x} \right) = \pi \lambda _{{\rm{BS}}}^{\left( {\rm{I}} \right)} \sum\limits_{r \in \left\{ {{\rm{LOS}},{\rm{NLOS}}} \right\}} {\left( {\Theta _r \widehat q_r^{\left[ {0,\widehat D_1 } \right]} \left( {\left( {\frac{x}{{\kappa _r }}} \right)^{2/\alpha _r } {\cal F}_{r,{\rm{in}}} \left( z \right) - \widehat D_1^2 {\cal F}_{r,{\rm{in}}} \left( {\frac{{xz}}{{\kappa _r \widehat D_1^{\alpha _r } }}} \right)} \right) + \Theta _r \widehat q_r^{\left[ {\widehat D_1 ,\infty } \right]} \widehat D_1^2 {\cal F}_{r,{\rm{in}}} \left( {\frac{{xz}}{{\kappa _r \widehat D_1^{\alpha _r } }}} \right)} \right)}  \\
 \widehat {\cal T}_{{\rm{out}}} \left( {z,x} \right) = \pi \lambda _{{\rm{BS}}}^{\left( {\rm{I}} \right)} \sum\nolimits_{r \in \left\{ {{\rm{LOS}},{\rm{NLOS}}} \right\}} {\left( {\Theta _r \widehat q_r^{\left[ {\widehat D_1 ,\infty } \right]} \left( {x/\kappa _r } \right)^{2/\alpha _r } {\cal F}_{r,{\rm{out}}} \left( z \right)} \right)}  \end{array}$ \\ \hline
$\begin{array}{l}
 \widehat\Lambda _{\Phi _{{\rm{in}}} } \left( {\left[ {0,x} \right)} \right) = \pi \lambda _{{\rm{BS}}} \sum\nolimits_{r \in \left\{ {{\rm{LOS}},{\rm{NLOS}}} \right\}} {\left( {\Theta _r \widehat q_r^{\left[ {0,\widehat D_1 } \right]} \left( {{x \mathord{\left/ {\vphantom {x {\kappa _r }}} \right. \kern-\nulldelimiterspace} {\kappa _r }}} \right)^{{2 \mathord{\left/  {\vphantom {2 {\alpha _r }}} \right. \kern-\nulldelimiterspace} {\alpha _r }}} } \right) \overline {\cal H}\left( {x - \kappa _r \widehat D_1^{\alpha _r } } \right)}
\\ \widehat\Lambda _{\Phi _{{\rm{out}}} } \left( {\left[ {0,x} \right)} \right) = \pi \lambda _{{\rm{BS}}} \sum\nolimits_{r \in \left\{ {{\rm{LOS}},{\rm{NLOS}}}\right\}} {\left( {\Theta _r \widehat q_r^{\left[ {0,\widehat D_1 } \right]} \widehat D_1^2  + \Theta _r \widehat q_r^{\left[ {\widehat D_1,\infty } \right]} \left( {\left( {x/\kappa _r } \right)^{2/\alpha _r }  - \widehat D_1^2 } \right)} \right){\cal H}\left( {x - \kappa _r \widehat D_1^{\alpha _r } } \right)}
 \end{array}$ \\ \hline
$\begin{array}{l}
 \widehat\Lambda _{\Phi _{s,{\rm{in}}} }^{\left( 1 \right)} \left( {\left[ {0,x} \right)} \right) = \pi \lambda _{{\rm{BS}}} \Theta _s \widehat q_s^{\left[ {0,\widehat D_1 } \right]} \left( {2/\alpha _s } \right)\kappa _s^{ - 2/\alpha _s } x^{2/\alpha _s  - 1}  \\
 \widehat\Lambda _{\Phi _{s,{\rm{out}}} }^{\left( 1 \right)} \left( {\left[ {0,x} \right)} \right) = \pi \lambda _{{\rm{BS}}} \Theta _s \widehat q_s^{\left[ {\widehat D_1 ,\infty } \right]} \left( {2/\alpha _s } \right)\kappa _s^{ - 2/\alpha _s } x^{2/\alpha _s  - 1}
 \end{array}$ \\ \hline
\end{tabular}
\label{Table_SimplifiedFramework} \vspace{-0.65cm}
\end{table}
Let us start by rewriting the ASE in \textit{Corollary \ref{Corollary__ASE_PT}} in an explicit manner, in order to make the physical meaning of its constituent elements more evident. With the aid of some algebra, the ASE is equal to ${\rm{ASE}} = \left( {{{\lambda _{{\rm{MT}}} p_{{\rm{sel}}} } \mathord{\left/ {\vphantom {{\lambda _{{\rm{MT}}} p_{{\rm{sel}}} } {\ln \left( 2 \right)}}} \right. \kern-\nulldelimiterspace} {\ln \left( 2 \right)}}} \right)\left( {{\mathcal{R}}_{{\rm{LOS}},{\rm{in}}}  + {\mathcal{R}}_{{\rm{LOS}},{\rm{out}}}  + {\mathcal{R}}_{{\rm{NLOS}},{\rm{in}}}  + {\mathcal{R}}_{{\rm{NLOS}},{\rm{out}}} } \right)$, where:
\begin{equation}
\label{Eq_32}
\begin{array}{l}
 {\mathcal{R}}_{s,{\rm{in}}}  = \int\nolimits_0^{\kappa _s \widehat D_1^{\alpha _s } } {\left( {\int\nolimits_0^\infty  {\exp \left( { - zx\frac{{\sigma _N^2 }}{{G^{\left( 0 \right)} P_{{\rm{RB}}} }}} \right)\exp \left( {\widehat {\cal T}_{{\rm{in}}} \left( {\frac{z}{{G^{\left( 0 \right)} }},x} \right)} \right)\bar {\cal M}_{g_s^{\left( 0 \right)} } \left( z \right)\frac{{dz}}{z}} } \right)}  \\
  \hspace{7.5cm} \times \widehat\Lambda _{\Phi _{s,{\rm{in}}} }^{\left( 1 \right)} \left( {\left[ {0,x} \right)} \right)\exp \left( { - \widehat\Lambda _{\Phi _{{\rm{in}}} } \left( {\left[ {0,x} \right)} \right)} \right)dx \\
 {\mathcal{R}}_{s,{\rm{out}}}  = \int\nolimits_{\kappa _s \widehat D_1^{\alpha _s } }^\infty  {\left( {\int\nolimits_0^\infty  {\exp \left( { - zx\frac{{\sigma _N^2 }}{{G^{\left( 0 \right)} P_{{\rm{RB}}} }}} \right)\exp \left( {\widehat {\cal T}_{{\rm{out}}} \left( {\frac{z}{{G^{\left( 0 \right)} }},x} \right)} \right)\bar {\cal M}_{g_s^{\left( 0 \right)} } \left( z \right)\frac{{dz}}{z}} } \right)}  \\
  \hspace{7.5cm} \times \widehat\Lambda _{\Phi _{s,{\rm{out}}} }^{\left( 1 \right)} \left( {\left[ {0,x} \right)} \right)\exp \left( { - \widehat\Lambda _{\Phi _{{\rm{out}}} } \left( {\left[ {0,x} \right)} \right)} \right)dx
\end{array}
\end{equation}
\noindent where $s \in \left\{ {{\rm{LOS}},{\rm{NLOS}}} \right\}$ and the rest of the functions are reported in Table \ref{Table_SimplifiedFramework}\footnote{Notation of Table \ref{Table_SimplifiedFramework} -- ${\cal F}_{r,{\rm{in}}} \left( z \right) = {\cal F}_r \left( z \right)\overline {\cal H} \left( {z - \kappa _s \widehat D_1^{\alpha _s } } \right)$, ${\cal F}_{r,{\rm{out}}} \left( z \right) = {\cal F}_r \left( z \right){\cal H}\left( {z - \kappa _s \widehat D_1^{\alpha _s } } \right)$.}.

The four terms that constitute the ASE have a clear physical interpretation: ${\mathcal{R}}_{s,t}$ for $s \in \left\{ {{\rm{LOS}},{\rm{NLOS}}} \right\}$ and $t \in \left\{ {{\rm{in},{\rm{out}}}} \right\}$ is the contribution to the ASE that originates when the serving BS is in state $s$ and is located either inside ($t=\rm{in}$) or outside ($t=\rm{out}$) the ball of radius $\widehat D_1$. It is worth mentioning, however, that the interfering BSs are not constrained to be located either inside or outside the ball of radius $\widehat D_1$ if $t=\rm{in}$ or $t=\rm{out}$, respectively.

The ASE in \eqref{Eq_32} is exact and holds for $\mathcal{S}=2$ and $\mathcal{B}=1$. In typical cellular network deployments, it can be further simplified. The condition $\widehat q_{{\rm{LOS}}}^{\left[ {\widehat D_1 ,\infty } \right]} \approx 0$, in fact, usually holds. This implies that ${{\mathcal{R}}_{{\rm{LOS}},{\rm{out}}} }$ is negligible compared to the other three addends. In the sequel, thus, we consider the approximation ${\rm{ASE}} \approx \left( {{{\lambda _{{\rm{MT}}} p_{{\rm{sel}}} } \mathord{\left/ {\vphantom {{\lambda _{{\rm{MT}}} p_{{\rm{sel}}} } {\ln \left( 2 \right)}}} \right. \kern-\nulldelimiterspace} {\ln \left( 2 \right)}}} \right)\left( {{\mathcal{R}}_{{\rm{LOS}},{\rm{in}}}  + {\mathcal{R}}_{{\rm{NLOS}},{\rm{in}}}  + {\mathcal{R}}_{{\rm{NLOS}},{\rm{out}}} } \right)$, which constitutes a tight estimate of the ASE. This is substantiated in Section \ref{Results} with the aid of empirical data. It is, however, still too complicated for gaining engineering insight. We propose, hence, four asymptotic approximations that correspond to four important operating regimes. In this section, we show that they shed light on key performance trends and provide (different) guidelines for the optimization of cellular networks. These findings are substantiated in Section \ref{Results}. For each case study, in particular, accurate and weak approximations are provided. The latter ones are useful for gaining deeper design insight and are denoted by using the symbol $\propto$.

For ease of exposition, wherever needed, $\lambda _{{\rm{BS}}}$ is replaced by its equivalent representation in terms of average cell radius ($R_{{\rm{cell}}}$), \textit{i.e.}, $\lambda _{{\rm{BS}}}  \leftrightarrow {1 \mathord{\left/ {\vphantom {1 {\left( {\pi R_{{\rm{cell}}}^2 } \right)}}} \right. \kern-\nulldelimiterspace} {\left( {\pi R_{{\rm{cell}}}^2 } \right)}}$ \cite{MDR_mmWave}. Also, the following short-hands are introduced: $\widehat \theta _{{\rm{LOS}}}^{\left[ {0,\widehat D_1 } \right]}  = \Theta _{{\rm{LOS}}} \widehat q_{{\rm{LOS}}}^{\left[ {0,\widehat D_1 } \right]} \widehat D_1^2$, $\widehat \theta _{{\rm{NLOS}}}^{\left[ {\widehat D_1 ,\infty } \right]}  = \Theta _{{\rm{NLOS}}} \widehat q_{{\rm{NLOS}}}^{\left[ {\widehat D_1 ,\infty } \right]} \widehat D_1^2$, $\widehat \phi _{{\rm{NLOS}}}^{\left[ {\widehat D_1 ,\infty } \right]}  = \Theta _{{\rm{NLOS}}} \widehat q_{{\rm{NLOS}}}^{\left[ {\widehat D_1 ,\infty } \right]}$, $\hat \kappa _D  = \left( {{{\kappa _{{\rm{LOS}}} } \mathord{\left/ {\vphantom {{\kappa _{{\rm{LOS}}} } {\kappa _{{\rm{NLOS}}} }}} \right. \kern-\nulldelimiterspace} {\kappa _{{\rm{NLOS}}} }}} \right)\widehat D_1^{\left( {\alpha _{{\rm{LOS}}}  - \alpha _{{\rm{NLOS}}} } \right)}$, $P_{N,D}  = {{P_{{\rm{RB}}} } \mathord{\left/ {\vphantom {{P_{{\rm{RB}}} } {\left( {\sigma _N^2 \kappa _{{\rm{LOS}}} \widehat D_1^{\alpha _{{\rm{LOS}}} } } \right)}}} \right.  \kern-\nulldelimiterspace} {\left( {\sigma _N^2 \kappa _{{\rm{LOS}}} \widehat D_1^{\alpha _{{\rm{LOS}}} } } \right)}}$, $P_N  = {{P_{{\rm{RB}}} } \mathord{\left/ {\vphantom {{P_{{\rm{RB}}} } {\left( {\sigma _N^2 \kappa _{{\rm{NLOS}}} } \right)}}} \right. \kern-\nulldelimiterspace} {\left( {\sigma _N^2 \kappa _{{\rm{NLOS}}} } \right)}}$.
\setcounter{paragraph}{0}
\paragraph{Very Dense (VD) Cellular Networks} This regime emerges if the following conditions are satisfied: i) ${{\lambda _{{\rm{BS}}} } \mathord{\left/ {\vphantom {{\lambda _{{\rm{BS}}} } {\lambda _{{\rm{MT}}} }}} \right. \kern-\nulldelimiterspace} {\lambda _{{\rm{MT}}} }} \gg 1$ and $R_{{\rm{cell}}}  \ll \widehat D_1$, ii) $p_{{\rm{sel}}}$ and $p_{{\rm{off}}}$ are those in \eqref{Eq_8}. Usually, in addition, $p_{{\rm{sel}}}$ in \eqref{Eq_8} is close to one, \textit{i.e.}, $p_{{\rm{sel}}} \to 1$. As a result, the ASE is dominated by ${\mathcal{R}}_{{\rm{LOS}},{\rm{in}}}$, \textit{i.e.}, ${\rm{ASE}} \to {\rm{ASE}}^{{(\rm{VD})}}  = \left( {{{\lambda _{{\rm{MT}}} } \mathord{\left/ {\vphantom {{\lambda _{{\rm{MT}}} } {\ln \left( 2 \right)}}} \right. \kern-\nulldelimiterspace} {\ln \left( 2 \right)}}} \right){\mathcal{R}}^{{(\rm{VD})}}_{{\rm{LOS}},{\rm{in}}}$, where ${\cal R}_{{\rm{LOS}}{\rm{,in}}}^{\left( {{\rm{VD}}} \right)}$ can be formulated as follows:
\begin{equation}
\label{Eq_33}
\begin{array}{l}
 \hspace{-0.40cm} {\cal R}_{{\rm{LOS}}{\rm{,in}}}^{\left( {{\rm{VD}}} \right)} \mathop  \to \limits^{\left( \rm{VD} \right)} \int\nolimits_0^\infty  {\int\nolimits_0^\infty  {\exp \left( {\pi \frac{{\lambda _{{\rm{MT}}} }}{{N_{{\rm{RB}}} }}\left( {y{\cal F}_{{\rm{LOS}}} \left( {\frac{z}{{G^{\left( 0 \right)} }}} \right) - \hat \theta _{{\rm{LOS}}}^{\left[ {0,\widehat D_1 } \right]} {\cal F}_{{\rm{LOS}}} \left( {\left( {\frac{y}{{\widehat \theta _{{\rm{LOS}}}^{\left[ {0,\widehat D_1 } \right]} }}} \right)^{{{\alpha _{{\rm{LOS}}} } \mathord{\left/
 {\vphantom {{\alpha _{{\rm{LOS}}} } 2}} \right.
 \kern-\nulldelimiterspace} 2}} \frac{z}{{G^{\left( 0 \right)} }}} \right)} \right)} \right)} }  \\
 \hspace{0.20cm} \times \pi \lambda _{{\rm{BS}}} \exp \left( { - \pi \lambda _{{\rm{BS}}} y} \right)\bar {\cal M}_{g_{{\rm{LOS}}}^{\left( 0 \right)} } \left( z \right)\frac{{dzdy}}{z}
  \propto \int\nolimits_0^\infty  {\left( {1 - \frac{{\lambda _{{\rm{MT}}} }}{{N_{{\rm{RB}}} \lambda _{{\rm{BS}}} }}{\cal F}_{{\rm{LOS}}} \left( {\frac{z}{{G^{\left( 0 \right)} }}} \right)} \right)^{ - 1} } \bar {\cal M}_{g_{{\rm{LOS}}}^{\left( 0 \right)} } \left( z \right)\frac{{dz}}{z}
\end{array}
\end{equation}
\noindent where (VD) is obtained by taking the following into account: i) $\widehat q_{{\rm{LOS}}}^{\left[ {\widehat D_1 ,\infty } \right]}  \approx 0$, ii) $\widehat D_1^2  \gg \left( {\left( {{{\kappa _{{\rm{LOS}}} } \mathord{\left/ {\vphantom {{\kappa _{{\rm{LOS}}} } {\kappa _{{\rm{NLOS}}} }}} \right. \kern-\nulldelimiterspace} {\kappa _{{\rm{NLOS}}} }}} \right)\widehat D_1^{\alpha _{{\rm{LOS}}} } } \right)^{2/\alpha _{{\rm{NLOS}}} }$ and $\left( {{{\kappa _{{\rm{LOS}}} } \mathord{\left/ {\vphantom {{\kappa _{{\rm{LOS}}} } {\kappa _{{\rm{NLOS}}} }}} \right. \kern-\nulldelimiterspace} {\kappa _{{\rm{NLOS}}} }}} \right)\widehat D_1^{\alpha _{{\rm{LOS}}}  - \alpha _{{\rm{NLOS}}} }  \ll 1$, since $\alpha _{{\rm{NLOS}}}  > \alpha _{{\rm{LOS}}}$, iii) ${\cal F}_s \left( z \right) \to 0$ if $z \to 0$, as well as, for very dense cellular networks, that iv) the noise is negligible compared to the other-cell interference and v) $\left( {\pi \lambda _{{\rm{BS}}} \widehat D_1^2 } \right)\exp \left( { - \pi \lambda _{{\rm{BS}}} \widehat D_1^2 y} \right) \approx 0$ if $y \in \left[ {1,\infty } \right)$. The weaker approximation in $\propto$ follows by noting that $\pi \lambda _{{\rm{BS}}} \widehat D_1^2 \exp \left( { - \pi \lambda _{{\rm{BS}}} \widehat D_1^2 y} \right) \to \pi \lambda _{{\rm{BS}}} \widehat D_1^2 \delta \left( y \right)$ if $\pi \lambda _{{\rm{BS}}} \widehat D_1^2  = \left( {{{\widehat D_1^2 } \mathord{\left/ {\vphantom {{\widehat D_1^2 } {R_{{\rm{cell}}}^2 }}} \right. \kern-\nulldelimiterspace} {R_{{\rm{cell}}}^2 }}} \right) \gg 1$, which implies ${\cal F}_{{\rm{LOS}}} \left( {y^{{{\alpha _{{\rm{LOS}}} } \mathord{\left/ {\vphantom {{\alpha _{{\rm{LOS}}} } 2}} \right. \kern-\nulldelimiterspace} 2}} z} \right) \approx y^{{{\alpha _{{\rm{LOS}}} } \mathord{\left/ {\vphantom {{\alpha _{{\rm{LOS}}} } 2}} \right. \kern-\nulldelimiterspace} 2}} {\cal F}_{{\rm{LOS}}} \left( z \right)$ and $y{\cal F}_{{\rm{LOS}}} \left( z \right) - y^{{{\alpha _{{\rm{LOS}}} } \mathord{\left/ {\vphantom {{\alpha _{{\rm{LOS}}} } 2}} \right. \kern-\nulldelimiterspace} 2}} {\cal F}_{{\rm{LOS}}} \left( z \right) \approx y{\cal F}_{{\rm{LOS}}} \left( z \right)$. $\propto$ is obtained by computing the resulting integral.
\paragraph{Dense (D) Cellular Networks} This regime emerges if the following conditions are satisfied: i) the network is sufficiently dense that the typical MT is served, almost surely, by a BS in LOS and located inside the ball of radius $\widehat D_1$, \textit{i.e.}, $R_{{\rm{cell}}}  < \widehat D_1$ and ${\rm{ASE}} \to {\rm{ASE}}^{({\rm{D}})}  = \left( {{{\lambda _{{\rm{MT}}} p_{{\rm{sel}}} } \mathord{\left/ {\vphantom {{\lambda _{{\rm{MT}}} p_{{\rm{sel}}} } {\ln \left( 2 \right)}}} \right. \kern-\nulldelimiterspace} {\ln \left( 2 \right)}}} \right){\mathcal{R}}^{({\rm{D}})}_{{\rm{LOS}},{\rm{in}}}$, but ii) the network is still sparse enough that there are (almost) no inactive BSs and some MTs are still blocked, \textit{i.e.}, $p_{{\rm{off}}} \to 0$, ${{\lambda _{{\rm{BS}}} } \mathord{\left/ {\vphantom {{\lambda _{{\rm{BS}}} } {\lambda _{{\rm{MT}}} }}} \right. \kern-\nulldelimiterspace} {\lambda _{{\rm{MT}}} }} < 1$, and $p_{{\rm{sel}}}$ is that in \eqref{Eq_9}. Thus, ${\rm{ASE}}^{\left( {\rm{D}} \right)}  = \left( {{{N_{{\rm{RB}}} \lambda _{{\rm{BS}}} } \mathord{\left/ {\vphantom {{N_{{\rm{RB}}} \lambda _{{\rm{BS}}} } {\ln \left( 2 \right)}}} \right. \kern-\nulldelimiterspace} {\ln \left( 2 \right)}}} \right){\cal R}_{{\rm{LOS}}{\rm{,in}}}^{\left( {\rm{D}} \right)}$ and ${\cal R}_{{\rm{LOS}}{\rm{,in}}}^{\left( {\rm{D}} \right)}$ can be formulated as follows:
\begin{equation}
\label{Eq_34}
\begin{array}{l}
 \hspace{-0.35cm}{\cal R}_{{\rm{LOS}}{\rm{,in}}}^{\left( {\rm{D}} \right)} \mathop  \to \limits^{\left( {\rm{D}} \right)} \pi \lambda _{{\rm{BS}}} \widehat \theta _{{\rm{LOS}}}^{\left[ {0,\widehat D_1 } \right]} \int\nolimits_0^1 {\int\nolimits_0^\infty  {\exp \left( { - \pi \lambda _{{\rm{BS}}} \widehat \theta _{{\rm{LOS}}}^{\left[ {0,\widehat D_1 } \right]} \left( {y - y{\cal F}_{{\rm{LOS}}} \left( {\frac{z}{{G^{\left( 0 \right)} }}} \right) + {\cal F}_{{\rm{LOS}}} \left( {\frac{{y^{{{\alpha _{{\rm{LOS}}} } \mathord{\left/
 {\vphantom {{\alpha _{{\rm{LOS}}} } 2}} \right.
 \kern-\nulldelimiterspace} 2}} z}}{{G^{\left( 0 \right)} }}} \right)} \right)} \right)} }  \\
 \hspace{0.75cm} \times \exp \left( {\pi \lambda _{{\rm{BS}}} \widehat \theta _{{\rm{NLOS}}}^{\left[ {\widehat D_1 ,\infty } \right]} {\cal F}_{{\rm{NLOS}}} \left( {\widehat \kappa _D \frac{{y^{{{\alpha _{{\rm{LOS}}} } \mathord{\left/
 {\vphantom {{\alpha _{{\rm{LOS}}} } 2}} \right.
 \kern-\nulldelimiterspace} 2}} z}}{{G^{\left( 0 \right)} }}} \right)} \right)\bar {\cal M}_{g_{{\rm{LOS}}}^{\left( 0 \right)} } \left( z \right)\frac{{dzdy}}{z}
\end{array}
\end{equation}
\begin{equation} \nonumber
%\label{Eq_34}
\begin{array}{l}
 \hspace{-0.35cm} \propto \int\nolimits_0^\infty  {\int\nolimits_0^\infty  {\exp \left( {y{\cal F}_{{\rm{LOS}}} \left( {\frac{z}{{G^{\left( 0 \right)} }}} \right) - \pi \lambda _{{\rm{BS}}} \hat \theta _{{\rm{LOS}}}^{\left[ {0,\widehat D_1 } \right]} {\cal F}_{{\rm{LOS}}} \left( {\left( {\frac{y}{{\pi \lambda _{{\rm{BS}}} \hat \theta _{{\rm{LOS}}}^{\left[ {0,\widehat D_1 } \right]} }}} \right)^{{{\alpha _{{\rm{LOS}}} } \mathord{\left/
 {\vphantom {{\alpha _{{\rm{LOS}}} } 2}} \right.
 \kern-\nulldelimiterspace} 2}} \frac{z}{{G^{\left( 0 \right)} }}} \right) - y} \right)} } \bar {\cal M}_{g_{{\rm{LOS}}}^{\left( 0 \right)} } \left( z \right)\frac{{dzdy}}{z}
\end{array}
\end{equation}
\noindent where (D) is obtained by taking the following into account: i) $\widehat q_{{\rm{LOS}}}^{\left[ {\widehat D_1 ,\infty } \right]}  \approx 0$, $\,$ ii) $\widehat D_1^2  \gg \left( {\left( {{{\kappa _{{\rm{LOS}}} } \mathord{\left/ {\vphantom {{\kappa _{{\rm{LOS}}} } {\kappa _{{\rm{NLOS}}} }}} \right. \kern-\nulldelimiterspace} {\kappa _{{\rm{NLOS}}} }}} \right)\widehat D_1^{\alpha _{{\rm{LOS}}} } } \right)^{2/\alpha _{{\rm{NLOS}}} }$, since $\alpha _{{\rm{NLOS}}}  > \alpha _{{\rm{LOS}}}$, as well as, for dense cellular networks, that iii) the noise is negligible compared to the other-cell interference. The weaker approximation in $\propto$ follows by noting that ${\cal F}_{{\rm{NLOS}}} \left( {\widehat \kappa _D y^{{{\alpha _{{\rm{LOS}}} } \mathord{\left/ {\vphantom {{\alpha _{{\rm{LOS}}} } 2}} \right. \kern-\nulldelimiterspace} 2}} z} \right) \ll {\cal F}_{{\rm{LOS}}} \left( {y^{{{\alpha _{{\rm{LOS}}} } \mathord{\left/ {\vphantom {{\alpha _{{\rm{LOS}}} } 2}} \right. \kern-\nulldelimiterspace} 2}} z} \right)$ since $\widehat \kappa _D  \ll 1$ for $\alpha _{{\rm{NLOS}}}  > \alpha _{{\rm{LOS}}}$, and $\left( {\pi \lambda _{{\rm{BS}}} \widehat D_1^2 } \right)\exp \left( { - \pi \lambda _{{\rm{BS}}} \widehat D_1^2 y} \right) \approx 0$ if $y \in \left[ {1,\infty } \right)$ and $R_{{\rm{cell}}}  < \widehat D_1$.
\paragraph{Sparse (S) Cellular Networks} This regime emerges if the following conditions are satisfied: i) the network is sufficiently sparse such that there are (almost) no inactive BSs, \textit{i.e.}, $p_{{\rm{off}}} \to 0$, but ii) the network is dense enough that \textit{both} ${\mathcal{R}}_{{\rm{LOS}},{\rm{in}}}$ and ${\mathcal{R}}_{{\rm{NLOS}},{\rm{out}}}$ contribute to the ASE, \textit{i.e.}, $R_{{\rm{cell}}}  > \widehat D_1$, some MTs are blocked, \textit{i.e.}, ${{\lambda _{{\rm{BS}}} } \mathord{\left/ {\vphantom {{\lambda _{{\rm{BS}}} } {\lambda _{{\rm{MT}}} }}} \right. \kern-\nulldelimiterspace} {\lambda _{{\rm{MT}}} }} < 1$, and $p_{{\rm{sel}}}$ is that in \eqref{Eq_9}. Thus, ${\rm{ASE}} \to {\rm{ASE}}^{\left( {\rm{S}} \right)}  = \left( {{{N_{{\rm{RB}}} \lambda _{{\rm{BS}}} } \mathord{\left/ {\vphantom {{N_{{\rm{RB}}} \lambda _{{\rm{BS}}} } {\ln \left( 2 \right)}}} \right. \kern-\nulldelimiterspace} {\ln \left( 2 \right)}}} \right)\left( {{\cal R}_{{\rm{LOS}}{\rm{,in}}}^{\left( {\rm{S}} \right)}  + {\cal R}_{{\rm{NLOS}}{\rm{,out}}}^{\left( {\rm{S}} \right)} } \right)$ with ${\cal R}_{{\rm{LOS}}{\rm{,in}}}^{\left( {\rm{S}} \right)}$, ${\cal R}_{{\rm{NLOS}}{\rm{,out}}}^{\left( {\rm{S}} \right)}$ equal to:
\begin{equation}
\label{Eq_35}
\begin{array}{l}
 {\cal R}_{{\rm{LOS}}{\rm{,in}}}^{\left( {\rm{S}} \right)} \mathop  \to \limits^{\left( {{\rm{S}}_1 } \right)} \pi \lambda _{{\rm{BS}}} \widehat \theta _{{\rm{LOS}}}^{\left[ {0,\widehat D_1 } \right]} \int\nolimits_0^1 {\int\nolimits_0^\infty  {\exp \left( {- \frac{{y^{{{\alpha _{{\rm{LOS}}} } \mathord{\left/
 {\vphantom {{\alpha _{{\rm{LOS}}} } 2}} \right.
 \kern-\nulldelimiterspace} 2}} }}{{P_{N,D} }}\frac{z}{{G^{\left( 0 \right)} }}} \right)\bar {\cal M}_{g_{{\rm{LOS}}}^{\left( 0 \right)} } \left( z \right)} }  \\
 \hspace{1.45cm} \times \exp \left( { - \pi \lambda _{{\rm{BS}}} \widehat \theta _{{\rm{LOS}}}^{\left[ {0,\widehat D_1 } \right]} \left( {y - y{\cal F}_{{\rm{LOS}}} \left( {\frac{z}{{G^{\left( 0 \right)} }}} \right) + {\cal F}_{{\rm{LOS}}} \left( {y^{{{\alpha _{{\rm{LOS}}} } \mathord{\left/
 {\vphantom {{\alpha _{{\rm{LOS}}} } 2}} \right.
 \kern-\nulldelimiterspace} 2}} \frac{z}{{G^{\left( 0 \right)} }}} \right)} \right)} \right)\frac{{dzdy}}{z} \\
 \hspace{1.45cm} \propto \pi \lambda _{{\rm{BS}}} \hat \theta _{{\rm{LOS}}}^{\left[ {0,\widehat D_1 } \right]} \int\nolimits_0^1 {\int\nolimits_0^\infty  {\exp \left( { - \frac{{y^{{{\alpha _{{\rm{LOS}}} } \mathord{\left/
 {\vphantom {{\alpha _{{\rm{LOS}}} } 2}} \right.
 \kern-\nulldelimiterspace} 2}} }}{{P_{N,D} }}\frac{z}{{G^{\left( 0 \right)} }}} \right)} } \bar {\cal M}_{g_{{\rm{LOS}}}^{\left( 0 \right)} } \left( z \right) \\
 \hspace{1.45cm} \times \left( {1 - \pi \lambda _{{\rm{BS}}} \hat \theta _{{\rm{LOS}}}^{\left[ {0,\widehat D_1 } \right]} \left( {y - y{\cal F}_{{\rm{LOS}}} \left( {\frac{z}{{G^{\left( 0 \right)} }}} \right) + {\cal F}_{{\rm{LOS}}} \left( {y^{{{\alpha _{{\rm{LOS}}} } \mathord{\left/
 {\vphantom {{\alpha _{{\rm{LOS}}} } 2}} \right.
 \kern-\nulldelimiterspace} 2}} \frac{z}{{G^{\left( 0 \right)} }}} \right)} \right)} \right)\frac{{dzdy}}{z}
\end{array}
\end{equation}
\begin{equation}
\label{Eq_36}
\begin{array}{l}
 {\cal R}_{{\rm{NLOS}}{\rm{,out}}}^{\left( {\rm{S}} \right)} \mathop  \to \limits^{\left( {{\rm{S}}_2 } \right)} \int\nolimits_0^\infty  {\int\nolimits_0^\infty  {\exp \left( { - \left( {\frac{y}{{\pi \lambda_{\rm{BS}} \widehat \phi _{{\rm{NLOS}}}^{\left[ {\widehat D_1 ,\infty } \right]} }}} \right)^{{{\alpha _{{\rm{NLOS}}} } \mathord{\left/
 {\vphantom {{\alpha _{{\rm{NLOS}}} } 2}} \right.
 \kern-\nulldelimiterspace} 2}} \frac{1}{{P_N }}\frac{z}{{G^{\left( 0 \right)} }}} \right)\bar {\cal M}_{g_{{\rm{NLOS}}}^{\left( 0 \right)} } \left( z \right)} }  \\
 \hspace{1.85cm} \times \exp \left( { - \left( {y - y{\cal F}_{{\rm{NLOS}}} \left( {\frac{z}{{G^{\left( 0 \right)} }}} \right)} \right)} \right)\frac{{dzdy}}{z}
\end{array}
\end{equation}
\noindent where (${\rm{S}}_1$) and (${\rm{S}}_2$) are obtained by taking the following into account: i) $\widehat q_{{\rm{LOS}}}^{\left[ {\widehat D_1 ,\infty } \right]}  \approx 0$, ii) $\widehat D_1^2  \gg \left( {\left( {{{\kappa _{{\rm{LOS}}} } \mathord{\left/ {\vphantom {{\kappa _{{\rm{LOS}}} } {\kappa _{{\rm{NLOS}}} }}} \right. \kern-\nulldelimiterspace} {\kappa _{{\rm{NLOS}}} }}} \right)\widehat D_1^{\alpha _{{\rm{LOS}}} } } \right)^{2/\alpha _{{\rm{NLOS}}} }$, since $\alpha _{{\rm{NLOS}}}  > \alpha _{{\rm{LOS}}}$, as well as, for sparse cellular networks, iii) $\pi \lambda _{{\rm{BS}}} \widehat D_1^2 \exp \left( { - \pi \lambda _{{\rm{BS}}} \widehat D_1^2 y} \right) \approx 0$ if $y \in \left[ {0,1 } \right]$ in (${\rm{S}}_2$), since $R_{{\rm{cell}}}  > \widehat D_1$. The weaker approximation in $\propto$ follows by noting that $f\left( {y,z} \right) = y  - y{\cal F}_{{\rm{LOS}}} \left( z \right) + {\cal F}_{{\rm{LOS}}} \left( {y^{{{\alpha _{{\rm{LOS}}} } \mathord{\left/ {\vphantom {{\alpha _{{\rm{LOS}}} } 2}} \right. \kern-\nulldelimiterspace} 2}} z} \right) \in \left[ {0,1} \right]$ for $y \in \left[ {0,1} \right]$, $z \ge 0$, so $\exp \left( { - \pi \lambda _{{\rm{BS}}} \hat \theta _{{\rm{LOS}}}^{\left[ {0,\widehat D_1 } \right]} f\left( {y,z} \right)} \right) \approx 1 - \pi \lambda _{{\rm{BS}}} \hat \theta _{{\rm{LOS}}}^{\left[ {0,\widehat D_1 } \right]} f\left( {y,z} \right)$ if $\pi \lambda _{{\rm{BS}}} \widehat D_1^2  = \left( {{{\widehat D_1^2 } \mathord{\left/ {\vphantom {{\widehat D_1^2 } {R_{{\rm{cell}}}^2 }}} \right. \kern-\nulldelimiterspace} {R_{{\rm{cell}}}^2 }}} \right) < 1$.
\paragraph{Very Sparse (VS) Cellular Networks} This regime emerges if the following conditions are satisfied: i) ${{\lambda _{{\rm{BS}}} } \mathord{\left/ {\vphantom {{\lambda _{{\rm{BS}}} } {\lambda _{{\rm{MT}}} }}} \right. \kern-\nulldelimiterspace} {\lambda _{{\rm{MT}}} }} \ll 1$ and $R_{{\rm{cell}}}  \gg \widehat D_1$, ii) $p_{{\rm{sel}}}$ and $p_{{\rm{off}}}$ are those in \eqref{Eq_9}.  Usually, in addition, $p_{{\rm{off}}}$ in \eqref{Eq_9} is close to zero, \textit{i.e.}, $p_{{\rm{off}}} \to 0$. As a result, the ASE is dominated by ${\mathcal{R}}_{{\rm{NLOS}},{\rm{out}}}$, \textit{i.e.}, ${\rm{ASE}} \to {\rm{ASE}}^{\left( {{\rm{VS}}} \right)}  = \left( {{{N_{{\rm{RB}}} \lambda _{{\rm{BS}}} } \mathord{\left/ {\vphantom {{N_{{\rm{RB}}} \lambda _{{\rm{BS}}} } {\ln \left( 2 \right)}}} \right. \kern-\nulldelimiterspace} {\ln \left( 2 \right)}}} \right){\mathcal{R}}_{{\rm{NLOS}},{\rm{out}}}$, where ${\mathcal{R}}^{(\rm{VS})}_{{\rm{NLOS}},{\rm{out}}}$ can be formulated as follows:
\begin{equation}
\label{Eq_37}
\begin{array}{l}
 {\cal R}_{{\rm{NLOS}}{\rm{,out}}}^{\left( {\rm{VS}} \right)} \mathop  \to \limits^{\left( {{\rm{S}}_2 } \right)} \int\nolimits_0^\infty  {\int\nolimits_0^\infty  {\exp \left( { - \left( {\frac{y}{{\pi \lambda_{\rm{BS}} \widehat \phi _{{\rm{NLOS}}}^{\left[ {\widehat D_1 ,\infty } \right]} }}} \right)^{{{\alpha _{{\rm{NLOS}}} } \mathord{\left/
 {\vphantom {{\alpha _{{\rm{NLOS}}} } 2}} \right.
 \kern-\nulldelimiterspace} 2}} \frac{1}{{P_N }}\frac{z}{{G^{\left( 0 \right)} }}} \right)\bar {\cal M}_{g_{{\rm{NLOS}}}^{\left( 0 \right)} } \left( z \right)} }  \\
 \hspace{1.85cm} \times \exp \left( { - \left( {y - y{\cal F}_{{\rm{NLOS}}} \left( {\frac{z}{{G^{\left( 0 \right)} }}} \right)} \right)} \right)\frac{{dzdy}}{z}
\end{array}
\end{equation}
\noindent where (${\rm{VS}}$) is obtained similar to (${\rm{S}}_2$) in \eqref{Eq_36}. In fact, ${\cal R}_{{\rm{NLOS}}{\rm{,out}}}^{\left( {{\rm{VS}}} \right)}  = {\cal R}_{{\rm{NLOS}}{\rm{,out}}}^{\left( {\rm{S}} \right)}$.

From \eqref{Eq_34}-\eqref{Eq_37}, the impact of important design parameters can be unveiled. A summary of the related performance trends is provided in Table \ref{Table_SummaryTrends}\footnote{Notation of Table \ref{Table_SummaryTrends} -- $X \, \nearrow$, $X \, \searrow$ and $X \, \leftrightarrow$ mean that $X$ increases with, decreases with, is independent of $X$, respectively; $X \, ?$ means that the trend is unpredictable and further details are provided in the main body of the text.}.

Before proceeding further, it is worth mentioning that the ASE in \eqref{Eq_32} is conveniently formulated in terms of a two-fold integral whose integrand function is the product of four terms, each one having a precise physical meaning: 1) $\exp \left( { - zx{{\sigma _N^2 } \mathord{\left/ {\vphantom {{\sigma _N^2 } {\left( {G^{\left( 0 \right)} P_{{\rm{RB}}} } \right)}}} \right. \kern-\nulldelimiterspace} {\left( {G^{\left( 0 \right)} P_{{\rm{RB}}} } \right)}}} \right)$ accounts for the noise, 2) $\exp \left( {\widehat {\cal T}_t \left( {{z \mathord{\left/ {\vphantom {z {G^{\left( 0 \right)} }}} \right. \kern-\nulldelimiterspace} {G^{\left( 0 \right)} }},x} \right)} \right)$ accounts for the other-cell interference, 3) $\bar {\cal M}_{g_s^{\left( 0 \right)} } \left( z \right)$ accounts for the fast-fading of the intended link, and 4) $\widehat\Lambda _{\Phi _{s,t} }^{\left( 1 \right)} \left( {\left[ {0,x} \right)} \right)\exp \left( { - \widehat\Lambda _{\Phi _t } \left( {\left[ {0,x} \right)} \right)} \right)$ accounts for the path-loss of the intended link. This helps interpreting, in the next sub-sections, the (approximated) mathematical expressions in \eqref{Eq_34}-\eqref{Eq_37}.

\begin{table}[!t] \footnotesize
\centering
\caption{Summary of performance trends. ``Rate'' and ``ASE'' are those in \eqref{Eq_34}-\eqref{Eq_37} for VD, D, S, and VS cellular networks, respectively. The notation is provided in footnote 4. \vspace{-0.25cm}}
\newcommand{\tabincell}[2]{\begin{tabular}{@{}#1@{}}#2\end{tabular}}
\begin{tabular}{|c|c|c|c|c|} \hline
 & Very Dense (VD) Networks & Dense (D) Networks & Sparse (S) Networks & Very Sparse (VS) Networks \\ \hline
${\lambda _{{\rm{BS}}} }$ $\nearrow$ & Rate $\nearrow$ $\;$ -- $\;$ ASE $\nearrow$ & Rate $\searrow$ $\;$ -- $\;$ ASE $?$ & Rate $\nearrow$ $\;$ -- $\;$ ASE $\nearrow$ & Rate $\nearrow$ $\;$ -- $\;$ ASE $\nearrow$ \\ \hline
${\lambda _{{\rm{MT}}} }$ $\nearrow$ & Rate $\searrow$ $\;$ -- $\;$ ASE $\nearrow$ & Rate $\leftrightarrow$ $\;$ -- $\;$ ASE $\leftrightarrow$ & Rate $\leftrightarrow$ $\;$ -- $\;$ ASE $\leftrightarrow$ & Rate $\leftrightarrow$ $\;$ -- $\;$ ASE $\leftrightarrow$ \\ \hline
${N_{{\rm{RB}}} }$ $\nearrow$ & Rate $\nearrow$ $\;$ -- $\;$ ASE $\nearrow$ & Rate $\leftrightarrow$ $\;$ -- $\;$ ASE $\nearrow$ & Rate $\searrow$ $\;$ -- $\;$ ASE $\nearrow$ & Rate $\searrow$ $\;$ -- $\;$ ASE $\nearrow$ \\ \hline
$P_{{\rm{BS}}}$ $\nearrow$ & Rate $\leftrightarrow$ $\;$ -- $\;$ ASE $\leftrightarrow$ & Rate $\leftrightarrow$ $\;$ -- $\;$ ASE $\leftrightarrow$ & Rate $\nearrow$ $\;$ -- $\;$ ASE $\nearrow$ & Rate $\nearrow$ $\;$ -- $\;$ ASE $\nearrow$ \\ \hline
${G^{\left( 0 \right)} }$ $\nearrow$ & Rate $\nearrow$ $\;$ -- $\;$ ASE $\nearrow$ & Rate $\nearrow$ $\;$ -- $\;$ ASE $\nearrow$ & Rate $\nearrow$ $\;$ -- $\;$ ASE $\nearrow$ & Rate $\nearrow$ $\;$ -- $\;$ ASE $\nearrow$ \\ \hline
$D_1$ $\nearrow$ & Rate $\searrow$ $\;$ -- $\;$ ASE $\searrow$ & Rate $\searrow$ $\;$ -- $\;$ ASE $\searrow$ & Rate $\nearrow$ $\;$ -- $\;$ ASE $\nearrow$ & Rate $\leftrightarrow$ $\;$ -- $\;$ ASE $\leftrightarrow$ \\ \hline
$\sigma _s$ $\nearrow$ & Rate $\nearrow$ $\;$ -- $\;$ ASE $\nearrow$ & Rate $\nearrow$ $\;$ -- $\;$ ASE $\nearrow$ & Rate $?$ $\;$ -- $\;$ ASE $?$ & Rate $\nearrow$ $\;$ -- $\;$ ASE $\nearrow$ \\ \hline
\end{tabular}
\label{Table_SummaryTrends} \vspace{-0.65cm}
\end{table}
The impact of several system parameters follows by direct inspection of \eqref{Eq_34}-\eqref{Eq_37}. These simple case studies are not explicitly discussed in the sequel. The impact of a few important parameters deserves, on the other hand, further comments and clarifications. In some cases, in addition, their impact in very dense, dense, sparse and very sparse cellular networks is different.
\setcounter{paragraph}{0}
\vspace{-0.5cm}
\subsection{Impact of the Density of Base Stations} \label{Trends_BSs} \vspace{-0.25cm}
Increasing the density of BSs has a different impact, depending on the operating regime being considered. The comments in what follows hold if the antennas are not very directive. In Section \ref{Trends_Antenna}, the impact of the antenna radiation pattern is discussed and elaborated in detail.
\paragraph{Very Dense Regime} Both approximations in \eqref{Eq_33} highlight that rate and ASE increase as $\lambda_{\rm{BS}}$ increases. The accurate  approximation in \eqref{Eq_33} shows that increasing $\lambda_{\rm{BS}}$ brings the BSs closer to the MTs (because of $\pi \lambda_{\rm{BS}} \exp(- \pi \lambda_{\rm{BS}}y)$) without increasing the other-cell interference (see the first exponential function in the integrand), which does not depend on $\lambda_{\rm{BS}}$.
\paragraph{Dense Regime} The weak approximation in \eqref{Eq_34} shows that increasing $\lambda_{\rm{BS}}$ decreases the rate. In fact, the function $ - \left( {{{\widehat D_1^2 } \mathord{\left/ {\vphantom {{\widehat D_1^2 } {R_{{\rm{cell}}}^2 }}} \right. \kern-\nulldelimiterspace} {R_{{\rm{cell}}}^2 }}} \right){\cal F}_{{\rm{LOS}}} \left( {\left( {\left( {{{R_{{\rm{cell}}}^2 } \mathord{\left/ {\vphantom {{R_{{\rm{cell}}}^2 } {\widehat D_1^2 }}} \right. \kern-\nulldelimiterspace} {\widehat D_1^2 }}} \right)y} \right)^{{{\alpha _{{\rm{LOS}}} } \mathord{\left/ {\vphantom {{\alpha _{{\rm{LOS}}} } 2}} \right. \kern-\nulldelimiterspace} 2}} \left( {{z \mathord{\left/ {\vphantom {z {G^{\left( 0 \right)} }}} \right. \kern-\nulldelimiterspace} {G^{\left( 0 \right)} }}} \right)} \right)$ is positive and monotonically decreases as $R_{{\rm{cell}}}$ decreases for $y \in \left[ {0,1} \right]$ and $z \ge 0$. Unlike \eqref{Eq_33}, in fact, \eqref{Eq_34} highlights that both the intended power and the other-cell interference depend on $\lambda_{\rm{BS}}$. The impact of $\lambda_{\rm{BS}}$ on the ASE depends, on the other hand, on the pair $\left( {\widehat D_1 ,G^{\left( 0 \right)} } \right)$. The derivative of the related integrand function with respect to $\lambda_{\rm{BS}}$ is, in fact, neither always positive nor always negative for every $y \in \left[ {0,1} \right]$ and $z \ge 0$. Further comments are provided in Sections \ref{Trends_Antenna}, \ref{Trends_Blockages}.
\paragraph{Sparse and Very Sparse Regimes} The weak approximation in \eqref{Eq_35} and \eqref{Eq_36} demonstrate that increasing $\lambda_{\rm{BS}}$ increases the rate. In fact, $\left( {\pi \lambda _{{\rm{BS}}} \hat \theta _{{\rm{LOS}}}^{\left[ {0,\widehat D_1 } \right]} } \right)^2  \ll \pi \lambda _{{\rm{BS}}} \hat \theta _{{\rm{LOS}}}^{\left[ {0,\widehat D_1 } \right]}$ if ${{\widehat D_1^2 } \mathord{\left/ {\vphantom {{\widehat D_1^2 } {R_{{\rm{cell}}}^2 }}} \right. \kern-\nulldelimiterspace} {R_{{\rm{cell}}}^2 }} < 1$, which implies that \eqref{Eq_35} increases with $\lambda_{\rm{BS}}$. The impact of $\lambda_{\rm{BS}}$ on the ASE is the same.
\vspace{-0.5cm}
\subsection{Impact of the Density of Mobile Terminals} \label{Trends_MTs} \vspace{-0.25cm}
The density of MTs has a noticeable impact on rate and ASE only in the very dense regime. In this case, in fact, $\lambda_{\rm{MT}}$ determines the other-cell interference, since many BSs are likely not to have MTs to serve and, thus, are inactive. In all the other regimes, on the other hand, all BSs are likely to be active and to contribute to the other-cell interference. Both approximations in \eqref{Eq_33} show that the rate decreases as $\lambda_{\rm{MT}}$ increases. The impact of $\lambda_{\rm{MT}}$ on the ASE needs deeper inspection. Based on the weak approximation in \eqref{Eq_33}, the ASE is a function of ${{\lambda _{{\rm{MT}}} } \mathord{\left/ {\vphantom {{\lambda _{{\rm{MT}}} } {\left( {1 + \lambda _{{\rm{MT}}} f\left( z \right)} \right)}}} \right. \kern-\nulldelimiterspace} {\left( {1 + \lambda _{{\rm{MT}}} f\left( z \right)} \right)}}$, where $f\left( z \right) = - (N_{{\rm{RB}}} \lambda _{{\rm{BS}}})^{-1} {\mathcal{F}}\left( {{z \mathord{\left/ {\vphantom {z {G^{\left( 0 \right)} }}} \right. \kern-\nulldelimiterspace} {G^{\left( 0 \right)} }}} \right) \ge 0$ for $z \ge 0$. Since its first derivative with respect to $\lambda_{\rm{MT}}$ is positive for $z \ge 0$, we conclude that the ASE increases as $\lambda_{\rm{MT}}$ increases.
\vspace{-0.5cm}
\subsection{Impact of the Number of Resource Blocks} \label{Trends_NRB} \vspace{-0.25cm}
In very dense and dense regimes, the impact of $N_{\rm{RB}}$ follows from \eqref{Eq_33} and \eqref{Eq_34}. In sparse and very sparse regimes, \eqref{Eq_35}-\eqref{Eq_37} highlight that the impact of $N_{\rm{RB}}$ on the ASE depends on two contrasting effects: on the one hand, the number of served MTs increases with $N_{\rm{RB}}$, and, on the other hand, the transmit power per RB decreases with $N_{\rm{RB}}$. The net impact of $N_{\rm{RB}}$ in these regimes deserves some additional comments. In the sparse regime, since $\left( {\pi \lambda _{{\rm{BS}}} \hat \theta _{{\rm{LOS}}}^{\left[ {0,\widehat D_1 } \right]} } \right)^2  \ll \pi \lambda _{{\rm{BS}}} \hat \theta _{{\rm{LOS}}}^{\left[ {0,\widehat D_1 } \right]}$ if ${{\widehat D_1^2 } \mathord{\left/ {\vphantom {{\widehat D_1^2 } {R_{{\rm{cell}}}^2 }}} \right. \kern-\nulldelimiterspace} {R_{{\rm{cell}}}^2 }} < 1$, the net impact of $N_{\rm{RB}}$ on ${\cal R}_{{\rm{LOS}}{\rm{,in}}}^{\left( {\rm{S}} \right)}$ is determined by:
\begin{equation}
\label{Eq_38}
\begin{array}{l}
N_{{\rm{RB}}} \int\nolimits_0^1 {\exp \left( { - N_{{\rm{RB}}} y^{{{\alpha _{{\rm{LOS}}} } \mathord{\left/
 {\vphantom {{\alpha _{{\rm{LOS}}} } 2}} \right.
 \kern-\nulldelimiterspace} 2}} f\left( z \right)} \right)dy}  = \frac{2}{{\alpha _{{\rm{LOS}}} }}f\left( z \right)^{ - {2 \mathord{\left/
 {\vphantom {2 {\alpha _{{\rm{LOS}}} }}} \right.
 \kern-\nulldelimiterspace} {\alpha _{{\rm{LOS}}} }}} N_{{\rm{RB}}}^{1 - {2 \mathord{\left/
 {\vphantom {2 {\alpha _{{\rm{LOS}}} }}} \right.
 \kern-\nulldelimiterspace} {\alpha _{{\rm{LOS}}} }}} \gamma \left( {\frac{2}{{\alpha _{{\rm{LOS}}} }},N_{{\rm{RB}}} f\left( z \right)} \right)
\end{array}
\end{equation}
\noindent where $f\left( z \right) = {{\left( {\sigma _N^2 \kappa _{{\rm{LOS}}} \widehat D_1^{\alpha _{{\rm{LOS}}} } z} \right)} \mathord{\left/ {\vphantom {{\left( {\sigma _N^2 \kappa _{{\rm{LOS}}} \widehat D_1^{\alpha _{{\rm{LOS}}} } z} \right)} {\left( {P_{{\rm{BS}}} G^{\left( 0 \right)} } \right)}}} \right. \kern-\nulldelimiterspace} {\left( {P_{{\rm{BS}}} G^{\left( 0 \right)} } \right)}} \ge 0$ and $\gamma \left( { \cdot , \cdot } \right)$ is the lower incomplete gamma function. From \eqref{Eq_38}, we conclude that, in the sparse regime, ${\cal R}_{{\rm{LOS}}{\rm{,in}}}^{\left( {\rm{S}} \right)}$ increases as $N_{\rm{RB}}$ increases. A similar study can be conducted for ${\cal R}_{{\rm{NLOS}}{\rm{,out}}}^{\left( {\rm{S}} \right)}$. In this case, the integral over $y$ can be expressed in closed-form in terms of the Meijer G-function, which can be shown to increase as $N_{\rm{RB}}$ increases. From \eqref{Eq_35}-\eqref{Eq_37}, we conclude that the ASE increases as $N_{\rm{RB}}$ increases.
\vspace{-0.5cm}
\subsection{Impact of the Antenna Radiation Pattern} \label{Trends_Antenna} \vspace{-0.25cm}
Since ${\cal F}_s \left( z \right) \to 0$ if $z \to 0$, \eqref{Eq_33}-\eqref{Eq_37} prove, in all regimes, that rate and ASE increase as the directivity of the antenna increases. In very dense and dense regimes, the other-cell interference is reduced. In sparse and very sparse regimes, the intended link is enhanced. If, \textit{e.g.}, the antennas are highly directive, increasing $\lambda_{\rm{BS}}$ increases the ASE in the dense regime (Section \ref{Trends_BSs}).
\vspace{-0.5cm}
\subsection{Impact of the Density of Blockages} \label{Trends_Blockages} \vspace{-0.25cm}
According to \cite{Heath__Blockage}, the parameter $b_{\rm{RS}}$ of the blockage model based on random shape theory (see Table \ref{LinkStateModels}) is directly related to the percentage of area covered by buildings. The higher the density of blockages is, more specifically, the larger $b_{\rm{RS}}$ is. In the single-ball model of \textit{Corollary \ref{Corollary__ASE_PT}}, the radius, $\widehat D_1$, of the LOS/NLOS ball plays the same role as $b_{\rm{RS}}$. By applying the matching criterion in \eqref{Eq_25}, in particular, it is possible to show that $\widehat D_1$ decreases as $b_{\rm{RS}}$ increases. Further details are provided in Section \ref{Results}. In other words, the higher the density of blockages is, the smaller $\widehat D_1$ is. This is in agreement with intuition: the more the buildings, the shorter the distance that a link is in LOS with high probability. By analyzing the impact of $\widehat D_1$ in \eqref{Eq_33}-\eqref{Eq_37}, as a result, the effect of blockages can be unveiled. Let us consider \eqref{Eq_33} and \eqref{Eq_34}. By direct inspection, it follows that $ - \widehat D_1^2 {\cal F}_s \left( { z/\widehat D_1^2} \right)$ is positive and that it monotonically decreases as ${\widehat D_1 } > 1$ increases. This implies that increasing ${\widehat D_1 }$ (\textit{i.e.}, fewer blockages are present), both rate and ASE decrease. In very dense and dense regimes, hence, the presence of blockages is useful for reducing the impact of the other-cell interference. The weaker approximation in \eqref{Eq_33}, however, is independent of ${\widehat D_1 }$. This implies that, in the very dense regime, the impact of blockages is expected to be limited. From \eqref{Eq_37}, we note that the impact of blockages is minor in the very sparse regime as well. From \eqref{Eq_35}, on the other hand, we note that the impact of blockages is determined by:
\begin{equation}
\label{Eq_39}
\begin{array}{l}
\Theta _{{\rm{LOS}}} \widehat q_{{\rm{LOS}}}^{\left[ {0,\widehat D_1 } \right]} \widehat D_1^2 \int\nolimits_0^1 {\exp \left( { - \frac{{\widehat D_1^{\alpha _{{\rm{LOS}}} } z}}{{G^{\left( 0 \right)} }}y^{{{\alpha _{{\rm{LOS}}} } \mathord{\left/
 {\vphantom {{\alpha _{{\rm{LOS}}} } 2}} \right.
 \kern-\nulldelimiterspace} 2}} } \right)dy}  = \Theta _{{\rm{LOS}}} \widehat q_{{\rm{LOS}}}^{\left[ {0,\widehat D_1 } \right]} \frac{2}{{\alpha _{{\rm{LOS}}} }}\left( {\frac{z}{{G^{\left( 0 \right)} }}} \right)^{ - {2 \mathord{\left/
 {\vphantom {2 {\alpha _{{\rm{LOS}}} }}} \right.
 \kern-\nulldelimiterspace} {\alpha _{{\rm{LOS}}} }}} \gamma \left( {\frac{2}{{\alpha _{{\rm{LOS}}} }},\frac{{\widehat D_1^{\alpha _{{\rm{LOS}}} } z}}{{G^{\left( 0 \right)} }}} \right)
\end{array}
\end{equation}
\noindent From \eqref{Eq_39}, we conclude that, in the sparse regime, rate and ASE decrease as $\widehat D_1$ decreases.
\vspace{-0.5cm}
\subsection{Impact of the Shadowing Severity} \label{Trends_Shadowing} \vspace{-0.25cm}
Based on Section \ref{IM_Methodology}, the standard deviation of shadowing, $\sigma_s$, affects $\widehat D_1$, $\widehat q_s^{[\cdot,\cdot]}$ and $\Theta_s$. By inspection of \eqref{Eq_33}-\eqref{Eq_37}, the impact of $\sigma_s$ implicitly emerges in ${\widehat \theta _{{\rm{LOS}}}^{\left[ {0,\widehat D_1 } \right]} }$, ${\widehat \phi _{{\rm{NLOS}}}^{\left[ {\widehat D_1, \infty } \right]} }$ and $P_{N,D}$. By applying the matching criterion in \eqref{Eq_25} to different blockage models, it is possible to show that ${\widehat \theta _{{\rm{LOS}}}^{\left[ {0,\widehat D_1 } \right]} }$ and $\widehat D_1$ both decrease and ${\widehat \phi _{{\rm{NLOS}}}^{\left[ {\widehat D_1, \infty } \right]} }$ increases as $\sigma_s$ increases. Further details are provided in Section \ref{Results}. From \eqref{Eq_33}-\eqref{Eq_37}, as a result, the impact of shadowing on rate and ASE is determined by $\widehat D_1$ in very dense and dense regimes. The trends, thus, follow from Section \ref{Trends_Blockages}. In sparse networks, \eqref{Eq_39} highlights that the impact of $\sigma_s$ highly depends on the blockage model being considered, \textit{i.e.}, the specific triplet of parameters $\left( {\Theta _{{\rm{LOS}}} ,\widehat q_{{\rm{LOS}}}^{\left[ {0,\widehat D_1 } \right]} ,\widehat D_1^{\alpha _{{\rm{LOS}}} } } \right)$ that appears in \eqref{Eq_39}. In very sparse networks, \eqref{Eq_37} shows that rate and ASE increase as $\sigma_s$ increases.
\vspace{-0.5cm}
\subsection{Existence of a Local Minimum and Maximum of the Rate} \label{Trends_MinimumMaximum} \vspace{-0.25cm}
Let us consider the expressions of the rate for very dense and dense cellular networks in \eqref{Eq_33} and \eqref{Eq_34}, respectively. They have an opposite trend as a function of the density of BSs: if $\lambda _{{\rm{BS}}}$ increases, the rate increases in the very dense regime and decreases in the dense regime, respectively. They, in addition, coincide with each other if $\lambda _{{\rm{BS}}}  = {{\lambda _{{\rm{MT}}} } \mathord{\left/ {\vphantom {{\lambda _{{\rm{MT}}} } {N_{{\rm{RB}}} }}} \right. \kern-\nulldelimiterspace} {N_{{\rm{RB}}} }}$. This implies that the rate is expected to have a local minimum when $\lambda _{{\rm{BS}}}  \approx {{\lambda _{{\rm{MT}}} } \mathord{\left/ {\vphantom {{\lambda _{{\rm{MT}}} } {N_{{\rm{RB}}} }}} \right. \kern-\nulldelimiterspace} {N_{{\rm{RB}}} }}$ approximatively holds.

Let us consider the expressions of the rate for dense and sparse cellular networks in \eqref{Eq_34} and \eqref{Eq_35}, respectively. They have an opposite trend as a function of the density of BSs: if $\lambda _{{\rm{BS}}}$ increases, the rate decreases in the dense regime and increases in the sparse regime, respectively. This implies that the rate is expected to have a local maximum. By direct inspection of the weaker approximation in \eqref{Eq_35}, this local maximum occurs when $\pi \lambda _{{\rm{BS}}} \hat \theta _{{\rm{LOS}}}^{\left[ {0,\widehat D_1 } \right]}  \approx \left( {\pi \lambda _{{\rm{BS}}} \hat \theta _{{\rm{LOS}}}^{\left[ {0,\widehat D_1 } \right]} } \right)^2$, which implies ${{\widehat D_1^2 } \mathord{\left/ {\vphantom {{\widehat D_1^2 } {R_{{\rm{cell}}}^2 }}} \right. \kern-\nulldelimiterspace} {R_{{\rm{cell}}}^2 }} \approx \left( {{{\widehat D_1^2 } \mathord{\left/ {\vphantom {{\widehat D_1^2 } {R_{{\rm{cell}}}^2 }}} \right. \kern-\nulldelimiterspace} {R_{{\rm{cell}}}^2 }}} \right)^2$ and $R_{{\rm{cell}}}  \approx \widehat D_1$. In other words, the local maximum depends on the density of blockages and the corresponding average cell radius is expected to be proportional to the radius, $\widehat D_1$, of the LOS/NLOS ball that models the blockages. From Section \ref{Trends_Blockages}, we know that the higher the density of blockages is, the smaller the radius of the LOS/NLOS ball. Moving from rural to urban scenarios, thus, increasing the density of BSs is needed, as expected, to enable cellular networks working close to such a local maximum.

In the rate, the local minimum and maximum are expected to be clearly visible if the condition ${{\lambda _{{\rm{MT}}} } \mathord{\left/ {\vphantom {{\lambda _{{\rm{MT}}} } {N_{{\rm{RB}}} }}} \right. \kern-\nulldelimiterspace} {N_{{\rm{RB}}} }} \gg {1 \mathord{\left/ {\vphantom {1 {\left( {\pi \widehat D_1^2 } \right)}}} \right. \kern-\nulldelimiterspace} {\left( {\pi \widehat D_1^2 } \right)}}$ holds. In this case, in fact, the dense regime emerges distinctly.
\vspace{-0.5cm}
\subsection{Guidelines for System-Level Optimization} \label{Trends_SystemLevelOptimization} \vspace{-0.25cm}
From an engineering standpoint, the existence of a local minimum and of a local maximum of the rate provides important design guidelines for system-level optimization. More specifically, a density of BSs in the range $\lambda _{{\rm{BS}}}  \in \left( {{1 \mathord{\left/ {\vphantom {1 {\left( {\pi \widehat D_1^2 } \right)}}} \right. \kern-\nulldelimiterspace} {\left( {\pi \widehat D_1^2 } \right)}},{{\lambda _{{\rm{MT}}} } \mathord{\left/ {\vphantom {{\lambda _{{\rm{MT}}} } {N_{{\rm{RB}}} }}} \right. \kern-\nulldelimiterspace} {N_{{\rm{RB}}} }}} \right)$ with ${{\lambda _{{\rm{MT}}} } \mathord{\left/ {\vphantom {{\lambda _{{\rm{MT}}} } {N_{{\rm{RB}}} }}} \right. \kern-\nulldelimiterspace} {N_{{\rm{RB}}} }} > {1 \mathord{\left/ {\vphantom {1 {\left( {\pi \widehat D_1^2 } \right)}}} \right. \kern-\nulldelimiterspace} {\left( {\pi \widehat D_1^2 } \right)}}$ should be avoided, since the rate decreases if $\lambda _{{\rm{BS}}}$ increases. Setups where $\lambda _{{\rm{BS}}}  > {{\lambda _{{\rm{MT}}} } \mathord{\left/ {\vphantom {{\lambda _{{\rm{MT}}} } {N_{{\rm{RB}}} }}} \right. \kern-\nulldelimiterspace} {N_{{\rm{RB}}} }}$ may be considered only if economically convenient and cost-effective. As a rule of thumb, system setups where the average cell radius is of the order of magnitude of the radius of the LOS/NLOS ball, which depends on the density of blockages and, so, on the specific environment, may represent a good trade-off between achievable performance and cost. A similar conclusion was somehow implied in \cite{Lopez-Perez__LosNlos} for fully-loaded cellular networks and for a specific set of system parameters. The connection with the density of blockages, however, was not explicitly made. Quoting \cite{Lopez-Perez__LosNlos}: ``\textit{Note that our conclusion is made from the investigated set of parameters, and it is of significant interest to further study the generality of this conclusion in other network models and with other parameter sets}''. The mathematical approach proposed in the present paper is applicable to a more general system setup and the impact of all system parameters clearly emerges from the asymptotic frameworks. It generalizes, in addition, the preliminary findings in \cite{Galiotto__LosNlos} on the impact of the density of BSs in partially-loaded cellular networks, which were obtained with the aid of linear regression analysis. The IM-based approach allows us to draw general conclusions and to perform accurate system-level optimization without the need of simplifying the system model.
\begin{table}[!t] \footnotesize
\centering
\caption{Simulation setup (compliant with 3GPP and Long Term Evolution Advance (LTE-A)). \vspace{-0.35cm}}
\newcommand{\tabincell}[2]{\begin{tabular}{@{}#1@{}}#2\end{tabular}}
\begin{tabular}{|c||c|} \hline
Path-Loss & $\alpha_{\rm{LOS}} = 2.6$, $\alpha_{\rm{NLOS}} = 3.8$, $\kappa_{\rm{LOS}}=\kappa_{\rm{NLOS}}= (4\pi f_0/ c_0)^2$ with $f_0 = 2.1$ GHz, $c_0 \approx 3 \cdot 10^8$ m/s \\ \hline
Shadowing, fading & $\sigma_{\rm{LOS}} = 4$ dB, $\sigma_{\rm{NLOS}} = 10$ dB, $\Omega_{\rm{LOS}} = \Omega_{\rm{NLOS}} = 1$, $m_{\rm{LOS}} = 2.8$, $m_{\rm{NLOS}} = 1$ \\ \hline
BS power, noise & $P_{\rm{BS}} = 20$ dBm, $\sigma_{N}^2 = -174 + 10 \log_{10}(\rm{B_W}) + \mathcal{F}$ dBm with $\rm{B_W} = 180$ kHz, $\mathcal{F} = 10$ dB \\ \hline
Link-state & 3GPP \cite{3GPP_pathloss}: $a_{\rm{3G}} = 18$, $b_{\rm{3G}} = 36$, $c_{\rm{3G}} = 1$; RS \cite{Kulkarni__Globecom2014}: $a_{\rm{RS}} = 1$, $b_{\rm{RS}} = 0.046$ ${\rm{m}^{-1}}$  \\ \hline
Empirical & BSs: O2 in \cite[Table 1]{ACM_MDR}, \cite[Fig. 1]{ACM_MDR}, $R_{{\rm{cell}}}  \approx 83.4$ m; buildings: London \cite[Fig. 1]{ACM_MDR}, \cite[Sec. 2.3.1]{ACM_MDR} \\ \hline
$\lambda _{{\rm{MT}}}  = \left( {\pi R_{{\rm{MT}}}^2 } \right)^{ - 1}$ & $R_{{\rm{MT}}}  \approx \left\{ {3.9,7.6,11.9,50,100} \right\}$ m is the population density of Paris, London, Rome, Pennsylvania, Texas \\ \hline
\end{tabular}
\label{Table_Setup} \vspace{-0.45cm}
\end{table}
\vspace{-0.5cm}
\section{Numerical and Simulation Results} \label{Results} \vspace{-0.25cm}
In this section, numerical results are illustrated and commented with the aim of validating the accuracy of the IM-based approach for various blockage, \textit{i.e.}, link state, models (see Table \ref{LinkStateModels}), and of substantiating the findings and performance trends identified in Section \ref{Trends}. The proposed approach is further compared against empirical data and numerical estimates of rate and ASE obtained for the actual locations of BSs and footprints of buildings corresponding to a dense urban area in downtown London. Information about this empirical dataset is available in \cite{ACM_MDR}. The details of the simulation setup are provided in Table \ref{Table_Setup}. For ease of illustration and for its practical relevance, a two-state blockage model is considered, \textit{i.e.}, with LOS and NLOS links.

In all figures, system-level (Monte Carlo) simulation results are obtained without enforcing any approximations on the antenna radiation pattern and on the blockage model. The IM-based approach is, on the other hand, based on their approximations in Tables \ref{AntennaRadiationPatterns_Fitting} and \ref{Table_IM_3ball}.
\begin{table}[!t] \footnotesize
\centering
\caption{Parameters of the IM-based approximation computed by using \eqref{Eq_25}: one-ball model. \vspace{-0.35cm}}
\newcommand{\tabincell}[2]{\begin{tabular}{@{}#1@{}}#2\end{tabular}}
\begin{tabular}{|c|c|c|c|} \hline
 & $\widehat D_1$ (m) & $\widehat q_{\rm{LOS}}^{[0, \widehat D_1]}$ & $\widehat q_{\rm{LOS}}^{[\widehat D_1, \infty]}$ \\ \hline
3GPP & 186.2083 & 0.4256 & $\approx 10^{-12}$ \\ \hline
Random shape & 38.7305 & 0.3999 & 0 \\ \hline
%Linear & 47.8632 & 0.7170 & 0.2007 \\ \hline
Empirical & 87.6027 & 0.3466 & 0 \\ \hline
\end{tabular}
\label{Table_IM_1ball} \vspace{-0.45cm}
\end{table}
\begin{table}[!t] \footnotesize
\centering
\caption{Parameters of the IM-based approximation computed by using \eqref{Eq_25}: three- and four-ball models. \vspace{-0.35cm}}
\newcommand{\tabincell}[2]{\begin{tabular}{@{}#1@{}}#2\end{tabular}}
\begin{tabular}{|c|c|c|c|c|c|c|c|c|c|} \hline
 & $\widehat D_1$ (m) & $\widehat D_2$ (m) & $\widehat D_3$ (m) & $\widehat D_4$ (m) & $\widehat q_{\rm{LOS}}^{[0, \widehat D_1]}$ & $\widehat q_{\rm{LOS}}^{[\widehat D_1, \widehat D_2]}$ & $\widehat q_{\rm{LOS}}^{[\widehat D_2, \widehat D_3]}$ & $\widehat q_{\rm{LOS}}^{[\widehat D_3, \widehat D_4]}$ & $\widehat q_{\rm{LOS}}^{[\widehat D_4, \infty]}$ \\ \hline
3GPP & 38.8639 & 187.0276 & 1708.6 & 23922 & 0.9119 & 0.2312 & 0.0241 & 0.0019 & $4.63 \cdot 10^{-5}$ \\ \hline
Random shape & 10.2020 & 30.4979 & 105.1919 & $\infty$ & 0.7666 & 0.3923 & 0.0588 & 0 & -- \\ \hline
%Linear & 16.6105 & 41.9872 & 86.2753 & 0.8287 & 0.6206 & 0.3360 & 0.2002 \\ \hline
Empirical & 15.9867 & 60.2296 & 242.2488 & $\infty$ & 0.7125 & 0.3299 & 0.0572 & 0 & -- \\ \hline
\end{tabular}
\label{Table_IM_3ball} \vspace{-0.45cm}
\end{table}
\vspace{-0.5cm}
\subsection{IM-based Approach: Approximations and Impact of Shadowing and Density of Blockages} \label{IM} \vspace{-0.25cm}
Tables \ref{Table_IM_1ball} and \ref{Table_IM_3ball} provide the input parameters of the IM-based approach for one- and three- \& four-ball approximations, respectively. Unless otherwise stated, the setup in Table \ref{Table_Setup} is used. The three- \& four-ball models offer a more accurate approximation of the actual blockage models, at the cost of a higher computational complexity. The reason why both approximations are considered is that the one-ball model, even though less accurate, provides similar performance trends as the three- \& four-ball models, which, in further text, are shown to be in agreement with the conclusions drawn in Section \ref{Trends} and based on \eqref{Eq_33}-\eqref{Eq_37}. All mathematical frameworks are generated by using the data in Tables \ref{Table_IM_1ball}, \ref{Table_IM_3ball} and \eqref{Eq_32}. The data reported in Tables \ref{Table_IM_Blockages} and \ref{Table_IM_Shadowing}, on the other hand, are useful for validating the conclusions drawn in Sections \ref{Trends_Blockages} and \ref{Trends_Shadowing}, respectively. They are shown only for three- \& four-ball approximations, but the same trends hold for the one-ball approximation and for link state models different from 3GPP and RS. They confirm that the radii of the balls decrease as the density of blockages increases (\textit{i.e.}, $b_{\rm{RS}}$ increases), and that they decrease as the shadowing standard deviation increases. Thus, the expected trends discussed in Section \ref{Trends_Shadowing} for the parameters $\widehat\theta _{{\rm{LOS}}}^{\left[ {0,\widehat D_1 } \right]}$ and $\widehat\varphi _{{\rm{NLOS}}}^{\left[ {\widehat D_1 ,\infty } \right]}$ are confirmed.
\begin{table}[!t] \footnotesize
\centering
\caption{Impact of blockages on the IM-based approximation of the RS link state model. \vspace{-0.35cm}}
\newcommand{\tabincell}[2]{\begin{tabular}{@{}#1@{}}#2\end{tabular}}
\begin{tabular}{|c|c|c|c|c|c|c|c|} \hline
 & $\widehat D_1$ (m) & $\widehat D_2$ (m) & $\widehat D_3$ (m) & $\widehat q_{\rm{LOS}}^{[0, \widehat D_1]}$ & $\widehat q_{\rm{LOS}}^{[\widehat D_1, \widehat D_2]}$ & $\widehat q_{\rm{LOS}}^{[\widehat D_2, \widehat D_3]}$ & $\widehat q_{\rm{LOS}}^{[\widehat D_3, \infty]}$ \\ \hline
$b_{\rm{RS}} = 0.01$ & 29.5080 &	112.7958 &	397.2890 & 	0.8711 &	0.4728 &	0.0767 &	0 \\ \hline
$b_{\rm{RS}} = 0.03$ & 13.2667 &	43.2311 &	152.0526 &	0.8057 &	0.4221 &	0.0630 &	0 \\ \hline
$b_{\rm{RS}} = 0.05$ & 9.7242 &	28.5634 &	98.0039 &	0.7577 &	0.3857 &	0.0579 &	0 \\ \hline
$b_{\rm{RS}} = 0.07$ & 8.0938 &	22.1310 &	74.0757 &	0.7171 &	0.3565 &	0.0543 &	0 \\ \hline
$b_{\rm{RS}} = 0.09$ & 7.144 &	18.5074 &	60.5836 &	0.6801 &	0.3316 &	0.0511 &	0 \\ \hline
\end{tabular}
\label{Table_IM_Blockages} \vspace{-0.45cm}
\end{table}
\begin{table}[!t] \footnotesize
\centering
\caption{Impact of shadowing on the IM-based approximation of 3GPP link state model ($\sigma_{\rm{LOS}} = \sigma_{\rm{NLOS}} = \sigma$). \vspace{-0.75cm}}
\newcommand{\tabincell}[2]{\begin{tabular}{@{}#1@{}}#2\end{tabular}}
\begin{tabular}{|c|c|c|c|c|c|c|c|c|c|} \hline
 & $\widehat D_1$ (m) & $\widehat D_2$ (m) & $\widehat D_3$ (m) & $\widehat D_4$ (m) & $\widehat q_{\rm{LOS}}^{[0, \widehat D_1]}$ & $\widehat q_{\rm{LOS}}^{[\widehat D_1, \widehat D_2]}$ & $\widehat q_{\rm{LOS}}^{[\widehat D_2, \widehat D_3]}$ & $\widehat q_{\rm{LOS}}^{[\widehat D_3, \widehat D_4]}$ & $\widehat q_{\rm{LOS}}^{[\widehat D_3, \infty]}$ \\ \hline
$\sigma = 1$ dB & 52.1531 &	292.4923 &	2958.8 &	39558 &	0.9368 &	0.1597 &	0.0158 &	0.0013 &	$4.83 \cdot 10^{-5}$ \\ \hline
$\sigma = 3$ dB & 45.9521 &	252.7484 &	2604.3 &	36179 &	0.9294 &	0.1699 &	0.0164 &	0.0013 &	$4.50 \cdot 10^{-5}$ \\ \hline
$\sigma = 5$ dB & 36.4947 &	199.8493 &	2094.5 &	30848 &	0.9110 &	0.180 &	0.0169 &	0.0013 &	$3.87 \cdot 10^{-5}$ \\ \hline
$\sigma = 7$ dB & 26.3860 &	148.8296 &	1568.1 &	24696 &	0.8808 &	0.1831 &	0.0169 &	0.0013 &	$3.08 \cdot 10^{-5}$ \\ \hline
$\sigma = 9$ dB & 17.4373 &	105.1683 &	1109.2 &	18702 &	0.8367 &	0.1774 &	0.0164 &	0.0012 &	$2.25 \cdot 10^{-5}$ \\ \hline
\end{tabular}
\label{Table_IM_Shadowing} \vspace{-0.65cm}
\end{table}
\vspace{-0.5cm}
\subsection{IM-based Approach: On the Importance of Modeling Assumptions} \label{DifferentTrends_Rev} 
\begin{figure}[!t]
\setlength{\captionmargin}{10.0pt}
\setlength{\abovecaptionskip}{4pt}
\centering
\begin{minipage}{0.5\columnwidth}
\centering
\includegraphics[width=1.0\columnwidth]{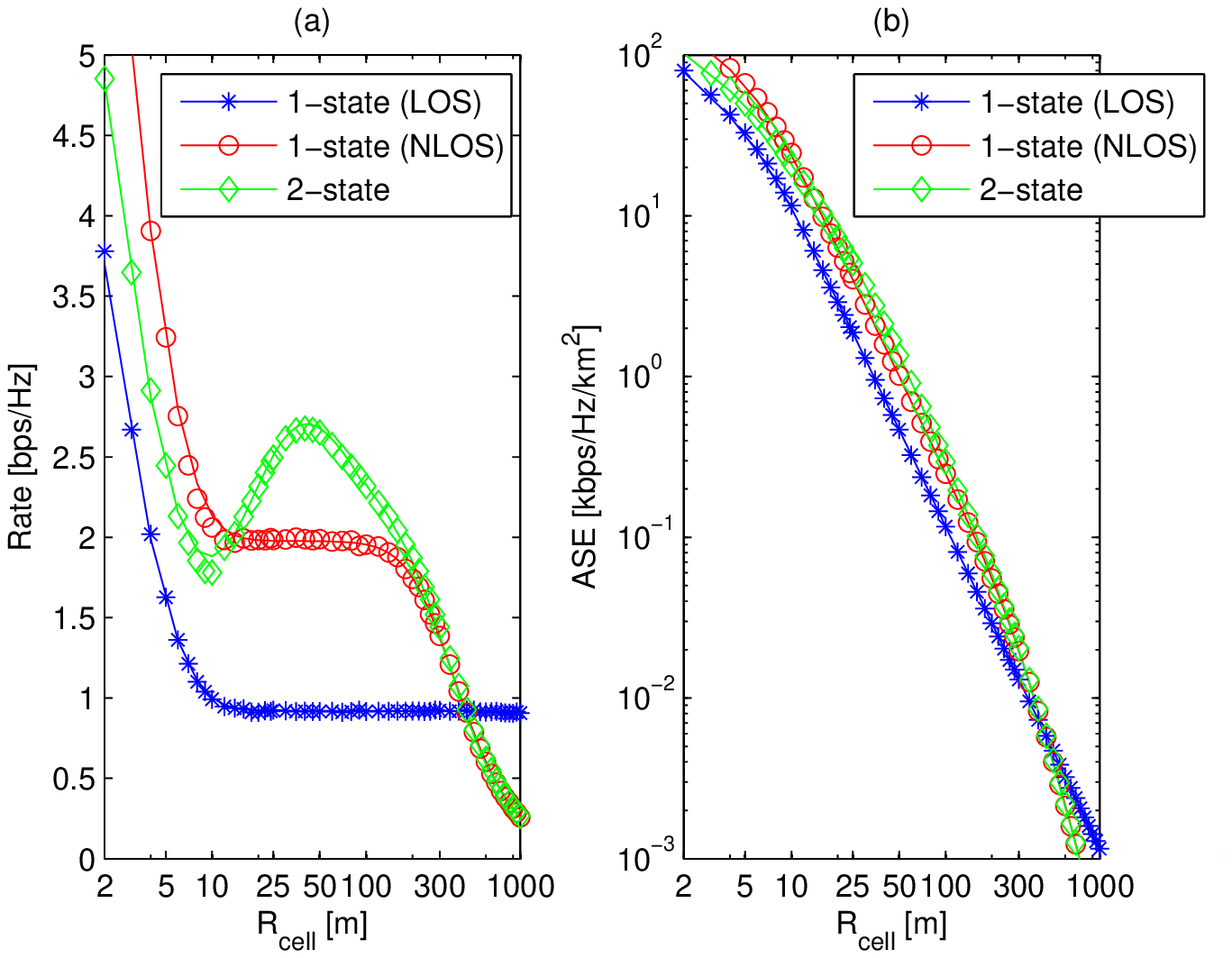}
\caption{${\mathcal{R}}/\ln \left( 2 \right)$ (a) and ASE (b) of one- and two-state blockage models. Markers: Monte Carlo simulations. Solid lines: IM-based approximation (three-ball). Setup: ``RS'' in Table \ref{Table_Setup}, PPP-distributed BSs, $N_{\rm{RB}} = 4$, $R_{\rm{MT}} = 3.9$ m, Omni antennas. \vspace{-0.5cm}} \label{Fig_0}
\end{minipage}%
\begin{minipage}{0.5\columnwidth}
\centering
\includegraphics[width=1.0\columnwidth]{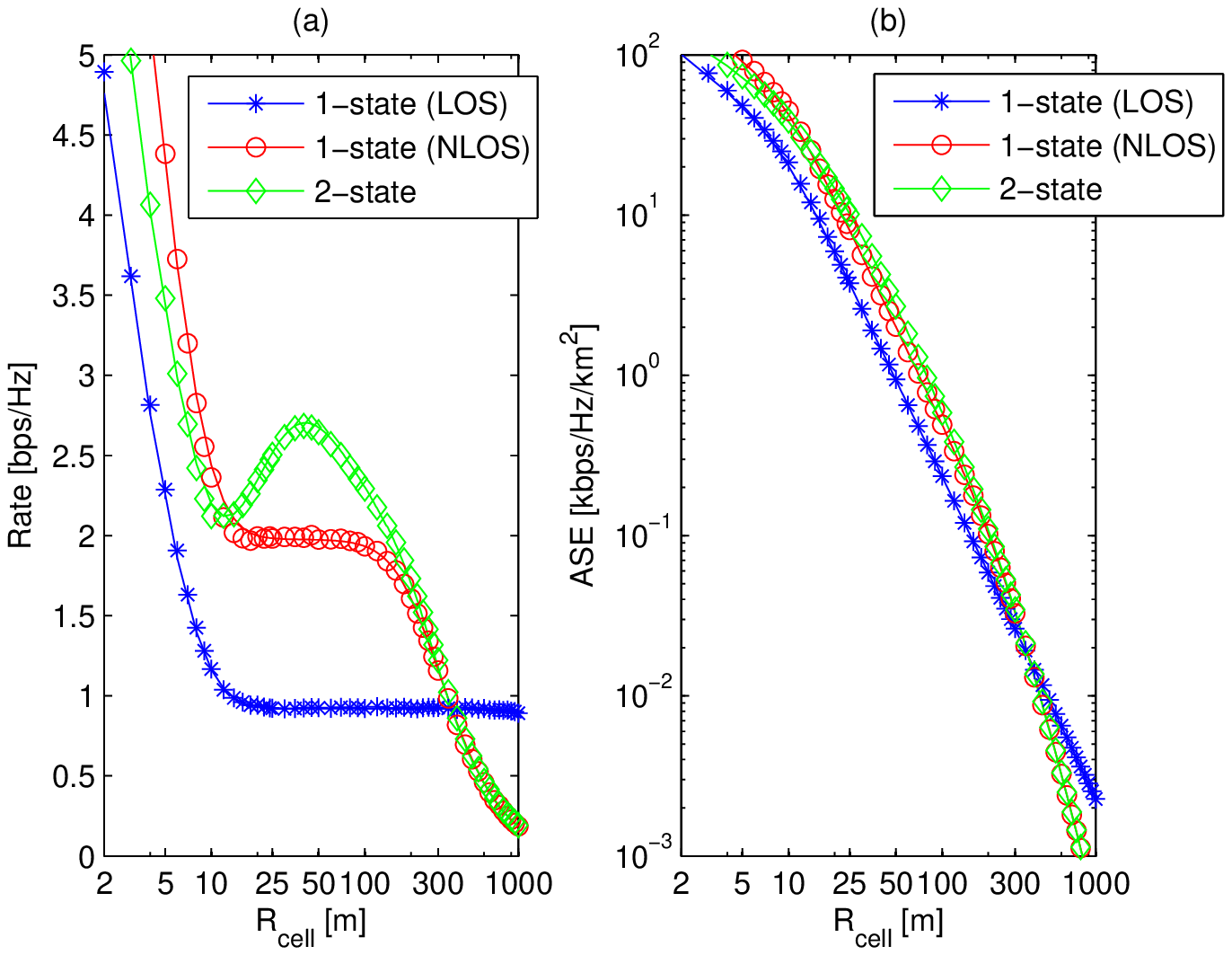}
\caption{${\mathcal{R}}/\ln \left( 2 \right)$ (a) and ASE (b) of one- and two-state blockage models. Markers: Monte Carlo simulations. Solid lines: IM-based approximation (three-ball). Setup: ``RS'' in Table \ref{Table_Setup}, PPP-distributed BSs, $N_{\rm{RB}} = 8$, $R_{\rm{MT}} = 3.9$ m, Omni antennas. \vspace{-0.5cm}} \label{Fig_00}
\end{minipage}
\end{figure}

\vspace{-0.25cm}
In Figs. \ref{Fig_0} and \ref{Fig_00}, rate and ASE of one-state (only LOS or NLOS links) and two-state (LOS and NLOS links) channel models are compared against each other. As far as the two-state channel model is concerned, the analytical results are obtained by using \eqref{Eq_32} and the approximation in Table \ref{Table_IM_3ball} (``RS''). As far as the one-state channel model is concerned, two case studies are considered: i) all links are in LOS and $\alpha = \alpha_{\rm{LOS}}$ and ii) all links are in NLOS and $\alpha = \alpha_{\rm{NLOS}}$, where $\alpha_{\rm{LOS}}$ and $\alpha_{\rm{NLOS}}$ are those in Table \ref{Table_Setup}. In this latter case, the analytical results are still obtained by using \eqref{Eq_32} and assuming $p_{\rm{LOS}}(r) = 1$ and $p_{\rm{NLOS}}(r) = 1$ for every $r$, respectively. Furthermore, PPP-distributed BSs are assumed. Figs. \ref{Fig_0} and \ref{Fig_00} highlight the importance of taking accurate blockage models into account. A far as the rate is concerned, we note that an optimal value of the density of BSs emerges if LOS and NLOS links are considered. As far as the ASE is concerned, we note that, depending on the operating regime (\textit{e.g.}, sparse vs. dense deployments), it may increase either sub-linearly or super-linearly as a function of the density of BSs.
\vspace{-0.5cm}
\subsection{Validation of the IM-based Approach Against the Empirical Dataset in \cite{ACM_MDR}} \vspace{-0.25cm}
In Fig. \ref{Fig_1}, rate and ASE obtained by using \eqref{Eq_32} and the approximations in Table \ref{Table_IM_3ball} (``Empirical'') and Table \ref{AntennaRadiationPatterns_Fitting} (antenna radiation pattern) are compared against system-level simulations of an actual deployment of BSs and buildings. The IM-based approach provides a good accuracy. Since $R_{{\rm{cell}}}  \approx 83.4$ m, $R_{{\rm{MT}}}  \approx 7.6$ m, $\widehat D_1  \approx 87.6$ m, the network operates close to its local optimum ($R_{{\rm{cell}}}  \approx \widehat D_1$) and it is in between a dense and a sparse regime. Figure \ref{Fig_1} shows that the rate is almost independent but slightly decreases with $N_{{\mathop{\rm RB}\nolimits} }$ and that the ASE increases with $N_{{\mathop{\rm RB}\nolimits} }$. This agrees with Table \ref{Table_SummaryTrends}. It confirms the important role played by the directivity of the antennas, in order to make the intended link stronger and to reduce the other-cell interference.

The rest of the figures are generated by assuming that the BSs are distributed according to a PPP. They, in fact, are aimed to illustrate the impact of $\lambda_{\rm{BS}}$ on rate and ASE, which, on the other hand, is fixed and given in \cite{ACM_MDR}. Different blockage models, however, are considered.
\vspace{-0.5cm}
\subsection{Validation of the IM-based Approach Against the RS and 3GPP Blockage Models} 
\begin{figure}[!t]
\setlength{\captionmargin}{10.0pt}
\setlength{\abovecaptionskip}{4pt}
\centering
\begin{minipage}{0.5\columnwidth}
\centering
\includegraphics[width=1.0\columnwidth]{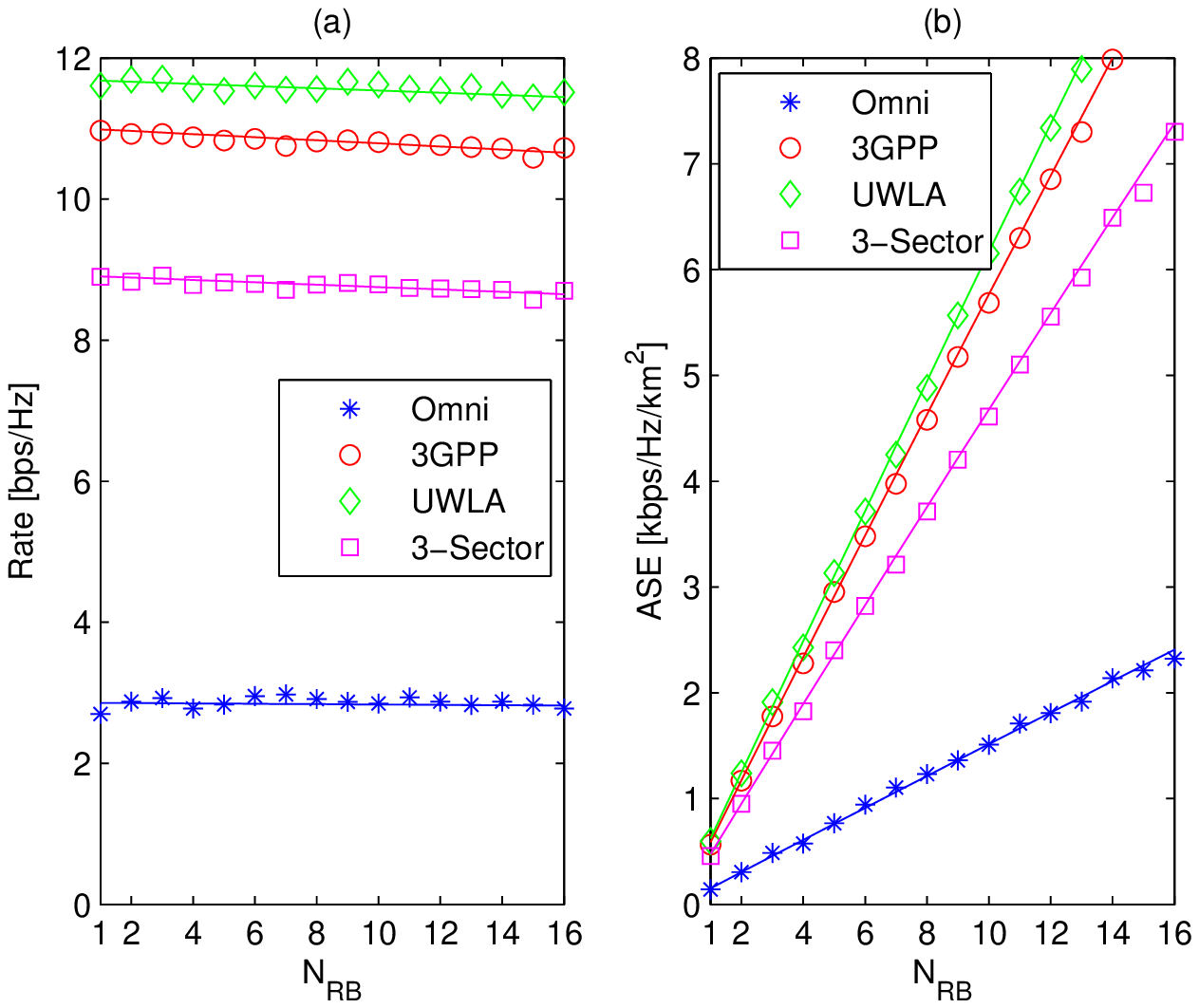}
\caption{${\mathcal{R}}/\ln \left( 2 \right)$ (a) and ASE (b). Markers: Monte Carlo simulations. Solid lines: IM-based approximation (three-ball). Setup: ``Empirical'' in Table \ref{Table_Setup}, $R_{\rm{MT}} = 7.6$ m. \vspace{-0.5cm}} \label{Fig_1}
\end{minipage}%
\begin{minipage}{0.5\columnwidth}
\centering
\includegraphics[width=1.0\columnwidth]{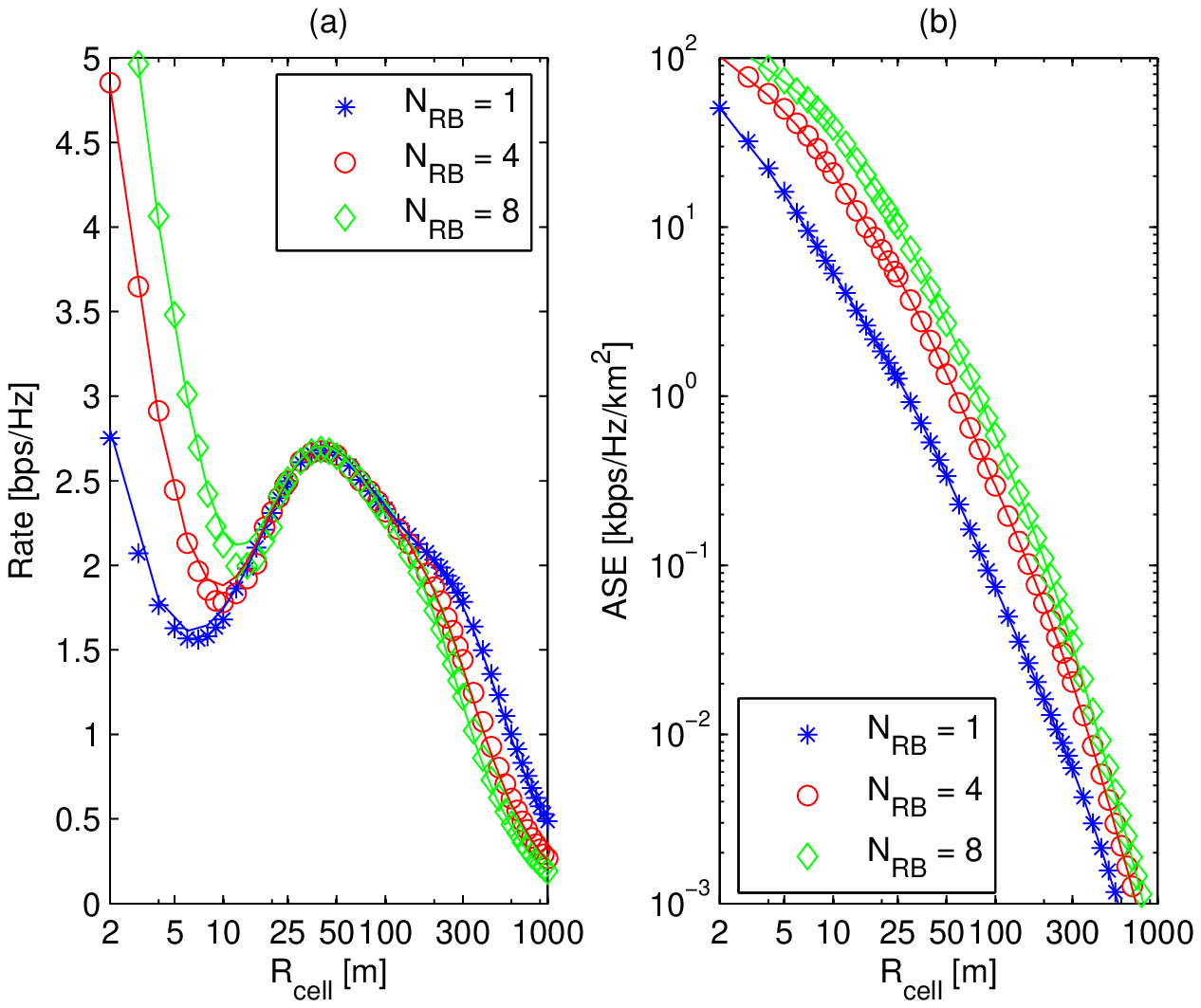}
\caption{${\mathcal{R}}/\ln \left( 2 \right)$ (a) and ASE (b). Markers: Monte Carlo simulations. Solid lines: IM-based approximation (three-ball). Setup: ``RS'' in Table \ref{Table_Setup}, PPP-distributed BSs, $R_{\rm{MT}} = 3.9$ m, Omni antennas. \vspace{-0.5cm}} \label{Fig_2}
\end{minipage}
\end{figure}

\begin{figure}[!t]
\setlength{\captionmargin}{10.0pt}
\setlength{\abovecaptionskip}{4pt}
\centering
\begin{minipage}{0.5\columnwidth}
\centering
\includegraphics[width=1.0\columnwidth]{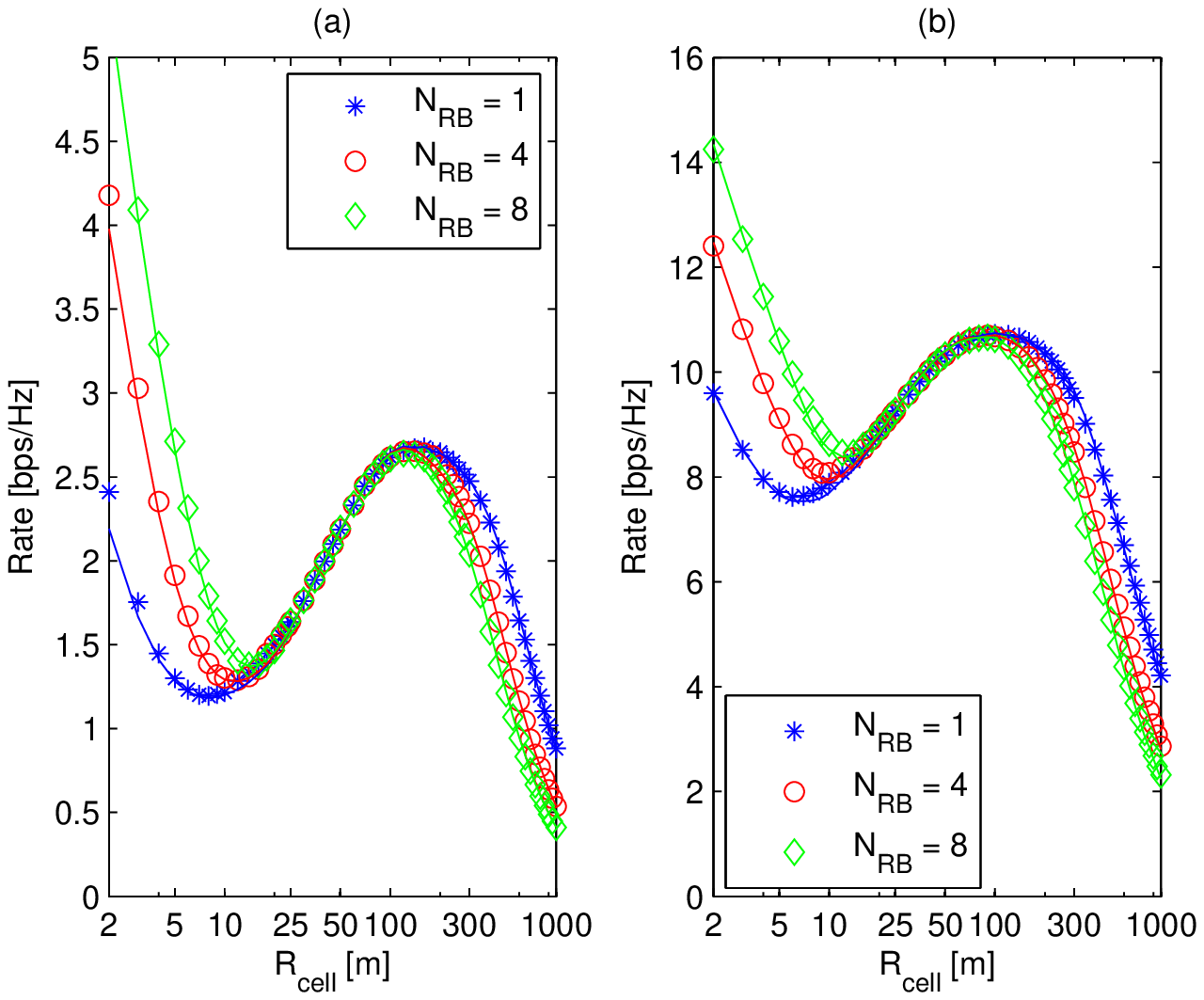}
\caption{${\mathcal{R}}/\ln \left( 2 \right)$ for Omni (a) and 3GPP (b) antennas. Markers: Monte Carlo simulations. Solid lines: IM-based approximation (three-ball). Setup: ``3GPP'' in Table \ref{Table_Setup}, PPP-distributed BSs, $R_{\rm{MT}} = 3.9$ m. \vspace{-0.5cm}} \label{Fig_3}
\end{minipage}%
\begin{minipage}{0.5\columnwidth}
\centering
\includegraphics[width=1.0\columnwidth]{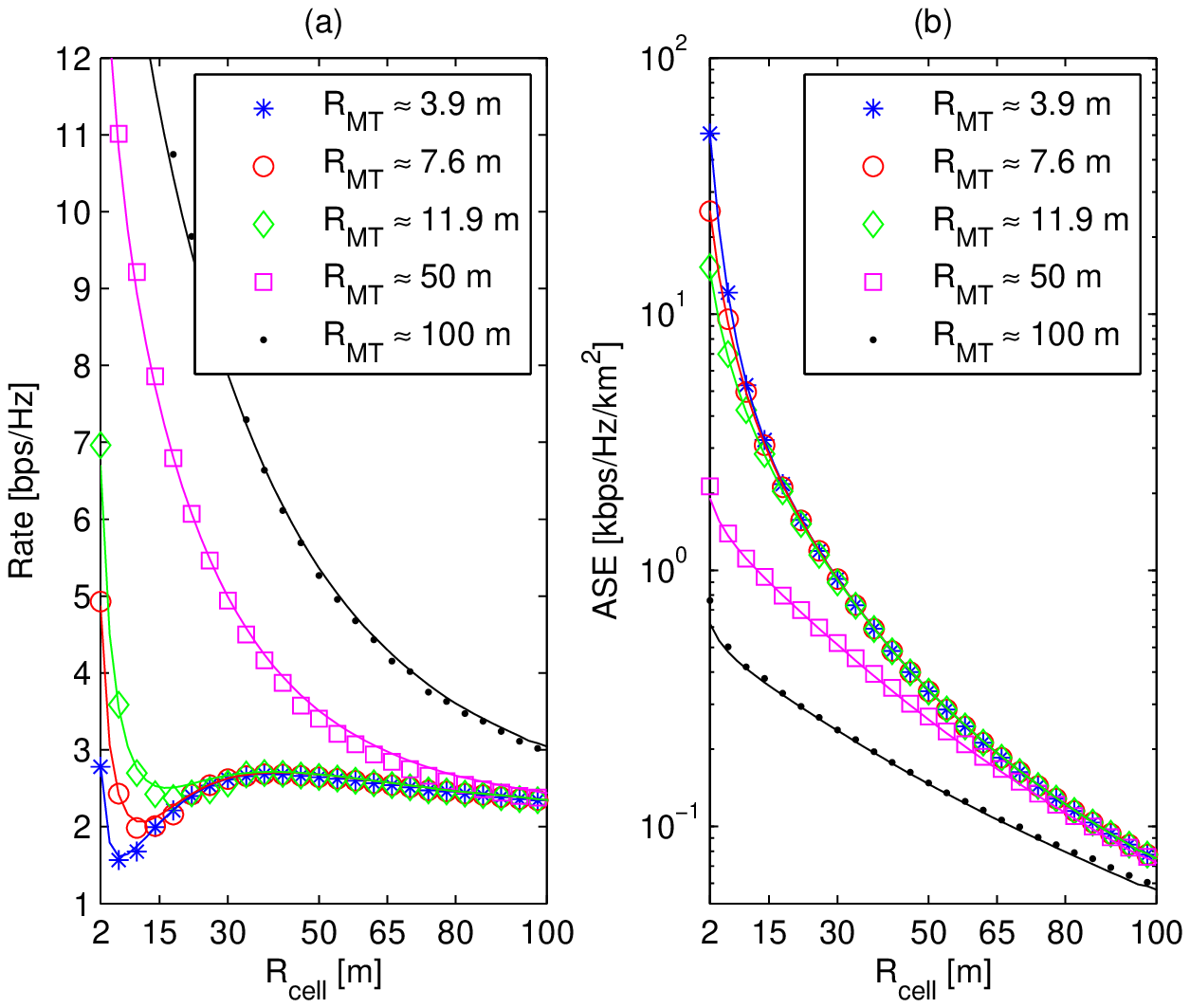}
\caption{${\mathcal{R}}/\ln \left( 2 \right)$ (a) and ASE (b). Markers: Monte Carlo simulations. Solid lines: IM-based approximation (three-ball). Setup: ``RS'' in Table \ref{Table_Setup}, PPP-distributed BSs, $N_{\rm{RB}} = 1$ m, Omni antennas. \vspace{-0.5cm}} \label{Fig_4}
\end{minipage}
\end{figure}

\begin{figure}[!t]
\setlength{\captionmargin}{10.0pt}
\setlength{\abovecaptionskip}{4pt}
\centering
\begin{minipage}{0.5\columnwidth}
\centering
\includegraphics[width=1.0\columnwidth]{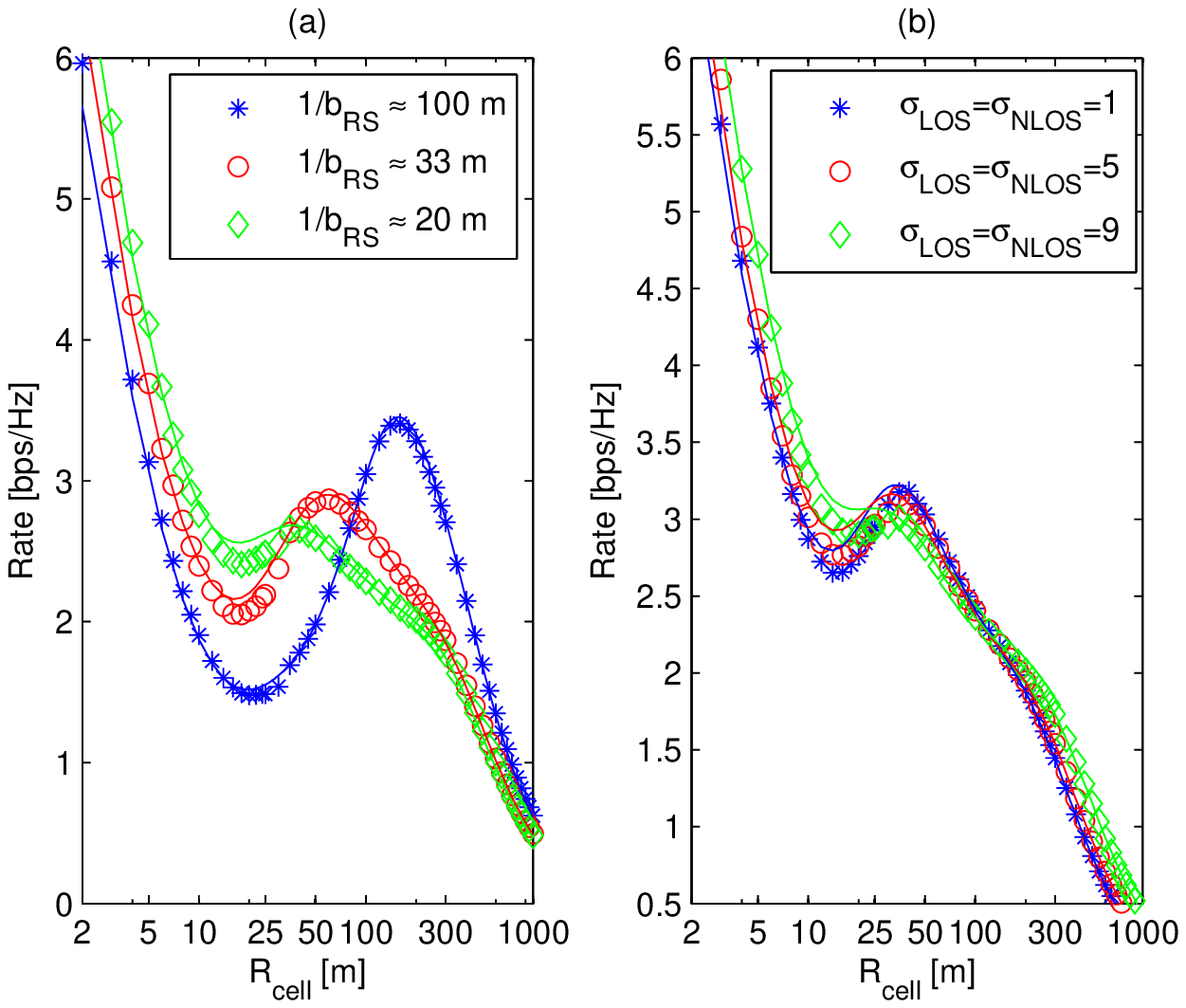}
\caption{${\mathcal{R}}/\ln \left( 2 \right)$ vs. density of blockages (a) and shadowing severity (b). Markers: Monte Carlo simulations. Solid lines: IM-based approximation (three-ball). Setup: ``RS'' in Table \ref{Table_Setup}, PPP-distributed BSs, $R_{\rm{MT}} = 11.9$ m, $N_{\rm{RB}} = 1$, Omni antennas. \vspace{-1.0cm}} \label{Fig_5}
\end{minipage}%
\begin{minipage}{0.5\columnwidth}
\centering
\includegraphics[width=1.0\columnwidth]{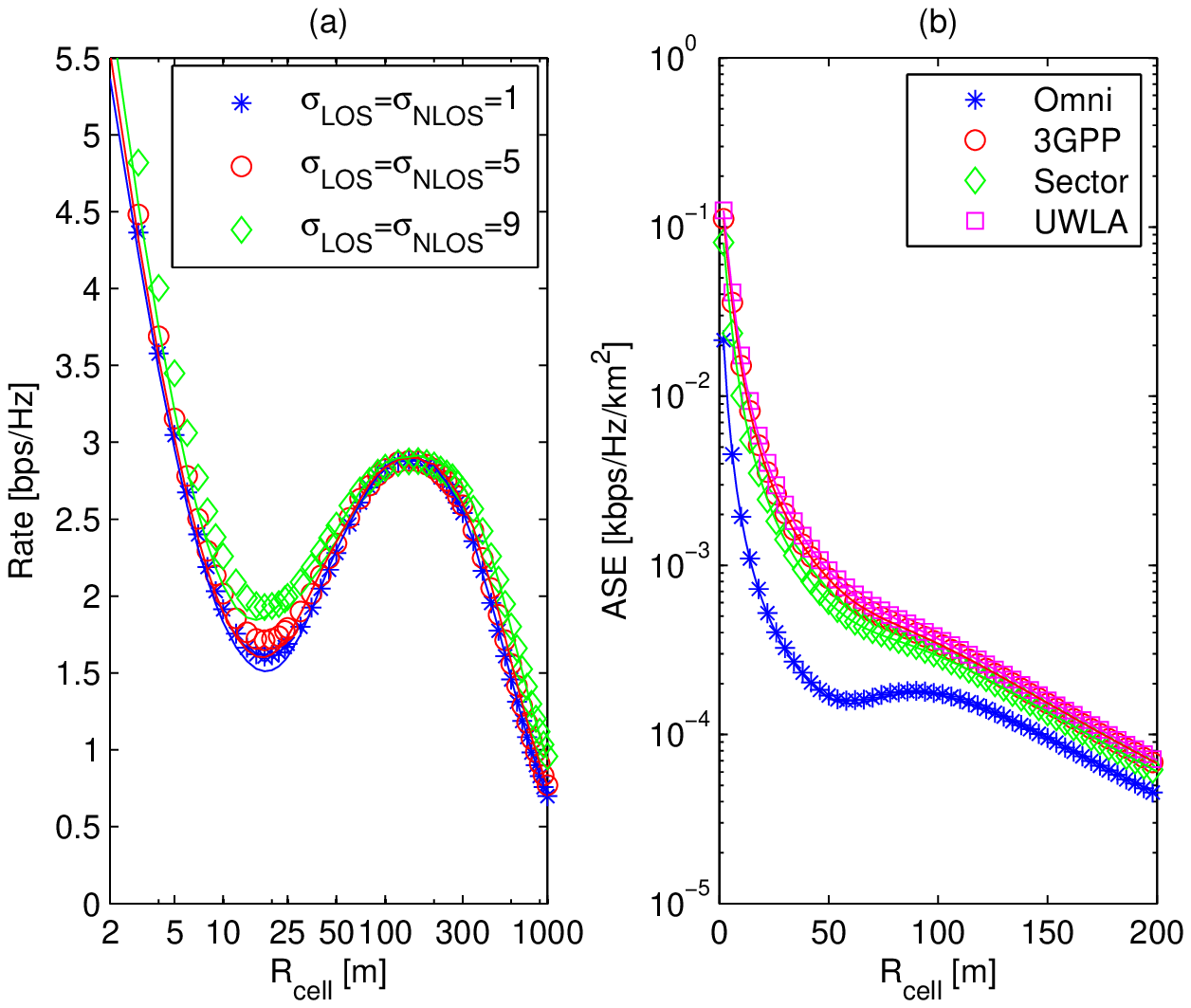}
\caption{(a) ${\mathcal{R}}/\ln \left( 2 \right)$ vs. shadowing severity. Markers: Monte Carlo simulations. Solid lines: IM-based approximation (three-ball). Setup: ``3GPP'' in Table \ref{Table_Setup}, PPP-distributed BSs, $R_{\rm{MT}} = 11.9$ m, $N_{\rm{RB}} = 1$, Omni antennas. (b) ASE for a special setup (see text). \vspace{-1.0cm}} \label{Fig_6}
\end{minipage}
\end{figure}

\vspace{-0.25cm}
In Figs. \ref{Fig_2} and \ref{Fig_3}, rate and ASE obtained by using \eqref{Eq_32} and the approximation in Table \ref{Table_IM_3ball} (``RS'' and ``3GPP'') are compared against system-level simulations. PPP-distributed BSs are assumed, as well as RS and 3GPP blockage models are considered, respectively. In both cases, the local minimum and maximum of the rate can be identified distinctly. The figures confirm that the local minimum is almost independent of the blockage model and increases with $N_{{\mathop{\rm RB}\nolimits} }$, while the local maximum increases with $\widehat D_1$, which, in turn, depends on the link state model. Qualitatively and quantitatively, the predictions in Section \ref{Trends_MinimumMaximum} are confirmed. Fig. \ref{Fig_3}(b) confirms that directive antennas significantly enhance the rate. In the dense regime, Fig. \ref{Fig_2}(b) shows that the ASE monotonically increases as $\lambda_{\rm{BS}}$ increases. This trend is preserved for all case studies based on the setup in Table \ref{Table_Setup}. In Fig. \ref{Fig_6}, a counter-example is shown, which highlights that, in the dense regime, the ASE may decrease as $\lambda_{\rm{BS}}$ increases. This confirms the unpredictability highlighted in Table \ref{Table_SummaryTrends} (see ``?'') and that the impact of some system parameters depends on the considered setup. We emphasize that we have analyzed several setups and that all the trends in Table \ref{Table_SummaryTrends} without ``?'' have been confirmed. This substantiates our mathematical analysis.
\vspace{-0.5cm}
\subsection{Validation of the IM-based Approach Against the Density of MTs} \vspace{-0.25cm}
In Fig. \ref{Fig_4}, rate and ASE obtained by using \eqref{Eq_32} and the approximation in Table \ref{Table_IM_3ball} (``RS'') are compared against system-level simulations. PPP-distributed BSs are assumed and the RS blockage model is considered. The impact of $\lambda_{\rm{MT}}$ on rate and ASE is in agreement with the predictions in Table \ref{Table_SummaryTrends}. In the rate, in particular, we note that the local minimum and the local maximum are not present if the condition ${{\lambda _{{\rm{MT}}} } \mathord{\left/ {\vphantom {{\lambda _{{\rm{MT}}} } {N_{{\rm{RB}}} }}} \right. \kern-\nulldelimiterspace} {N_{{\rm{RB}}} }} \gg {1 \mathord{\left/ {\vphantom {1 {\left( {\pi \widehat D_1^2 } \right)}}} \right. \kern-\nulldelimiterspace} {\left( {\pi \widehat D_1^2 } \right)}}$ is not satisfied. More precisely, they are present only if $R_{\rm{MT}} < \widehat D_1 = 38.7305$ m (see Table \ref{Table_IM_1ball}). This confirms the findings in Sections \ref{Trends_MinimumMaximum} and \ref{Trends_SystemLevelOptimization}. Furthermore, it is interesting to note that, by adopting a realistic load model, the ASE monotonically increases with both $\lambda_{\rm{MT}}$ and $\lambda_{\rm{BS}}$.
\vspace{-0.5cm}
\subsection{Validation of the IM-based Approach Against the Density of Blockages and Shadowing} \vspace{-0.25cm}
In Figs. \ref{Fig_5} and \ref{Fig_6}(a), the rate obtained by using \eqref{Eq_32} and the approximation in Table \ref{Table_IM_3ball} (``RS'' and ``3GPP'') is compared against system-level simulations. PPP-distributed BSs are assumed and the RS blockage model is considered. Both figures confirm the performance trends predicted in Table \ref{Table_SummaryTrends}. The density of blockages, in particular, has a noticeably different impact in dense and sparse regimes. By comparing Figs. \ref{Fig_5}(b) and \ref{Fig_6}(a), in the sparse regime, the unpredictable impact of the shadowing severity is confirmed: the rate decreases and increases as the shadowing standard deviation increases for RS and 3GPP link state models, respectively.
\vspace{-0.5cm}
\subsection{ASE in the Dense Regime: On the Unpredictable Impact of the Density of BSs} \vspace{-0.25cm}
In Fig. \ref{Fig_6}(b), we consider a special case study, which is aimed to show that, in the dense regime, the ASE may decrease as $\lambda_{\rm{BS}}$ increases. The following setup is considered: $R_{{\rm{MT}}}  = 3.9$ m, $\alpha _{{\rm{LOS}}}  = 2.01$, $\alpha _{{\rm{NLOS}}}  = 5.5$, $\sigma _{{\rm{LOS}}}  = \sigma _{{\rm{NLOS}}}  = 1$ dB, $\widehat D_1 = 29.5080$ m, $\widehat D_2 = 112.7958$ m, $\widehat D_3 = 397.2890$ m, $\widehat q_{{\rm{LOS}}}^{\left[ {0,\widehat D_1 } \right]} = 0.99$, $\widehat q_{{\rm{LOS}}}^{\left[ {\widehat D_1 ,\widehat D_2 } \right]} = 0.8711$, $\widehat q_{{\rm{LOS}}}^{\left[ {\widehat D_2 ,\widehat D_3 } \right]} = 0.0767$ and $\widehat q_{{\rm{LOS}}}^{\left[ {\widehat D_3 ,\infty } \right]} = 0$. This figure confirms that it may happen that the ASE decreases even if a practical load model is used (omni antennas setup). This is somehow in agreement with the findings in \cite{Lopez-Perez__LosNlos}, where no load is considered. We emphasize that in all the other case studies analyzed in the present paper, however, we have obtained that, for the considered load model, the ASE monotonically increases as $\lambda_{\rm{BS}}$ increases. The figure, in addition, confirms that the use of directional antennas provides a monotonic increase of the ASE, which is in agreement with the trends discussed in Section \ref{Trends_Antenna}. This confirms the benefits of densification under practical operating conditions.
\vspace{-0.5cm}
\section{Conclusion} \label{Conclusion} \vspace{-0.25cm}
In this paper, the IM-based approach has been introduced. It is a mathematically tractable approximation conceived for accurate system-level analysis of PPP-based cellular networks. The accuracy of the proposed approach has been substantiated with the aid of empirical data and for various blockage models. The approach is shown to provide insightful mathematical expressions for spectral efficiency and rate of cellular networks, and, in particular, several conclusions on the impact of network densification, blockage model and directivity of the antennas can be drawn.

Currently, the authors are working on the generalization of the proposed approach for application to more general load models, to non-PPP models for the locations of cellular BSs, to take into account spatial correlations originating from the presence of blockages, and to the design and optimization of inter-operator cloud radio access networks and resources sharing.
\end{document}